\documentclass{vldb}
\usepackage{listings}
\usepackage{epstopdf}

\usepackage{amsmath}
\usepackage{amssymb}
\usepackage{amsfonts}
\usepackage{epsfig}
\usepackage{graphicx}
\graphicspath{ {images/} }
\usepackage{subfigure}
\usepackage{url}
\usepackage[bottom]{footmisc}
\usepackage{balance}
\usepackage{algorithmic}
\usepackage{algorithm}
\usepackage{multirow}
\usepackage{xspace}
\usepackage{varwidth}
\usepackage{color}
\usepackage{times}
\usepackage{cite}
\usepackage{capt-of}
\usepackage{lipsum}
\usepackage{wrapfig,lipsum}

\newtheorem{exple}{Example}

\newcommand{\spara}[1]{\smallskip\noindent{\bf #1}}

\newcommand{\squishlist}{
 \begin{list}{$\bullet$}
  {  \setlength{\itemsep}{0pt}
     \setlength{\parsep}{3pt}
     \setlength{\topsep}{3pt}
     \setlength{\partopsep}{0pt}
     \setlength{\leftmargin}{2em}
     \setlength{\labelwidth}{1.5em}
     \setlength{\labelsep}{0.5em}
} }
\newcommand{\squishlisttight}{
 \begin{list}{$\bullet$}
  { \setlength{\itemsep}{0pt}
    \setlength{\parsep}{0pt}
    \setlength{\topsep}{0pt}
    \setlength{\partopsep}{0pt}
    \setlength{\leftmargin}{2em}
    \setlength{\labelwidth}{1.5em}
    \setlength{\labelsep}{0.5em}
} }

\newcommand{\squishdesc}{
 \begin{list}{}
  {  \setlength{\itemsep}{0pt}
     \setlength{\parsep}{3pt}
     \setlength{\topsep}{3pt}
     \setlength{\partopsep}{0pt}
     \setlength{\leftmargin}{1em}
     \setlength{\labelwidth}{1.5em}
     \setlength{\labelsep}{0.5em}
} }

\newcommand{\squishend}{
  \end{list}
}

\newcommand{\To}{\mbox{\bf to}\ }

\newcommand{\Each}{\mbox{\bf each}\ }

\newcommand{\Or}{\mbox{\bf or}\ }

\newcommand{\eat}[1]{}

\newcommand{\sharpP}{\ensuremath{\mathbf{\#P}}\xspace}

\newcounter{ccc}

\newcommand{\bigO}{\mathcal{O}}

\vldbTitle{An In-Depth Comparison of s-t Reliability Algorithms over Uncertain Graphs}
\vldbAuthors{X. Ke, A. Khan and L. Lim}
\vldbDOI{https://doi.org/TBD}
\vldbVolume{12}
\vldbNumber{xxx}
\vldbYear{2019}

\begin{document}

\title{An In-Depth Comparison of s-t Reliability Algorithms \\ over Uncertain Graphs}

\numberofauthors{3} 
%
\author{
\alignauthor Xiangyu Ke\\
\affaddr{NTU Singapore}\\
\email{xiangyu001@e.ntu.edu.sg}
\alignauthor Arijit Khan\\
\affaddr{NTU Singapore}\\
\email{arijit.khan@ntu.edu.sg}
\alignauthor Leroy Lim Hong Quan\\
\affaddr{NTU Singapore}\\
\email{llim031@e.ntu.edu.sg}
}


\maketitle

\begin{abstract}
Uncertain, or probabilistic, graphs have been increasingly used to represent noisy linked data in many emerging applications, and have recently attracted the attention of the database research community. A fundamental problem on uncertain graphs is the $s$-$t$ reliability, which measures the probability that a target node $t$ is reachable from a source node $s$ in a probabilistic (or uncertain) graph, i.e., a graph where every edge is assigned a probability of existence.

Due to the inherent complexity of the $s$-$t$ reliability estimation problem (\sharpP-hard), various sampling and indexing based efficient algorithms were proposed in the literature. However, since they have not been thoroughly compared with each other, it is not clear whether the later
algorithm outperforms the earlier ones. More importantly, the comparison framework, datasets, and metrics were often not consistent
(e.g., different convergence criteria were employed to find the optimal number of samples) across these works. We address
this serious concern by re-implementing six state-of-the-art $s$-$t$ reliability estimation methods in a common system
and code base, using several medium and large-scale, real-world graph datasets, identical evaluation metrics, and query workloads.

Through our systematic and in-depth analysis of experimental results, we report surprising findings, such as many follow-up algorithms can actually be several orders of magnitude inefficient, less accurate, and more memory intensive compared to the ones that were proposed earlier. We conclude by discussing our recommendations on the road ahead.
\end{abstract}

\vspace{-2mm}
\section{Introduction}
\label{sec:intro}
\emph{Uncertain graphs}, i.e., graphs whose edges are assigned a probability of existence, have attracted a great
deal of attention \cite{KC15,KYC18}, due to their rich expressiveness and given that uncertainty is inherent in the data in a wide range of applications,
including noisy measurements \cite{A09}, inference and prediction models \cite{AdarR07}, and explicit manipulation, e.g.,
for privacy purposes~\cite{Boldietal12}. A fundamental problem in uncertain graphs is the so-called \emph{reliability},
which asks to measure the probability that two given nodes are reachable \cite{AMG75}.
Reliability has been well-studied in the context of device networks, i.e.,
networks whose nodes are electronic devices and the (physical) links between such devices have a probability of failure \cite{AMG75}.
Recently, the attention has been shifted to other types of networks that can be represented as uncertain graphs, such as social and
biological networks \cite{PBGK10,JLDW11}. Specific problem formulations in this class ask to measure the probability that a certain reliability event occurs, e.g., what is the probability that two given nodes are connected (\emph{two-terminal reliability} \cite{AMG75}), all nodes in the network are pairwise connected (\emph{all-terminal reliability} \cite{SM09}), or all nodes in a given subset are pairwise connected (\emph{$k$-terminal reliability} \cite{HLL07}).
\begin{figure}[tb!]
\centering
\subfigure {
\includegraphics[scale=0.37]{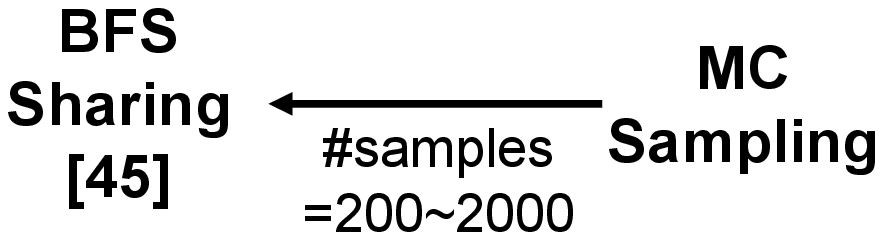}
}
$\quad$
\subfigure {
\includegraphics[scale=0.37]{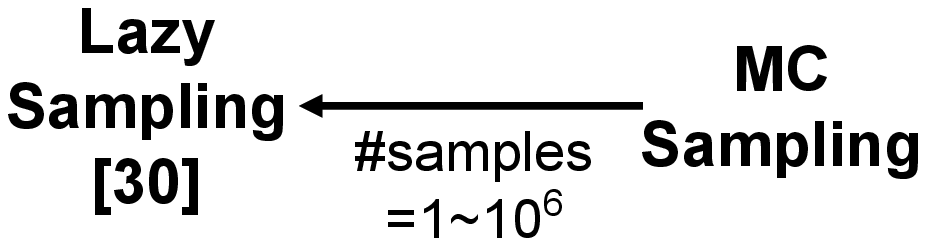}
}
\subfigure {
\includegraphics[scale=0.37]{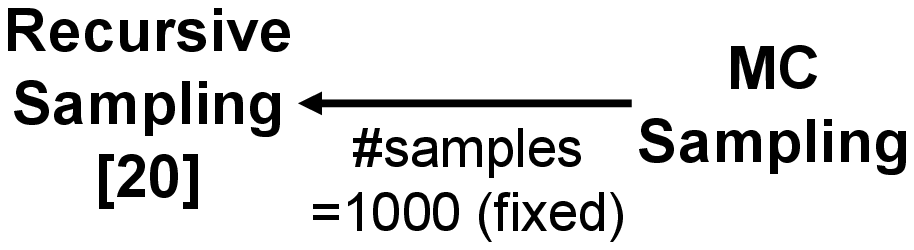}
}
$\quad$
\subfigure {
\includegraphics[scale=0.37]{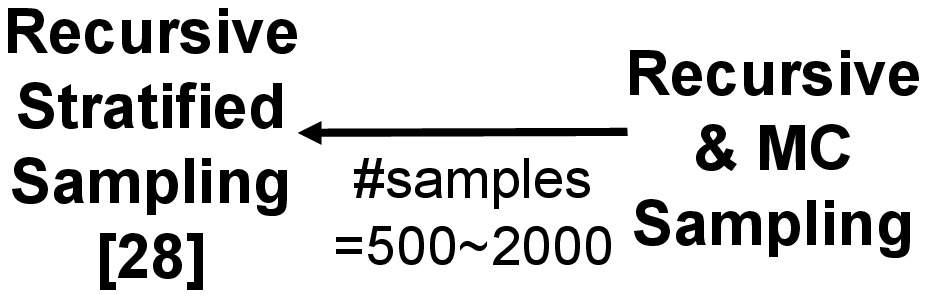}
}
\subfigure {
\includegraphics[scale=0.37]{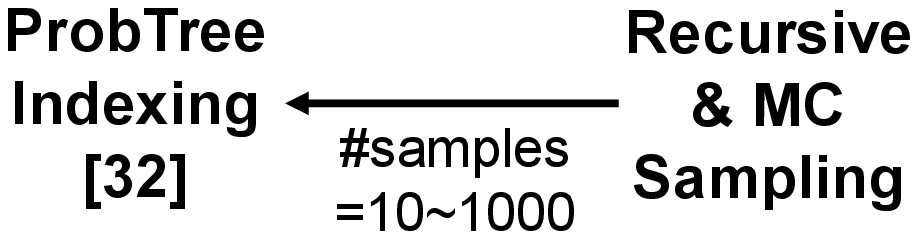}
}
\vspace{-4.5mm}
\caption{\small State-of-the-art reliability estimation algorithms in uncertain graphs: A directed
arrow depicts reported superiority in prior works. All algorithms have not been thoroughly compared
with each other. Moreover, previous works did not employ identical frameworks, datasets, and metrics for comparison. Thus, it is
critical to investigate their trade-offs and superiority over each other.}
\label{fig:comparison}
\vspace{-6mm}
\end{figure}
\begin{figure*}[tb!]
	\vspace{-4mm}
	\centering
	\includegraphics[scale=0.385]{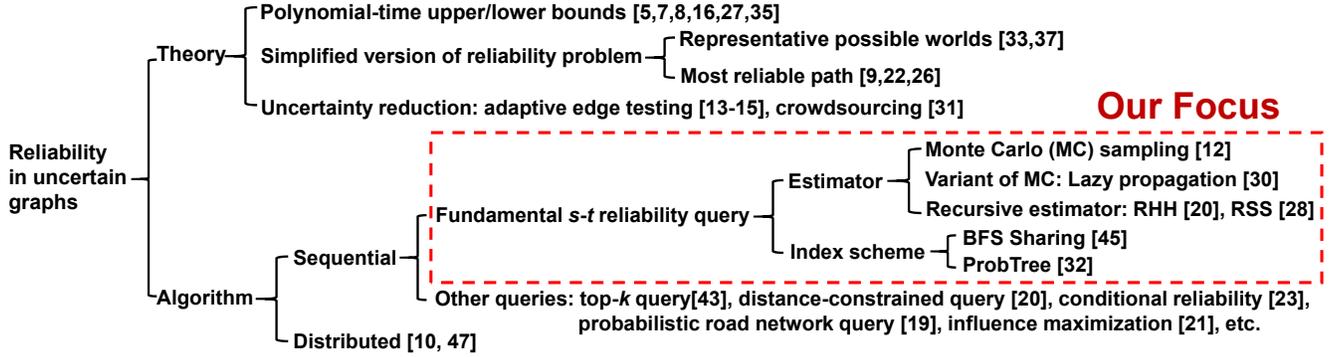}
	\vspace{-7mm}
	\caption{\small The broad spectrum of reliability problem over uncertain graphs}
	\label{fig:wp}
	\vspace{-4mm}
\end{figure*}

In this work, we shall investigate two-terminal reliability: The probability that a target node $t$ is reachable from
a source node $s$ in an uncertain graph, also denoted as the $s$-$t$ reliability.
This $s$-$t$ reliability estimation has been used in many applications such as measuring the quality of connections between two terminals
in a sensor network \cite{Ghosh2007Routing}, finding other proteins that are highly probable to be connected with a specific protein in a protein-protein
interaction (PPI) network \cite{JLDW11}, identifying highly reliable peers containing some file to transfer in a peer-to-peer (P2P) network,
probabilistic path queries in a road network \cite{HP10}, and evaluating information diffusions in a social influence network \cite{KBGN18}.

Due to the inherent complexity of the problem (\sharpP-hard) \cite{B86}, although the exact reliability detection has
received attention in the past \cite{AMG75}, the focus nowadays has mainly been on
approximate and heuristic solutions over large-scale graphs \cite{KC15}. The large spectrum of the reliability problem is categorized in Figure~\ref{fig:wp}. In this paper, we shall focus on {\em sequential algorithms for the fundamental $s$-$t$ reliability query}.
Notice that we would not consider distributed algorithms \cite{ZLLL17,ChengYCW15}, other simplified versions of the $s$-$t$ reliability problem \cite{PGPB14,SZL16,CWW10,KS06}, neither the reduction of uncertainty of a graph (e.g., by crowdsourcing) before $s$-$t$ reliability estimation \cite{FXPFW17,FFXPWL17,FWK14,LPCX17}. If a method was designed for a specific kind of reliability query (e.g. distance-constrained \cite{JLDW11}), it can be generalized to the fundamental $s$-$t$ query, thus we include the algorithm in our study \cite{JLDW11,ZZL15}. In particular, various sampling and indexing-based efficient algorithms were proposed in the literature. Estimation of reliability in uncertain graphs has its beginnings with the usage of Monte Carlo (MC)
sampling \cite{F86}. Subsequently, more advanced sampling methods were proposed in the form of recursive samplings \cite{JLDW11,LYMJ14}
and shared possible worlds \cite{ZZL15}, as well as other indexing methods \cite{ManiuCS17}. {\em With the wide range of algorithms available for
estimating the $s$-$t$ reliability over uncertain graphs, there is an urgent need to realize their trade-offs, and to employ the best algorithm
for a given scenario}.

As depicted in Figure~\ref{fig:comparison}, we find serious concerns in the existing experimental comparisons of
state-of-the-art reliability estimation algorithms over uncertain graphs. {\bf (1)} There is no prior work that compared all state-of-the-art
methods with each other. It is, therefore, difficult to draw a general conclusion on the superiority and trade-offs of different methods.
{\bf (2)} As shown in Figure~\ref{fig:comparison}, with the exception
of \cite{LFZT17}, other experimental studies in \cite{JLDW11,LYMJ14,ManiuCS17,ZZL15} either considered
a fixed number of samples (e.g., 1\,000), or the maximum number of samples was limited by 2\,000.
However, we observe in our experiments that the number
of samples necessary for the convergence of reliability estimation varies a lot depending on the
specific algorithm used (e.g., for {\em Recursive Stratified Sampling} \cite{LYMJ14}, \#samples required for convergence is 250$\sim$1\,000,
while for {\em Lazy Propagation} \cite{LFZT17}, it is 500$\sim$1\,500), and also on the underlying characteristics of the uncertain graph dataset.
Therefore, the running time necessary to achieve convergence (and hence, good-quality results) should be reported differently,
as opposed to using the same number of samples in all experiments.
{\bf (3)} The metrics used for empirically comparing these techniques were
not consistent in the past literature, thereby making them {\em apple-to-orange} comparisons in the larger context.
For example, \cite{LYMJ14} measured relative variance of different estimators and their running times for the same number
of samples (2\,000). In both \cite{ZZL15,ManiuCS17}, the authors reported accuracy (with respect to baseline MC sampling)
and running times of different algorithms using a maximum of 1\,000 samples. On the other hand, \cite{JLDW11}
compared relative variance, accuracy, and running times of various estimators by considering a fixed (1\,000) number
of samples. In addition, surprisingly none of these studies reported the online memory usage which, according to our experiments,
varied a great extent. {\bf (4)} Last but not least, we find certain errors and further optimization scopes
in past algorithms (e.g., accuracy of \cite{LFZT17}, time complexity analysis of \cite{ZZL15}),
and by correcting (or, updating) them we significantly improve their performance.

\spara{Our contribution and roadmap.} Our contributions can be summarized
as follows.
\vspace{-2mm}
\begin{itemize}
\setlength\itemsep{0.001em}
\item We investigate the $s$-$t$ reliability estimation problem and summarize six state-of-the-art sequential algorithms in Section \ref{sec:algorithms},
together with their time complexity and sampling variance.
\item We correct certain issues in past algorithms (e.g., accuracy of \cite{LFZT17},
time complexity analysis of \cite{ZZL15}), which significantly improve their performance (Section \ref{sec:algorithms}).
\item We implemented five state-of-the-art algorithms \cite{F86,ZZL15,JLDW11,LYMJ14,LFZT17} in C++, and obtained C++ source code
of \cite{ManiuCS17} from respective authors. We compare them in a common environment, using same convergence criteria, and
present empirical comparisons of six $s$-$t$ reliability estimation algorithms over six real-world, uncertain graph datasets in Section \ref{sec:experiments}.
 {\em Our datasets and source code are available at: https://github.com/\\5555lan/RelComp}.
\item We report the accuracy, efficiency, and memory usage of six referred methods both at convergence and at \#samples= 1\,000, summarize their trade-offs, and provide guidelines
for researchers and practitioners (Sections \ref{sec:experiments} and \ref{sec:conclusion}).
\end{itemize}

\vspace{-3mm}
\section{Reliability Estimation Methods}
\label{sec:algorithms}
\vspace{-3mm}
\subsection{\lowercase{s-t} Reliability in Uncertain Graphs}
\label{sec:reliability_Def}
An uncertain graph $\mathcal{G}$ is a triple $(V, E, P)$, where $V$ is a set of $n$
nodes, $E \subseteq V \times V$ is a set of $m$ directed edges, and
$P: E \rightarrow (0,1]$ is a probability function that assigns a probability of existence
to each edge in $E$.

The bulk of the literature on uncertain graphs and device networks reliability assumes the existence of
the edges in the graph independent from one another, and interprets uncertain graphs according to the
well-known {\em possible-world semantics}~\cite{JLDW11,PBGK10,ZZZZL11,BD94,B85,PB83,HLL07,SM09}:
an uncertain graph $\mathcal{G}$ with $m$ edges yields $2^m$ possible deterministic graphs,
which are derived by sampling independently each edge $e \in E$ with probability $P(e)$.
More precisely, a possible graph $G \sqsubseteq \mathcal{G}$ is a pair $(V, E_{G})$,
where $E_{G} \subseteq E$, and its sampling probability is:
\vspace{-2.5mm}
\begin{equation}\label{equ:sampling_probability}
Pr(G) = \prod_{e \in E_{G}} P(e) \prod_{e \in E \setminus E_{G}} (1 - P(e))
\end{equation}
For a possible deterministic graph $G$, we define an indicator function
$I_{G}(s,t)$ to be 1 if there is a path in $G$ from
a source node $s \in V$ to a target node $t \in V$, and 0 otherwise.
The probability that $t$ is reachable from $s$ in the uncertain graph $\mathcal{G}$, denoted by $R(s,t)$, is computed as:
\vspace{-3mm}
\begin{equation}
\label{eq:R}	
R(s,t) = \sum_{G \sqsubseteq \mathcal{G}} I_{G}(s,t) \, Pr(G)
\end{equation}
The number of possible worlds $G \sqsubseteq \mathcal{G}$ is exponential in the number of edges,
which makes the exact computation of $R(s,t)$ infeasible even for modestly-sized networks.
In fact, the $s$-$t$ reliability computation is a prototypical \sharpP-complete problem~\cite{B86,V79}.

Due to intrinsic hardness, we tackle the reliability estimation problem from approximation and
heuristic viewpoints. In particular, we shall examine six sampling and indexing-based efficient algorithms
that were proposed in recent literature as follows:
{\bf (1)} Monte Carlo (MC) sampling \cite{F86}; {\bf (2)} Indexing via BFS sharing \cite{ZZL15};
{\bf (3)} Recursive sampling \cite{JLDW11}; {\bf (4)} Recursive stratified sampling \cite{LYMJ14};
{\bf (5)} Lazy propagation sampling \cite{LFZT17}; and {\bf (6)} Indexing via probabilistic trees \cite{ManiuCS17}.
\begin{algorithm}[tb!]
	\caption{\small MC Sampling with BFS}
	\label{algo:montecarlooptimised}
	\begin{algorithmic}[1]
		\REQUIRE source node $s$ and target node $t$ in uncertain graph $\mathcal{G}=(V,E,P)$, \#samples=$K$
		\ENSURE  reliability $R(s,t)$
		\STATE $R(s,t)\gets0$, $i\gets0$
		\WHILE{$i<K$}
		\STATE set all nodes not visited, $Q \gets \phi$
		\STATE $Q$.enqueue$(s)$     $\qquad$ /\/* set of visited nodes */\
		\STATE mark $s$ as visited
		\IF {$t=s$}
		\STATE $R(s,t)\gets R(s,t)+1$
		\STATE goto line 27    $\quad$ /\/* early stop at current round */\
		\ENDIF
		\WHILE {$Q$ not empty}
		\STATE $v$ = $Q$.dequeue$()$
		\FOR {\Each outgoing edge $e$ from $v$}
		\STATE sample $e$ according to $P(e)$
		\IF {$e$ exists after sampling}
		\STATE $w \gets$ target node of $e$
		\IF {$w$ not visited}
		\STATE $Q$.enqueue$(w)$
		\STATE mark $w$ as visited
		\IF {$t=w$}
		\STATE $R(s,t)\gets R(s,t)+1$
		\STATE goto line 27 $\quad$ /\/* early stop at current round */\
		\ENDIF
		\ENDIF
		\ENDIF
		\ENDFOR
		\ENDWHILE
		\STATE $i\gets i+1$
		\ENDWHILE
		\RETURN $R(s,t)/K$;
	\end{algorithmic}
\end{algorithm}

\vspace{-3mm}
\subsection{MC Sampling}
\label{sec:mc_sampling}
In the basic Monte Carlo (MC) sampling, we first sample $K$ possible worlds $G_1,G_2, \ldots, G_K$
of the uncertain graph $\mathcal{G}$ according to independent edge probabilities.
We then compute the reachability in each sampled graph $G_i$, and define $I_{G_i}(s,t)=1$ if $t$
is reachable from $s$ in $G_i$, and $0$ otherwise. Given this, we have the MC sampling estimator:
\vspace{-3mm}
\begin{equation}
\label{eq:MC_reliability}	
\displaystyle R(s,t) \approx \hat{R}(s,t) = \frac{1}{K}\sum_{i=1}^K I_{G_i}(s,t)
\vspace{-2.5mm}
\end{equation}
This is also known as the hit-and-miss Monte Carlo. The basic sampling estimator $\hat{R}(s,t)$ is an {\em unbiased estimator} of the
$s$-$t$ reliability, i.e., $E(\hat{R}(s,t)) = R(s,t)$, and its variance can be determined due to Binomial distribution $\sim B(K, R(s,t))$ \cite{F86,JLDW11}.
\vspace{-2.5mm}
\begin{eqnarray}
\label{eq:MC_variance}	
\displaystyle Var\left(\hat{R}(s,t)\right) &=& \frac{1}{K}\cdot R(s,t)\cdot \left(1-R(s,t)\right) \nonumber \\
&\approx& \frac{1}{K}\cdot \hat{R}(s,t)\cdot \left(1-\hat{R}(s,t)\right)
\vspace{-2.5mm}
\end{eqnarray}
It is possible to derive bounds on the number of MC samples needed to provide a good estimate for the $s$-$t$ reliability problem.
It was shown in \cite{PBGK10} by applying the
Chernoff bound that with number of samples $K \ge \frac{3}{\epsilon^2R(s,t)}\ln\left(\frac{2}{\lambda}\right)$, we can ensure the following.
\vspace{-2mm}
\begin{equation}
\label{eq:MC_accuracy}	
\displaystyle Pr\left(\left|\hat{R}(s,t)-R(s,t)\right|\ge\epsilon R(s,t)\right) \le \lambda
\end{equation}
The time complexity to generate $K$ possible worlds is $\bigO(mK)$. In each possible world, the reachability can be
determined by performing a breadth-first search (BFS) from
the source node. Each BFS requires $\bigO(m+n)$ time. Therefore, the overall time complexity of MC sampling based
reliability estimation is $\bigO(K(m+n))$. In essence, one may combine MC sampling with BFS from the source node
for improved efficiency \cite{JLDW11,KBGG14}. It means that an edge in the current possible world
is sampled only upon request. This avoids sampling of many edges in parts of the graph that are not reached
with the current BFS, thus increasing the chance of an early termination (lines 8, 21).
Our pseudocode is given in Algorithm~\ref{algo:montecarlooptimised}.

In practise, MC sampling can be inefficient over large-scale networks due to two reasons.
\vspace{-2mm}
\begin{itemize}
\setlength\itemsep{0.01em}
\item For each $s$-$t$ reliability query, we need to generate $K$ possible worlds via sampling.
Based on empirical evidences from state-of-the-art works \cite{PBGK10,JLDW11,KKT03}
as well as according to our own experimental results, $K$ can be in the order of thousands to
achieve a reasonable accuracy. However, as correctly pointed out in \cite{ZZL15,PGPB14}, this
sampling procedure does not contribute to the reliability estimation process directly.
For example, one can pre-compute these $K$ possible worlds in an offline manner
to further improve the efficiency of online $s$-$t$ reliability estimation.
\item There could be a significant overlap in structures of different possible worlds
\cite{ZZL15,JLDW11,ZZZZL11}. Unfortunately, the reliability estimation via basic MC sampling
performs a separate BFS over each possible world, therefore it cannot take advantage of the common
substructure across various possible worlds.
\end{itemize}
\vspace{-1mm}
\subsection{Indexing via BFS Sharing}
\label{sec:bfs_sharing}
\begin{figure}
\vspace{-2mm}
\centering
\includegraphics[scale=0.3]{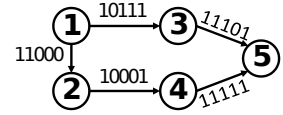}
\vspace{-5mm}
\caption{\small Compact structure with five possible worlds: BFS sharing}
\label{fig:bfscompactstructure}
\vspace{-5mm}
\end{figure}

Zhu et al. \cite{ZZL15} developed an offline sampling method to generate $K$ possible worlds: $G_1, G_2, \ldots, G_K$.
In order to minimize the storage overhead, they proposed a bit-vector based compact structure, as depicted in
Figure~\ref{fig:bfscompactstructure}. It essentially stores only one graph $G = (V,E)$ with the same set of  nodes and edges
as the input uncertain graph $\mathcal{G}$. However, each edge $e$ in $G$ has a bit-vector of size $K$ --- its $i$-th
bit represents whether the edge $e$ is present in the sampled graph $G_i$ or not.
\begin{algorithm}[tb!]
\caption{\small BFS Sharing}
\label{algo:bfssharing}
\begin{algorithmic}[1]
\REQUIRE source node $s$, target node $t$ in uncertain graph $\mathcal{G}=(V,E,P)$,
$K$ possible worlds $G_1, G_2, \ldots, G_K$ computed offline and stored in compact format (with bit vectors)
\ENSURE  reliability $R(s,t)$
\STATE  $U \gets \emptyset$  $\qquad$ /\/* set of visited nodes */\
\STATE  $U\gets \{s\}$
\STATE $\bf I_s \gets \textbf{1}$
\STATE $\bf I_t \gets \textbf{0}$
\STATE worklist $\gets \emptyset$
\STATE worklist.enqueue(all out-neighbors of $s$)
\WHILE {worklist is not empty}
\STATE $v \gets$ worklist.dequeue()
\IF {$v\in U$}
\STATE  goto line 28
\ENDIF
\IF {$v\not\in U$}
\STATE  $U\gets$ $U\cup\{v\}$
\STATE $\bf I_v \gets \textbf{0}$
\ENDIF
\FOR {\Each in-neighbor $in$ of $v$}
\IF {$in\in U$}
\STATE $\bf I_v = \bf I_v$ OR $(\bf I_{in}$ AND $\bf I_{e_{(in,v)}})$
\ENDIF
\ENDFOR
\FOR {\Each out-neighbor $out$ of $v$}
\IF {$out\not \in U$}
\STATE worklist.enqueue($out$)
\ELSE
\STATE update($v$, $out$, $U$)
\ENDIF
\ENDFOR
\ENDWHILE
\RETURN $\frac{||\bf I_t||_1}{K}$
\end{algorithmic}
\end{algorithm}

\begin{algorithm}[tb!]
	\caption{\small update($v$, $u$, $U$)}
	\label{algo:bfssharingupdate}
	\begin{algorithmic}[1]
		\STATE $\bf I_u \gets \bf I_u$ OR ($\bf I_v$ AND $\bf I_{e_{(v,u)}}$)
		\STATE mark $u$ as updated
		\STATE $Q$.enqueue$(u)$
		\WHILE {$Q \neq \emptyset$}
		\STATE $w \gets Q$.dequeue$()$;
		\FOR {\Each outgoing visited neighbor $x \in U$ of $w$}
		\IF {$x$ is not marked as updated}
		\STATE $\bf I_x^{'} \gets \bf I_x$ OR $(\bf I_w$ AND $\bf I_{e_{(w,x)}})$
		\STATE mark $x$ as updated
		\IF {$\bf I_x^{'} \neq \bf I_x$}
		\STATE $\bf I_x \gets \bf I_x^{'}$
		\STATE $Q$.enqueue$(x)$
		\ENDIF
		\ENDIF
		\ENDFOR
		\ENDWHILE
	\end{algorithmic}
\end{algorithm}

Given an $s$-$t$ reliability query, \cite{ZZL15} performs BFS over this compact graph structure (Algorithm~\ref{algo:bfssharing}),
which is equivalent to doing BFS traversals in parallel across the pre-computed possible worlds.
We attach an additional bit vector $I_v$ with each node $v$ that keeps track of the possible worlds
in which $v$ is reachable from $s$. Initially, $I_s=[1\,1\,\ldots\,1]$ and $I_v=[0\,0\,\ldots\,0]$
for all $v \ne s$. Let us also denote by $U$ the set of visited nodes based on BFS. Initially
$U=\{s\}$ (line 3). At each step, when we find an unexplored node $v$ that is an out-neighbor of at least one node
$u$ in $U$, we insert $v$ into $U$. We update the bit vector $I_v$ to include the possible worlds
where all such $v$'s are reachable from $s$ (line 18). Before proceeding to the next step of BFS, one may note
that if $v$ has some out-neighbor $w$ that is already in $U$, we may need to update $I_w$ (line 25).
Specifically, let $G_i$ be a possible world that is currently in $I_v$, but not in $I_w$.
Then, we should also include $G_i$ in $I_w$. Such changes in
$I_w$ may in turn affect $I_z$, where $z$ is an out-neighbor of $w$, and $z$ is also in $U$.
In general, we proceed to the next step of BFS only after finishing these
cascading updates. This updating procedure is demonstrated in Algorithm~\ref{algo:bfssharingupdate}.
Finally, the number of $1$'s in $I_t$, divided by $K$, provides
the MC estimation of $s$-$t$ reliability.

Clearly, BFS sharing has the same variance as the basic MC sampling. However,
by generating the $K$ possible worlds in an offline manner, \cite{ZZL15} reduces the online $s$-$t$ reliability estimation time.
On the other hand, {\em unlike the MC sampling approach described in Algorithm~\ref{algo:montecarlooptimised}, in this case
no early termination of BFS is possible}, even when the target node $t$ is reached earlier in Algorithm~\ref{algo:bfssharing}.
This is due to performing the required cascading updates (Algorithm~\ref{algo:bfssharingupdate}) in this method.
{\em It often makes BFS Sharing even more time consuming than MC sampling, which is evident in our experimental results}.

We note that the original algorithm in \cite{ZZL15} was developed to identify the top-$k$ target nodes having
the maximum reliability from a given source node $s$. However, as we discussed in Algorithm~\ref{algo:bfssharing}, one can
trivially update their technique to estimate the reliability of a given $s$-$t$ pair.

\spara{Our correction in complexity analysis.} The offline index building time complexity is $\bigO(Km)$, and index storage space is $\bigO(n+Km)$,
where $\bigO(Km)$ is the storage due to edge bit vectors. We load all edge bit vectors in memory during online query processing
to improve efficiency. Besides, $n$ node bit vectors of size $\bigO(Kn)$ are created during online query processing.
BFS sharing method has online reliability estimation time complexity $\bigO(K(m+n))$: Due to cascading updates (Algorithm~\ref{algo:bfssharingupdate}),
each node and edge can be visited at most $K$ times. Note that in the original paper \cite{ZZL15}, it was stated that the online
query time is independent of $K$. As we reasoned, however, the running time increases linearly with $K$ due to cascading updates.
This is also confirmed by our empirical results (e.g., notice in Tables~\ref{tab:EffHepU}, \ref{tab:EffDBLP02}, \ref{tab:EffDBLP005}, \ref{tab:EffBio},
the running times of BFS sharing increases for larger $K$ over the same dataset).
\vspace{-1mm}
\subsection{Recursive Sampling}
\label{sec:recursive_sampling}
Recursive sampling, which was proposed in \cite{JLDW11}, improves on MC sampling by considering the two following factors.
\vspace{-1.5mm}
\begin{itemize}
\setlength\itemsep{0.001em}
\item When some edges are missing in a possible world, the presence of other edges might no longer
be relevant with respect to certain $s$-$t$ reliability queries. Hence, those edges can be skipped
from sampling and query evaluation process.
\item Many possible worlds share a significant portion of existing or missing edges. Hence, the reachability
checking cost could be shared among them.
\end{itemize}
The basic approach, which follows a divide-and-conquer technique, is given below.
A very similar algorithm, called the Dynamic MC sampling, was developed in \cite{ZZZZL11}.

We start from the source node $s$, and say that $s$ is already {\em reached}.
An edge $e$ is {\em expandable} if it starts from a {\em reached} node. We randomly pick
an extendable edge $e$, then sample the existence of $e$ for $K$ iterations. The next step is to divide the
samples into two groups: one group with $e$ existing and another group with $e$ not existing. In the first group,
we may reach a new node $w$ via $e$, and in that case, more edges become {\em expandable}.
For both groups, we repeat the process of picking a random {\em expandable} edge, sampling its existence, and dividing
the group into smaller batches.

Formally, assume that $E_1\subseteq E$ be the set of included edges and $E_2 \subseteq E$ be the set of
not-included edges in one group (referred to as a {\em prefix group} in \cite{JLDW11})
at some intermediate stage of our method. Let us denote this group by $\mathcal{G}(E_1,E_2)$, i.e.,
the set of possible worlds of $\mathcal{G}=(V,E,p)$ which contains all edges in $E_1$,
and no edges in $E_2$. Clearly, $E_1\cup E_2 \subseteq E$ and $E_1\cap E_2 = \phi$.
The {\em generating probability} of the group  $\mathcal{G}(E_1,E_2)$ can be defined
as:
\vspace{-1mm}
\begin{equation}\label{equ:prefix_group}
Pr\left(\mathcal{G}(E_1,E_2)\right) = \prod_{e \in E_1} P(e) \prod_{e' \in E_2} (1 - P(e'))
\end{equation}

The $s$-$t$ reliability of a group $\mathcal{G}(E_1,E_2)$ is defined as the probability that $t$ is reachable from $s$
conditioned on the existence of the group $\mathcal{G}(E_1,E_2)$, i.e.,
\vspace{-2mm}
\begin{equation}\label{equ:reliability_prefix_group}
R_{\mathcal{G}(E_1,E_2)}(s,t) = \sum_{G \sqsubseteq \mathcal{G}(E_1,E_2)} I_{G}(s,t) \times \frac{Pr(G)}{Pr(\mathcal{G}(E_1,E_2))}
\end{equation}

Next, one may verify that the following holds.
\begin{equation}\label{equ:base_recursion}
R(s,t) = R_{\mathcal{G}(\phi,\phi)}(s,t)
\end{equation}

Also, for any edge $e \in E \setminus (E_1\cup E_2)$,
\begin{eqnarray}
\label{equ:reliability_prefix_factor}
&& R_{\mathcal{G}(E_1,E_2)}(s,t) \nonumber \\
&&= P(e)R_{\mathcal{G}(E_1\cup\{e\},E_2)}(s,t) + (1-P(e))R_{\mathcal{G}(E_1,E_2\cup\{e\})}(s,t)\nonumber \\
\end{eqnarray}

We terminate aforementioned recursive procedure when either $E_1$ contains an $s$-$t$ path with
$R_{\mathcal{G}(E_1,E_2)}(s,t)=1$, or $E_2$ contains an $s$-$t$ cut
with $R_{\mathcal{G}(E_1,E_2)}(s,t)=0$.

The efficiency can further be improved by selecting the ``best'' {\em expendable} edge
(i.e., edge $e$ in Equation~\ref{equ:reliability_prefix_factor})
at each iteration. In particular, by following the experimentally optimal strategy in \cite{JLDW11},
we employ depth-first search (DFS) for the next edge expansion. We also find that this strategy works
well in our experiments.
Starting from the source node $s$, we start to explore its first neighbor (its next neighbor is explored only if there is no path
to $t$ which can be found going through the earlier ones), and then recursively visit the neighbors of this neighbor.

\begin{algorithm}[tb!]
\caption{\small Recursive Sampling($\mathcal{G}$, $s$, $t$, $E_1$, $E_2$, $K$)}
\label{algo:recursivesampling}
\begin{algorithmic}[1]
\REQUIRE source node $s$, target node $t$ in uncertain graph $\mathcal{G}=(V,E,P)$,
$K$ = \#samples, $E_1$ = inclusion edge list (initially $\phi$), $E_2$ = exclusion edge list (initially $\phi$),
$Threshold$
\ENSURE  reliability $R(s,t)$
\IF { $K \le Threshold$}
\RETURN $\hat{R}$($\mathcal{G}$, $E_1$, $E_2$, $s$, $t$, $K$) /\/* non-recursive sampling */\
\ENDIF
\IF{$E_1$ contains a path from $s$ to $t$}
\RETURN 1
\ELSIF {$E_2$ contains a cut from $s$ to $t$}
\RETURN 0
\ENDIF
\STATE select an edge $e\in E\setminus(E_1\cup E_2)$
\RETURN $P(e)\cdot$ Recursive-Sampling($\mathcal{G}$, $s$, $t$, $E_1\cup \{e\}$, $E_2$, $\lfloor K\cdot P(e) \rfloor$) $+(1-P(e))\cdot$ Recursive-Sampling($\mathcal{G}$, $s$, $t$, $E_1$, $E_2\cup \{e\}$, $K-\lfloor K\cdot P(e) \rfloor$)
\end{algorithmic}
\end{algorithm}

The aforementioned recursive sampling process has the same variance as the basic MC sampling.
The variance can be reduced by eliminating the ``uncertainty'' of the existence of
edge $e$ in Equation~\ref{equ:reliability_prefix_factor}.
Let $\pi = R_{\mathcal{G}(E_1,E_2)}(s,t)$,
$\pi_1=R_{\mathcal{G}(E_1\cup\{e\},E_2)}(s,t)$, and $\pi_2=R_{\mathcal{G}(E_1,E_2\cup\{e\})}(s,t)$.
Now, instead of directly sampling both the children nodes from the root
(as suggested in Equation~\ref{equ:reliability_prefix_factor}), we consider to estimate both $\pi_1$
and $\pi_2$ independently, and then combine them together to estimate $\pi$.
Specifically, for $K$ total samples in the root, we deterministically allocate
$K_1$ of them to the left subtree (prefix group that includes edge $e$), and $K_2$ of
them to the right subtree (prefix group that excludes edge $e$). It was shown in \cite{JLDW11}
that when the sample size allocation is proportional to the edge inclusion probability,
i.e., $K_1$ = $P(e)\cdot K$ and $K_2 = (1 - P(e))\cdot K$, the variance of the earlier recursive estimator
can be reduced. Moreover, when a prefix group size $K$ is below a pre-defined threshold, we use a non-recursive
sampling method (lines 1-2), such as the basic Monte Carlo or more sophisticated Hansen-Hurwitz estimator to sample the remaining edges \cite{JLDW11}.
The ultimate recursive sampling procedure is given in Algorithm~\ref{algo:recursivesampling}.

The time complexity of recursive sampling estimator
is $\bigO(na)$, where $a$ is the average recursion depth, and is bounded by the diameter of the graph.
Note that in the original paper by Jin et al. \cite{JLDW11}, recursive sampling was proposed to estimate
distance-constraint reliability, that is, the probability that $s$ is reachable to $t$ within an input
distance $d$. In this work, we adapted the proposed approach to compute the $s$-$t$ reliability without any distance constraint.
\begin{table} [tb!]
	\vspace{-5mm}
	\caption{\small Stratum design for recursive stratified sampling}
	\scriptsize	
	\centering
	\begin{tabular} { |c|c|c| }
		{Stratum} & {$e_1\quad e_2\quad e_3\quad \ldots\quad e_r\quad e_{r+1}\quad\ldots\quad e_m$}  & Prob space \\ \hline
		Stratum 0 & 0\, $\quad$ 0\, $\quad$ 0\, $\quad$ \ldots $\quad$ 0\, $\quad$ *\,\,\,\,\,\,\,\,\,\, $\quad$ \ldots $\quad$ *\,  & $\Omega_0$ \\
		Stratum 1 & 1\, $\quad$ *\, $\quad$ *\, $\quad$ \ldots $\quad$ *\, $\quad$ *\,\,\,\,\,\,\,\,\,\, $\quad$ \ldots $\quad$ *\,  & $\Omega_1$ \\
		Stratum 2 & 0\, $\quad$ 1\, $\quad$ *\, $\quad$ \ldots $\quad$ *\, $\quad$ *\,\,\,\,\,\,\,\,\,\, $\quad$ \ldots $\quad$ *\,  & $\Omega_2$ \\
		\ldots    &                \ldots                                                                                            & \ldots \\
		Stratum r & 0\, $\quad$ 0\, $\quad$ 0\, $\quad$ \ldots $\quad$ 1\, $\quad$ *\,\,\,\,\,\,\,\,\,\, $\quad$ \ldots $\quad$ *\,  & $\Omega_r$ \\
		\hline
	\end{tabular}
	\vspace{-3mm}
	\label{tab:rssstratum}
\end{table}
\begin{algorithm}[tb!]
	\caption{\small  Sampling: RSS($\mathcal{G}$, $K$, $s$, $t$)}
	\label{algo:rss}
	\begin{algorithmic}[1]
		\REQUIRE source node $s$, target node $t$ in uncertain graph $\mathcal{G}=(V,E,P)$,
		$K$ = \#samples, $Threshold$, stratum parameter $r$
		\ENSURE  reliability $R(s,t)$
		\STATE $\hat{R} \gets 0$
		\IF {$K < Threshold$ \Or $|E| < r$}
		\FOR {$j=1$ to $K$}
		\STATE MC sampling to compute $I_j$ (indicator function)
		\STATE $\hat{R} \gets \hat{R}+I_j$
		\ENDFOR
		\RETURN $\frac{\hat{R}}{K}$
		\ELSE
		\STATE $T \gets $ select $r$ edges by BFS from $s$
		\FOR {$i=0$ to $r$}
		\STATE Let $X_i$ be the status vector of $T$ in stratum $i$
		\STATE $\mathcal{G}_i \gets$ simplify graph $\mathcal{G}$ based on $X_i$
		\STATE $K_i \gets \pi_i\cdot K$ $\quad$ /\/* $\pi_i$ computed in Equation~\ref{equ:rss}  */\
		\STATE $\mu_i \gets RSS(\mathcal{G}_i, K_i, s, t)$
		\STATE $\hat{R} \gets \hat{R}+\pi_i\mu_i$
		\ENDFOR
		\RETURN $\hat{R}$
		\ENDIF
	\end{algorithmic}
\end{algorithm}

\vspace{-1mm}
\subsection{Recursive Stratified Sampling}
\label{sec:rss}
Li et al. \cite{LYMJ14} developed an alternative divide-and-conquer approach to measure reliability,
called the recursive stratified sampling (RSS). By using a stratification method that partitions the probability
space $\Omega$ into $r+1$ non-overlapping subspaces $(\Omega_0,\ldots,$ $\Omega_r)$ via selecting $r$ edges,
they proceed to fix the states for certain edges in each stratum.
For example, as seen in Table~\ref{tab:rssstratum}, in stratum 0, we set the status of $r$ selected edges as 0,
while leaving the rest of edges as undetermined. Subsequently, in stratum $i$ ($1\le i \le r$), we set the status of edge $i$ to 1,
the status of those before it as 0, and those after it as undetermined.

Let $T$ be the set of selected $r$ edges via BFS from the source node $s$ (line 9, Algorithm~\ref{algo:rss}), and $X_{i,j}$ be
the corresponding status vector of the $j$-th ($1\le j \le r$) selected edge in stratum $i$.
Then, the probability of a possible graph in stratum $i$ is given by:
\vspace{-2mm}
\begin{eqnarray}
\pi_i&=&Pr\left[G_P\in\Omega_i\right] \nonumber \\
&=&\prod_{e_j\in T\bigwedge{X_{i,j}=1}} P(e_j)\prod_{e_j\in T\bigwedge{X_{i,j}=0}}{(1-P(e_j))} \quad
\label{equ:rss}
\end{eqnarray}
We then set the sample size of stratum $i$ to $K_i=(\pi_i\cdot K)$, where $K$ being the total sample size. The algorithm
recursively applies the sample size to each stratum and simplifies the graph. It terminates when the sample size of a stratum
is smaller than a given threshold. Reliability is then calculated by finding the sum of the reliabilities in all subspaces.
The complete procedure in given in Algorithm~\ref{algo:rss}.

It was shown in \cite{LYMJ14} that the time complexity of recursive stratified sampling is same as that of the MC sampling,
i.e., $\bigO(K(m+n))$, while the variance of the estimator is significantly reduced.

\begin{algorithm}[tb!]
	\caption{\small Lazy Propagation Sampling}
	\label{algo:lazypropagationsampling}
	\begin{algorithmic}[1]
		\REQUIRE source node $s$, target node $t$ in uncertain graph $\mathcal{G}=(V,E,P)$,
		$K$ = \#samples
		\ENSURE  reliability $R(s,t)$
		\STATE $r \gets 0$, set all nodes not initialized
		\FOR {$i=1$ to $K$}
		\IF {$s==t$}
		\STATE $r \gets r+1$
		\STATE goto line 29
		\ENDIF
		\STATE $h \gets \phi$, set all nodes not visited
		\STATE $h \gets \{s\}$  $\qquad$ /\/* set of visited nodes */\
		\STATE mark $s$ visited
		\WHILE {$h \neq \emptyset$}
		\STATE $v \gets h.pop()$
		\IF {$v$ is not initialized}
		\STATE $c_v \gets 0$, $h_v \gets \emptyset$, mark $v$ initialized
		\FOR {$nbr \in$ $v$'s neighbor sets}
		\STATE $X(nbr) \gets$ geometric r.v. instance
		\STATE $h_v \gets h_v \cup \{\langle nbr,X(nbr)+c_v\rangle\}$
		\ENDFOR
		\ENDIF
		\WHILE {$h_v.top()==c_v$}
		\STATE $\langle nbr, X(nbr)\rangle \gets h_v.pop()$
		\STATE $h \gets h \cup \{nbr\}$ if $nbr$ is not visited
		\STATE mark $nbr$ visited
		\STATE $X'(nbr) \gets$ geometric r.v. instance
		\STATE $h_v \gets h_v \cup \{\langle nbr, X'(nbr)+c_v{\bf +1}\rangle\}$ $\quad$ /\/* our correction in bold  */\
		\IF {$nbr==t$}
		\STATE $r \gets r+1$
		\STATE goto line 29
		\ENDIF
		\ENDWHILE
		\STATE $c_v \gets c_v+1$;
		\ENDWHILE
		\ENDFOR
		\RETURN $\frac{r}{K}$
	\end{algorithmic}
\end{algorithm}
\vspace{-1mm}
\subsection{Lazy Propagation Sampling}
\label{sec:lazy_sampling}
Li et al. proposed the lazy propagation sampling \cite{LFZT17} that
aims to bypass MC sampling's requirement of probing a large number of edges.
In particular, if an edge has a low probability, it will remain un-activated
in most of the samples; and therefore, those probings could be
``unnecessary''. In contrast, lazy propagation sampling estimates reliability by avoiding unnecessary
probing of edges as much as possible. To achieve this, the algorithm employs a geometric distribution
per edge, and probes an edge only when it will be activated.
The geometric random instance determines, based on the original edge probability,
the number of sampling instances before the edge is activated.
This means that the algorithm will predict how many $k$ worlds are sampled before an edge exists.
In doing so, the number of times an edge is probed is reduced by a factor of $\frac{1}{p\left(e\right)}$ in expectation.
It was proved in \cite{LFZT17} that there is no statistical difference between using lazy sampling
and the classic MC sampling. In other words, lazy sampling has the same variance as that of MC sampling,
while improving on the efficiency.

As the original algorithm by Li et al. was developed for personalized social influential tags exploration,
we adapt the algorithm to measure reliability over uncertain graphs. We show the complete
procedure in Algorithm~\ref{algo:lazypropagationsampling}. It initializes a heap $h_v$
for every visited node $v$, and pushes $v$'s each out-neighbor $nbr$ with a geometric
random instance, i.e., $\langle nbr, X(nbr)\rangle$ into $h_v$ (lines 12-18).
It also maintains a counter $c_v$ to keep track of the number of times $v$
has been visited. Once a random instance in $h_v$ is equal to $c_v$, the
corresponding neighbor will be probed, and then a new random instance
is generated to decide the next time to probe the neighbor
(lines 19-29).

\vspace{-1mm}
\spara{Our correction in the algorithm.}
In the original paper \cite{LFZT17}, it is $h_v \gets h_v \cup \{\langle nbr, X'(nbr) +c_v\rangle\}$ at line 24,
which, in fact, incurs error in reliability estimation \footnote{\scriptsize Yuchen Li,
first author of \cite{LFZT17}, acknowledged this issue in email communication.}. This is because node $nbr$ is probed in the current round
(since $h_v.top()==c_v$ was true at line 19), and now line 23 aims at assigning a new geometric random instance $X'(nbr)$ with $nbr$.
This new geometric random instance $X'(nbr)$ indicates that after how many times of failure, $nbr$ will be visited again following node $v$, starting from the next round.
Therefore, at line 24, the value of counter $c_v$ shall be that of the next round, i.e., one needs to add $c_v{\bf +1}$ with $X'(nbr)$
(as opposed to adding only $c_v$ specified in the original paper \cite{LFZT17}).

\begin{figure}[tb!]
	\centering
	\vspace{-2mm}
	\begin{minipage}[t]{0.15\textwidth}
		\centering
		\includegraphics[scale=0.36]{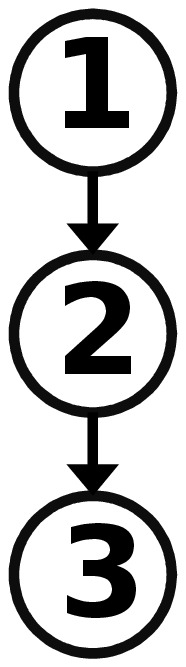}
		\caption{\small {Toy graph for Example \ref{ex:lz}}}
        \vspace{-6mm}
		\label{fig:lz}
	\end{minipage}
	$\quad$
	\begin{minipage}[t]{0.28\textwidth}
		\centering
		\vspace{-30mm}
		\includegraphics[scale=0.16,angle=270]{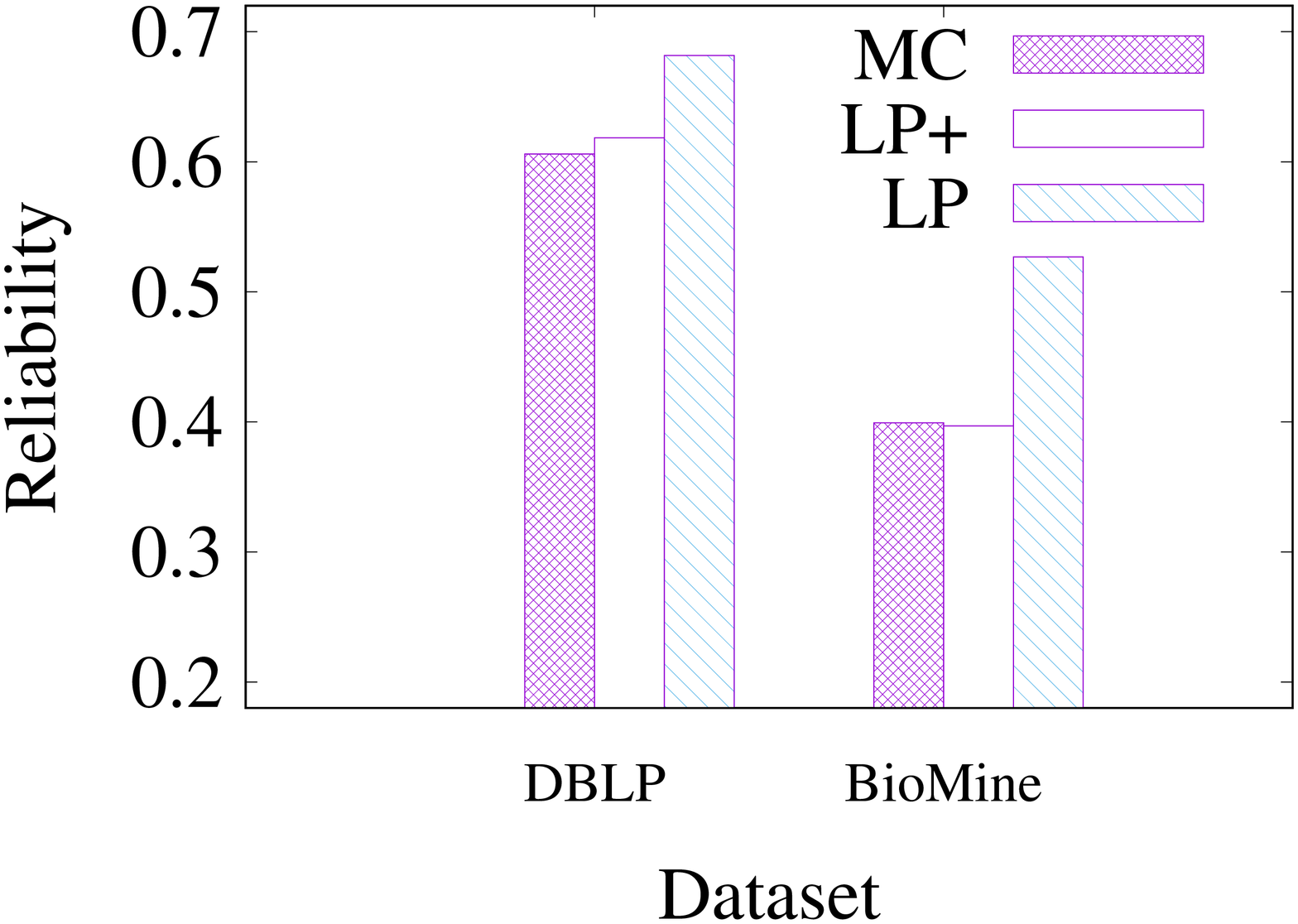}
		\vspace{-4.5mm}
		\caption{\small {Reliability estimated by MC, Lazy Propagation ({\sf LP}), and Lazy Propagation+ ({\sf LP+}) at convergence. Reliability
is reported as the average over 100 $s$-$t$ pairs. For experimental setting, see Sec.~\ref{sec:experiments}.}}
		\label{fig:lz+_rel_com}
        \vspace{-6mm}
	\end{minipage}
\end{figure}

We shall demonstrate the error in \cite{LFZT17} and our correction with the following example.
\begin{exple}
\vspace{-2mm}
Considering the graph in Figure \ref{fig:lz}: Node $1$ is the source node, and node $3$ the target node.
In the first round, $c_1=0$ and we initialize node $1$, push its neighbor node $2$ into $h_1$.
Suppose that the geometric random instance $X(2)$ for $2$ is generated as 0. Then,
node $2$ exists in the first possible world (i.e., the condition in line 19 holds true for node $2$),
and the algorithm proceeds to line 20-24. Now, if we follow $h_v \gets h_v \cup \{\langle nbr, X'(nbr) +c_v\rangle\}$ at line 24 as in the original paper \cite{LFZT17},
we shall encounter at least one of the two following errors.
	
(1) Assume that at line 24, the new geometric random instance $X'(2)>0$, e.g., $X'(2)=1$.
It indicates that in the next possible world, edge $1\rightarrow 2$ must not exist.
However, when following the original algorithm as in \cite{LFZT17}, $\langle Node\ 2,1\rangle$ will be stored in $h_1$, since $c_1=0$ at present.
When sampling the next possible world, $c_1=1$ and node $2$ will be probed again, which is incorrect. Clearly, this is an overestimation, and it will
result in higher estimated reliability from source to target.
	
(2) In the other case, when the new geometric random instance
$X'(2)=0$ at line 24, it means that in the next possible world, edge $1\rightarrow 2$ must exist.
By following the original algorithm as in \cite{LFZT17}, $\langle Node\ 2,0\rangle$ will be pushed in the heap $h_1$ and would be placed at the top.
However, when sampling all subsequent possible worlds, $c_1\geq 1$ and $\langle Node\ 2,0\rangle$ at the top of $h_1$ will always stop the algorithm to enter lines 20-24,
which makes node $1$ not expandable anymore, and thus incurs an underestimation error.
\label{ex:lz}
\vspace{-2mm}
\end{exple}
In practice, the first type of error, i.e., overestimation happens in most cases, and the original method in \cite{LFZT17} estimates much higher reliability.
Thus, we update the algorithm at line 24, as shown in Algorithm \ref{algo:lazypropagationsampling} to avoid such errors.
We denote it as {\sf LP+}, and the original one as {\sf LP}. Our experimental results in Figure \ref{fig:lz+_rel_com} confirms that
{\sf LP} indeed estimates much higher reliability than Monte Carlo (MC), whereas {\sf LP+} estimation is close to that of MC.
\vspace{-1mm}
\subsection{Indexing via Probabilistic Trees}
\label{sec:prob_tree}
Maniu et al. \cite{ManiuCS17} proposed ProbTree index to improve the efficiency of $s$-$t$ reliability queries. The method
builds a tree index structure, called ProbTree, from the input uncertain graph $G$. Next, given an
$s$-$t$ reliability query $q$, an equivalent graph $G(q)$ is created from the ProbTree index structure, and
MC sampling is performed on $G(q)$. Clearly, if $G(q)$ can be generated quickly,
and if $G(q)$ is smaller than $G$, then executing the query on $G(q)$ would be faster.

In \cite{ManiuCS17}, three index structures were developed. Among them, we employ FWD tree
(an abbreviation for fixed-width tree decomposition) because (a) its index building time,
space, and query processing time are linear in the input graph size, and
(b) the index structure is lossless (i.e., produces good-quality results) for a tree decomposition width
$w\leq 2$.  In fact, as reported in \cite{ManiuCS17}, as well as found in our experiments, FWD tree with w=2
produces high-quality results, and exhibits good efficiency.
\begin{algorithm}[tb!]
	\caption{\small FWD (Fixed Width) ProbTree Index Construction}
	\label{algo:probtreeindex}
	\begin{algorithmic}[1]
		\REQUIRE Uncertain graph $\mathcal{G}=(V,E,P)$, tree width $w$
		\ENSURE ProbTree index $(\mathcal{T},\mathcal{R})$
		\STATE $G \gets$ undirected, unweighted version of $\mathcal{G}$
		\STATE $\mathcal{T}=\emptyset,\mathcal{S}=\emptyset$
		\FOR {$d \gets 1$ \To $w$}
		\WHILE {there exists a node $v$ with degree $d$ in $G$}
		\STATE create new bag $B$
		\STATE $V(B) \gets v$ and all its neighbors
		\FOR {all unmarked edge $e$ in $G$ between nodes of $V(B)$}
		\STATE $E(B) \gets E(B) \cup \{e\}$; mark $e$
		\ENDFOR
		\STATE covered$(B)\gets\{v\}$
		\STATE remove $v$ from $G$ and add to $G$ a $(d-1)$-clique between $v$'s neighbors
		\STATE $\mathcal{S} \gets \mathcal{S} \cup \{B\}$
		\ENDWHILE
		\ENDFOR
		\STATE $V(\mathcal{R}) \gets$ all nodes in $G$ not in covered $(B)$
		\STATE $E(\mathcal{R}) \gets$ all unmarked edges in $G$
		\STATE root $\mathcal{T}$ at $\mathcal{R}$
		\FOR {bag $B$ in $S$}
		\STATE mark $B$
		\IF {$\exists$ an unmarked bag $B'$ s.t. $V(B) \backslash $covered$(B)\subseteq B'$ }
		\STATE update$(\mathcal{T},\mathcal{R})$ so $B'$ is parent of $B$
		\ELSE
		\STATE update$(\mathcal{T},\mathcal{R})$ so $\mathcal{R}$ is parent of $B$
		\ENDIF
		\ENDFOR
		\FOR {$h \gets$ height$(\mathcal{T})$ \To 0}
		\FOR {bag $B$ s.t. $level(B)=h$}
		\STATE collect pre-computed reliability from children (if any)
		\STATE compute pairwise reliability in current bag $B$
		\ENDFOR
		\ENDFOR
		\RETURN $(\mathcal{T},\mathcal{R})$
	\end{algorithmic}
\end{algorithm}
\begin{algorithm}[tb!]
	\caption{\small Reliability Estimation Using FWD ProbTree}
	\label{algo:query_probtree}
	\begin{algorithmic}[1]
		\REQUIRE ProbTree index $(\mathcal{T},\mathcal{R})$, source node $s$, target node $t$, $K=\#$samples
		\ENSURE Reliability $R(s,t)$
		\IF {$s \in \mathcal{R}$ and $t \in \mathcal{R}$}
		\STATE go to line 13
		\ENDIF
		\FOR {$h \gets$ height$(\mathcal{T})$ \To 0}
		\FOR {bag $B$ s.t. $level(B)=h$}
		\IF {covered$(B)\cap \{s, t\} \neq \emptyset$}
		\STATE delete the reliability in $parent(B)$ resulting from B
		\STATE $E(parent(B)) \gets E(parent(B)) \cup E(B)$
		\STATE $V(parent(B)) \gets V(parent(B)) \cup V(B)$
		\ENDIF
		\ENDFOR
		\ENDFOR
		\STATE estimate $R(s,t)$ via MC sampling ($K$ samples) at root $\mathcal{R}$
		\RETURN $R(s,t)$
	\end{algorithmic}
\end{algorithm}

FWD tree index building (Algorithm \ref{algo:probtreeindex}) has three phases: fixed-width tree decomposition, building of the FWD index tree, and pre-computation of reliability. The first phase (lines 2-17) is an adaption of the algorithms in \cite{W2010,ASK2012},
which performs a {\em relaxed} tree decomposition with a fixed width $w$. At each step (lines 4-13),
a node having degree at most $w$ is chosen, marked as \emph{covered}. A bag is created to contain this node and its neighbors, along with the probabilistic edges between them in $G$.
Then, the covered node is removed from $G$ and a clique between its neighbors is added into $G$.
This process repeats until there are no such nodes left. Finally, the rest of the uncovered nodes
and the remaining edges are copied in the root of the tree (lines 15-17).

The second phase (lines 18-25) is the creation of the FWD index tree. Each bag is visited in its creation order,
and their parents are defined as the bag whose node set contains all uncovered nodes of the visited bag.
If no such bag exists, then the parent of the bag will be the root of the tree.
In the final phase (lines 26-31), we need to compute $R(v_1, v_2)$ for each pair $(v_1,v_2)$ in each bag. It follows a bottom-up manner.
For each bag $B$, it collects the computed
reliability from $B$'s children, and combine them with information within current bag to compute the
current reliability for all node pairs in $B$.
When $w\leq2$, it is possible to pre-compute the correct probability distributions between node pairs, thereby
making the index structure lossless.

When conducting an $s$-$t$ query (Algorithm \ref{algo:query_probtree}), if the root contains both
$s$ and $t$, the root is the query graph. Otherwise, it searches from the bottom to find the bags
containing $s$ or $t$ as the covered node, respectively. Then it propagates them up all the way to the root,
and merges them as a combined graph for query answering. By doing so, all irrelevant branches will not be included
in the graph returned. An example of both index building and query processing is presented below.
\begin{figure}[tb!]
	\vspace{-2mm}
	\centering
	\includegraphics[scale=0.29]{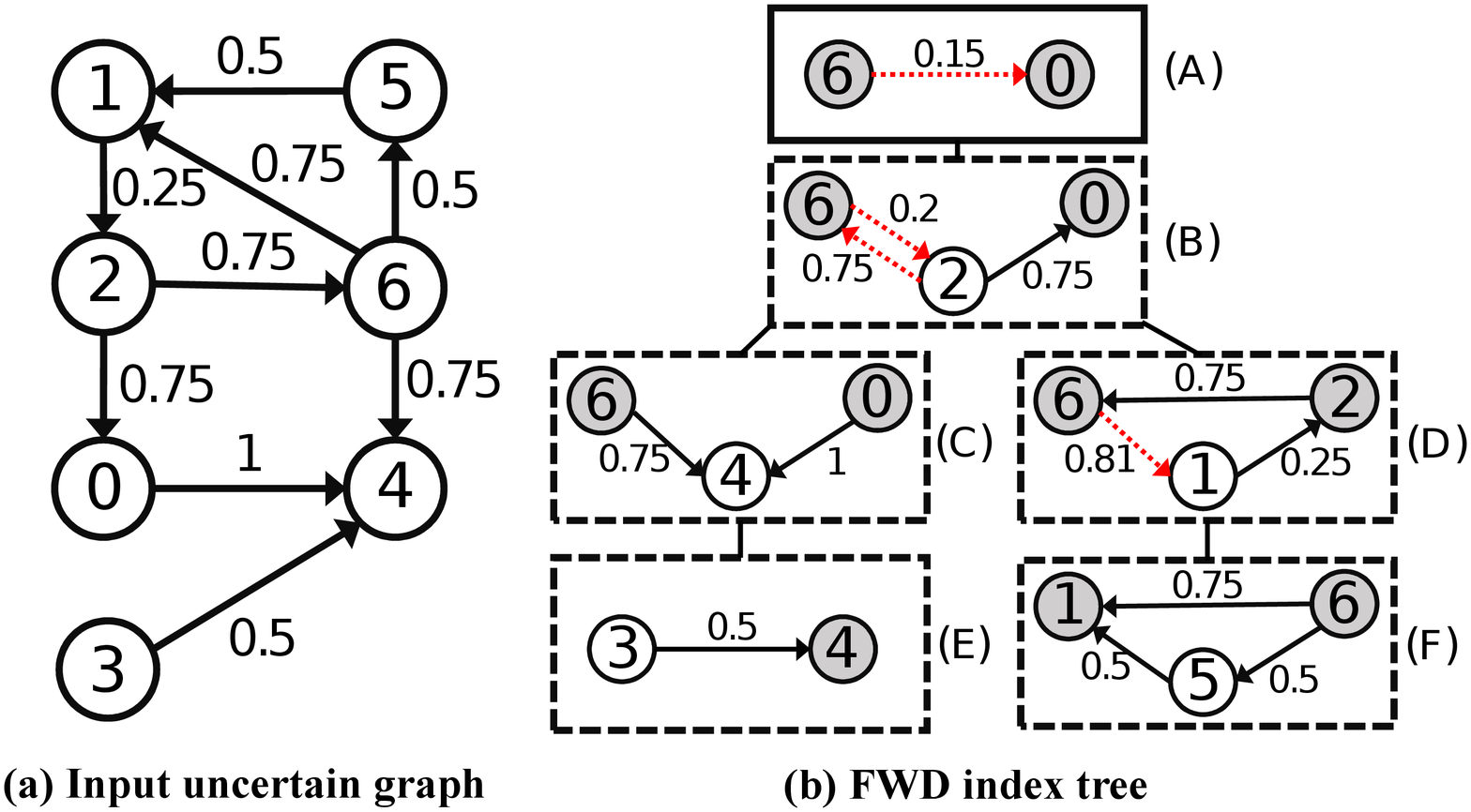}
	\vspace{-5mm}
	\caption{\small Example of FWD ProbTree (w=2) construction }
	\label{fig:probtree}
	\vspace{-4mm}
\end{figure}
\begin{exple}
\vspace{-2mm}
In Figure~\ref{fig:probtree}(b), we show the FWD index tree for the input uncertain graph in Figure~\ref{fig:probtree}(a). At first, node 3 has the smallest degree 1 and is chosen. A bag $(E)$ is created to contain it, together with its neighbor, node 4. Then, we put all the unmarked edges between them into this bag and mark them. After the construction of this bag, we remove node 3 (the selected node) from the original graph. Now, no node has degree 1, then we consider degree 2. Due to the removal of node 3, node 4 has degree 2 and can be selected. We treat it in the same way and obtain the second bag (labelled as $(C)$ in Figure~\ref{fig:probtree}(b)). Since there is no edge between node 0 and node 6 (neighbors of node 4) in the original graph, a new edge (0,6) is added. We repeat the aforementioned procedure until no more bag can be created. Then, we traverse them in the creation order $(E), (C), (F), (D), (B), (A)$, and decide their parent-child relationships using lines 20-24. For computing reliability in each bag, it follows a bottom-up manner as described before. For example, in bag $(D)$, the reliability from node 6 to node 1 is collected from its child, bag $(F)$, and is computed as: $1-(1-0.75)*(1-0.5*0.5)=0.8125$.
	
For reliability from
node 1 to node 2, we find bags $(B)$ and $(D)$ (which have these nodes as cover), and finally bags $(A)$, $(B)$, and $(D)$ are merged as the query graph. Bags $(C)$ and $(E)$ are irrelevant to this query. Bag $(F)$'s information about the query node is already contained in bag $(D)$. Thus, it will not be involved as well.
\vspace{-2mm}
\label{ex:probtree}
\end{exple}
\vspace{-2mm}
Following \cite{ManiuCS17}, the time complexity of FWD probtree index construction is linear in the number of nodes in the graph, and the complexity of pairwise reliability computations in each bag is $\bigO(w^2d)$, where $w$ is the tree width ($w=2$ to ensure loseless index), and $d$ is the diameter of graph.
The space complexity of FWD probtree index is $\bigO(|E|)$. During online $s$-$t$ query processing, the cost of retrieving $s$ and $t$ from the FWD probtree is linear in the tree depth. The reliability estimating cost is the same as MC sampling, $\bigO(K(n'+m'))$, but
with smaller number of nodes and edges.

\spara{Our adaptation in complexity.} The original {\em ProbTree} index is designed to support multiple kinds of $s$-$t$ queries, e.g., reliability query, shortest path query, etc. Therefore, it pre-computes the distance probability distribution between each node in every bag. However, if only considering $s$-$t$ reliability query, we can just compute and store the edge probability information, regardless of distance. Since $w=2$, for each pair of nodes in the current bag, there can exist at most 2 paths in its children. One can easily collect and aggregate them by $1-(1-p_1)(1-p_2)$, where $p_1$ and $p_2$ are the probability of these two paths. An illustration can be found in Example \ref{ex:probtree}. The complexity of pairwise reliability computations in each bag can be reduced from $\bigO(w^2d)$ to $\bigO(w^2)$.
Empirically, on our largest {\em BioMine} dataset, we can reduce the index building time from 4062 seconds to 2482 seconds.
%


\vspace{-1mm}
\subsection{Horizontal Comparison of Algorithms: \\Strengths and Weaknesses}
\label{sec:hcomp}
\vspace{-1mm}
\spara{Estimator variance.} The mean squared error (MSE) measures the quality of an estimator. The lower the MSE, the better is an estimator. For unbiased estimator, the MSE is equal to the variance of this estimator. {\em All estimators compared in this work are proved to be unbiased in original papers.  Statistically, BFS Sharing and Lazy Propagation shares same variance as basic MC Sampling. The two recursive methods, RHH and RSS, reduce the estimator variance by dividing the sample size according to either edge probability (e.g, RHH), or stratum probability (e.g., RSS)}. We refer to Theorem 2 in \cite{JLDW11}, and Theorems 4.2 and 4.3 in \cite{LYMJ14} as the theoretical basis for variance reduction in RHH and RSS, respectively.

\spara{Running time per sample.} For each sample, MC applies a BFS search from the source node $s$, each edge encountered exists with the probability on that edge. The BFS search terminates when the target node $t$ is found (early termination), or no new node can be expanded.
In contrast, while BFS Sharing follows the same BFS search strategy; no early termination is allowed.
The existence of each edge in every sample has been determined offline (via indexing).
Therefore, {\em BFS Sharing tends to sample more edges than MC in each sample, while the time cost of checking the existence of each edge is saved due to indexing}.

Lazy propagation employs geometric distribution to determine the existence of an edge in samples, that is, each time when an edge is probed during the BFS search in one sample, it decides after how many samples this edge will exist again. Therefore, unnecessary edge probing is avoided.

ProbTree decomposes the graph and pre-computes the reliability information beforehand, thus allowing running online BFS on a smaller graph to improve efficiency. RHH and RSS recursively simplify the graph, and apply MC sampling on the resulted graph when sample size is smaller than a threshold, or the graph is fully simplified. The improvement of the running time per sample depends on how much the graph can be simplified. Additional time cost may arise from graph simplification task and the recursive procedure. In all, {\em Lazy Propagation and ProbTree can reduce the time cost of investigating a sample when compared with MC, while BFS Sharing, RHH and RSS have both improvements and additional costs}. The detailed experimental study can be found in Section~\ref{sec:eff}.

\spara{Total running time.} To achieve the same estimator variance (e.g., same quality), MC, BFS Sharing, Lazy Propagation, and ProbTree would require same sample size, thus the Lazy Propagation and ProbTree shall improve the efficiency (due to lower running time per sample)
compared to MC. For RHH and RSS, less samples are needed, thereby also improving the efficiency compared to MC.
{\em With all these, the final efficiency comparison for Lazy Propagation, ProbTree, RHH, and RSS is unknown. Therefore, our experimental analysis is critical}.

\spara{Memory usage.} In addition to the original graph, the only memory cost of MC is the current sample, and the count of samples where $t$ is reachable from $s$. For BFS Sharing, instead of original graph, the whole index shall be kept in main memory. Each edge in BFS Sharing index is a vector of Boolean values, while in the original graph only a single value is stored. Compared with MC, Lazy Propagation additionally requires a global counter for each node, and a geometric random instance heap for its neighbors. The ProbTree index size is linear in the original graph size. Both RHH and RSS store the whole recursive stack, and simplified graph instances in main memory, which makes them more memory intensive.

\spara{Indexing cost, re-computation, and adaptability.} Both BFS Sharing and ProbTree index sizes are linear in the original graph size. ProbTree index is generally comparable to the original graph size, and is independent of the sample size $K$. However, BFS Sharing is also about linear in the sample size $K$. Thus, {\em BFS Sharing index is usually larger than ProbTree index, and BFS Sharing also requires more loading time into main memory. On the other hand, BFS Sharing index is easier to build than ProbTree index}. Only $K$ boolean values shall be generated based on the edge probability for each edge. The ProbTree index requires decomposing the original graph and pre-computing the reliability information.

{\em To ensure the independence among queries, index shall be updated between two successive queries for BFS Sharing, this is not necessary for ProbTree. Moreover, ProbTree has the adaptability to support estimators other than MC sampling, while BFS Sharing has no such potential}.

\vspace{-2mm}
\subsection{Other Related Work}
\label{sec:related}
\vspace{-1mm}
The large spectrum of the reliability problem is categorized in Figure~\ref{fig:wp}. Further research on reliability emphasized on polynomial-time upper/lower bounds to reliability problems
\cite{PB83,B85,BP88,BD94,Konak98animproved,GLP05}. Recently, efficient {\em distributed} algorithms have been developed for
reliability estimation \cite{ZLLL17,ChengYCW15}. Some orthogonal directions include finding one ``good''
possible world \cite{PGPB14,SZL16}, considering the most probable path \cite{CWW10,KS06}, as well as adaptive edge testing
\cite{FXPFW17,FFXPWL17,FWK14} and crowdsourcing \cite{LPCX17} for reducing uncertainty.
As discussed earlier in Sections~\ref{sec:mc_sampling} -- \ref{sec:prob_tree}, in this work
we focus on sampling and indexing based sequential algorithms for $s$-$t$ reliability estimation.

Other related queries over uncertain graphs include nearest neighbors \cite{PBGK10,LCFHM18}, shortest paths \cite{YuanCW10,ZPZ11},
reliable set \cite{KBGG14,ZZL15}, conditional reliability \cite{KBGN18},
distance-constrained reachability \cite{JLDW11}, probabilistic road network queries \cite{HP10},
and influence maximization \cite{KKT03}, among several others. Many of the efficient sampling and indexing strategies that we
investigate in this work can also be employed to answer such advanced queries. 

\vspace{-2mm}
\section{Experimental Results}
\label{sec:experiments}
\vspace{-1mm}
We conduct experiments to compare six state-of-the-art reliability estimation algorithms,
and report the number of samples required for their convergence, estimator accuracy, variance,
running time, and memory usage using medium and large-scale, real-world network datasets.
We obtain the source code of ProbTree \cite{ManiuCS17} from the authors, which is written in C++,
and we further optimize their index building method as discussed in the paragraph ``Our improvement in complexity''
in Section~\ref{sec:prob_tree}. We implement other five algorithms in C++, and perform experiments on
a single core of a 100GB, 2.40GHz Intel Xeon E5-4650 v2 server. {\em Our datasets and source code are available at: https://github.com/5555lan/RelComp}.
\begin{table} [tb!]
	\vspace{-3mm}
	\caption{\small Properties of datasets}
	\vspace{1mm}
	\scriptsize	
	\centering
	\begin{tabular} { |l|c|c|c|c| }
		\hline
		{\textsf{Dataset}}        & {\textsf{\#Nodes}}  & {\textsf{\#Edges}}  &  {\textsf{Edge Prob: Mean, SD, Quartiles}} \\ \hline
		{\em LastFM}              & 6\,899	            &  23\,696	          & 0.29 $\pm$ 0.25, \{0.13, 0.20, 0.33\} \\
		{\em NetHept\_uniform}         & 15\,233	            &  62\,774	          & 0.04 $\pm$ 0.04, \{0.001, 0.01, 0.10\} \\
		{\em AS\_Topology} & 45\,535 & 172\,294 & 0.23 $\pm$ 0.20, \{0.08, 0.21, 0.31\} \\
		{\em DBLP\_0.2}	      & 1\,291\,298	        &  7\,123\,632	      & 0.33 $\pm$ 0.18, \{0.18, 0.33, 0.45\} \\
		{\em DBLP\_0.05}	      & 1\,291\,298	        &  7\,123\,632	      & 0.11 $\pm$ 0.09, \{0.05, 0.10, 0.14\} \\
		{\em BioMine}	          & 1\,045\,414	        &  6\,742\,939	      & 0.27 $\pm$ 0.21, \{0.12, 0.22, 0.36\} \\ \hline
	\end{tabular}
	\vspace{-5mm}
	\label{tab:data}
\end{table}

\vspace{-2mm}
\subsection{Environment Setup}
\vspace{-3mm}
\subsubsection{Datasets}
\vspace{-1mm}
We downloaded five real-world networks (Table~\ref{tab:data}).
Many of them have been extensively used in past research on uncertain graphs,
including reliability estimation \cite{ZZL15,JLDW11,LYMJ14,LFZT17,PBGK10,ZZZZL11,KBGG14}.
$\bullet$ \emph{LastFM} (www.last.fm). Last.FM is a musical social network, where users listen to their favorite tracks, and communicate
with each other based on their musical preferences. We crawled a local network of Last.FM, and formed a bi-directed graph by connecting
two users if they communicated at least once.
$\bullet$ \emph{NetHEPT} (www.arXiv.org). This graph was extracted from the ``High Energy Physics – Theory'' section of the e-print arXiv with papers from 1991 to 2003. The nodes are connected by bi-directed edges if they co-authored at least once.
$\bullet$ \emph{AS Topology} (http://data.caida.org/datasets/topology/ ark/ipv4/). An autonomous system (AS) is a collection of connected Internet Protocol (IP) routing prefixes under the control of one or more network operators on behalf of a single administrative entity, e.g., a university. The AS connections are established with BGP protocol. It may fail due to various reasons, e.g., failure of physical links when one AS updates its connection configuration to ensure stricter security setting, while some of its peers can no longer satisfy it, or some connections are cancelled manually by the AS administrator. We downloaded one network snapshot per month, from January 2008 to December 2017.
$\bullet$ \emph{DBLP} (www.informatik. uni-trier.de/~ley/db/). This is a well-known collaboration network. We downloaded it on March 31,
2017. Each node is an authorship, and bi-directed edges denote their co-authorship relations.
$\bullet$ \emph{BioMine} (www.cs. helsinki.fi/group/ biomine/). This is the database of the BIOMINE project \cite{Eronen2012}. The graph is
constructed by integrating cross-references from several biological data -bases. Nodes represent biological concepts
such as genes, proteins, etc., and directed edges denote real-world phenomena between two nodes, e.g., a gene ``codes'' for a protein.

\vspace{-1.5mm}
\subsubsection{Edge Probability Models}
\vspace{-1mm}
By following bulk of the literature on reliability estimation over uncertain graphs \cite{ZZL15,JLDW11,LYMJ14,LFZT17,PBGK10,ZZZZL11,KBGG14},
we adopt the following widely-used edge probability models.
$\bullet$ \emph{LastFM}. The probability on any edge corresponds to the inverse of the out-degree of the node from which
that arc is outgoing.
$\bullet$ \emph{NetHEPT}. Each edge is assigned with a probability, chosen uniformly at random, from (0.1, 0.01, 0.001).
$\bullet$ \emph{AS\_Topology}. Once an AS connection (i.e., an edge) is observed for the first time, we calculate the ratio of snapshots containing this connection within all follow-up snapshots as the probability of existence for this edge.
$\bullet$ \emph{DBLP\_0.2} and \emph{DBLP\_0.05}. The edge probabilities are derived from an exponential
cdf of mean $\mu$ to the number of collaborations between two respective authors; hence, if
two authors collaborated $c$ times, the corresponding probability
is $1 - \exp^{-c/\mu}$. We consider $\mu =5, 20$ in our experiments,
and generate two uncertain networks, \emph{DBLP\_0.2} and \emph{DBLP\_0.05},
respectively. Clearly, higher values of $\mu$ generate
smaller edge probabilities.
$\bullet$ \emph{BioMine}. Edge probabilities, which quantify the existence of
a phenomenon between the two endpoints of that edge (a gene ``codes'' for
a protein), were determined in \cite{Eronen2012} as a combination of three criteria:
relevance (i.e., relative importance of that relationship type), informativeness (e.g., degrees of the nodes adjacent to that edge),
and confidence on the existence of a specific relationship (e.g., conformity with the biological {\sf STRING} database).

\begin{figure*}[t!]
	\vspace{-5mm}
	\centering
	\subfigure[\scriptsize {{\em lastFM}}]
	{\includegraphics[scale=0.19,angle=270]{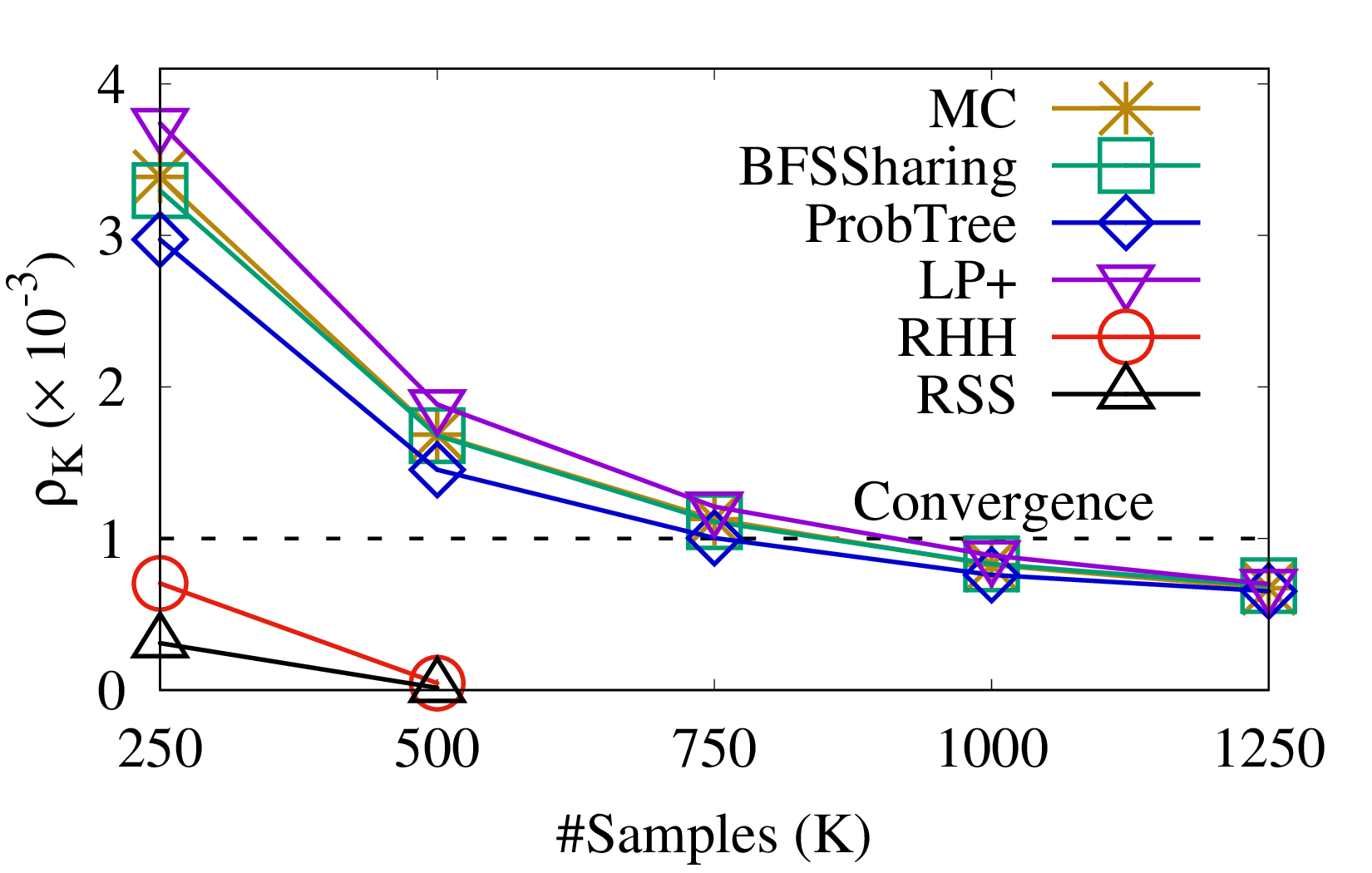}
		\label{fig:var_last}}
	\subfigure[\scriptsize {{\em NetHept}}]
	{\includegraphics[scale=0.19,angle=270]{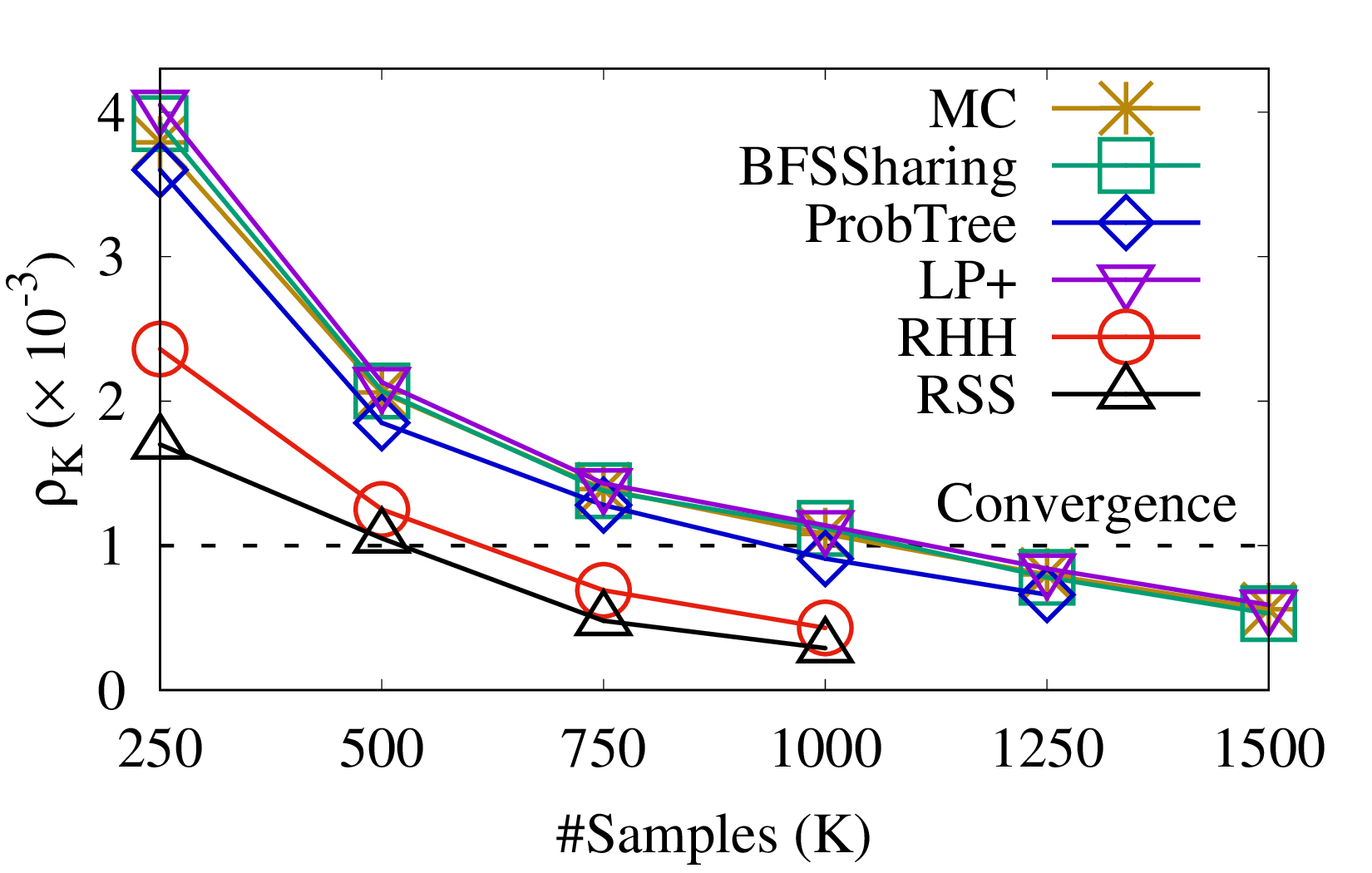}
		\label{fig:var_hep}}
	\subfigure[\scriptsize {{\em AS\_Topology}}]
	{\includegraphics[scale=0.19,angle=270]{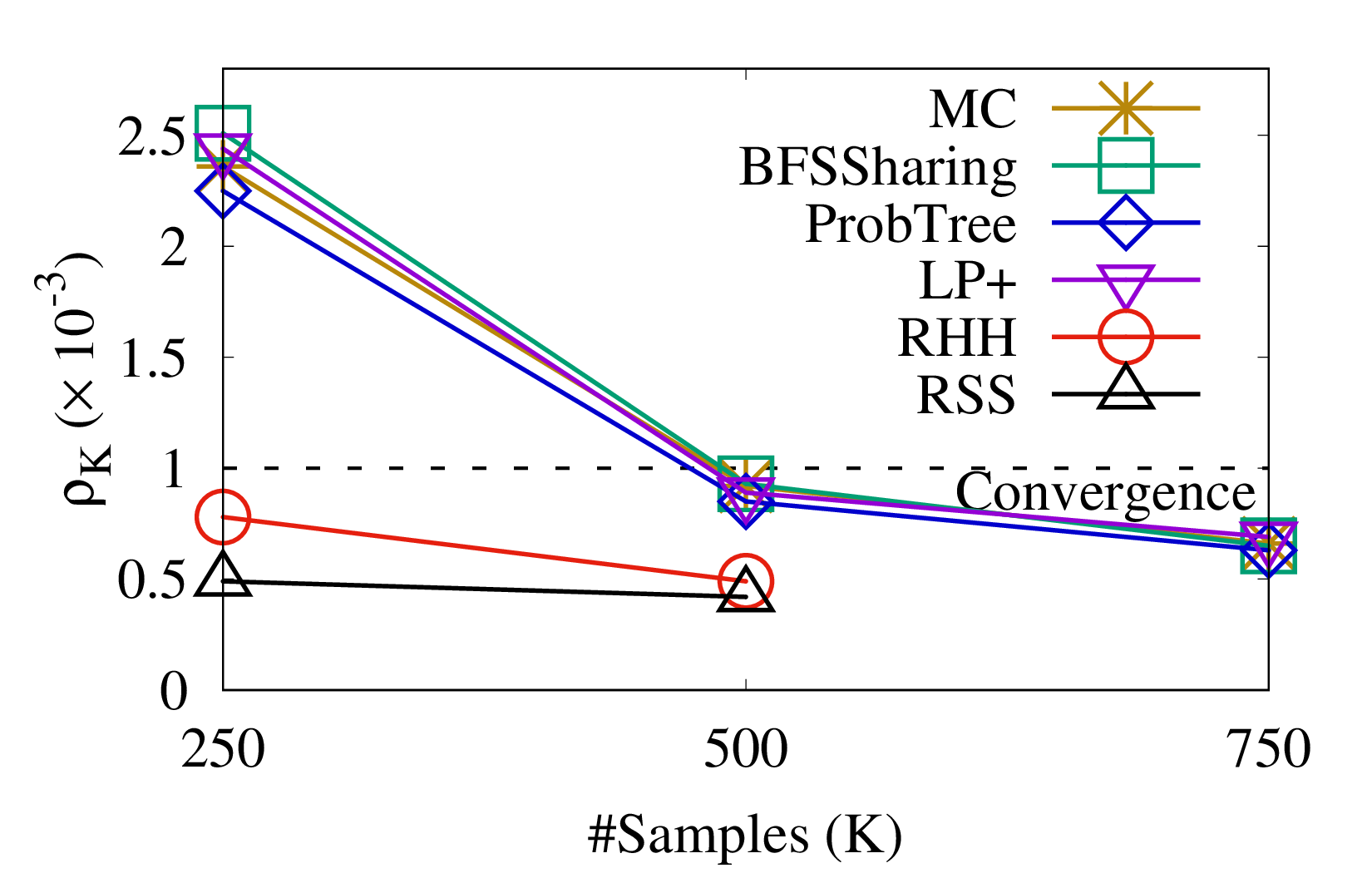}
		\label{fig:var_as}}
	\subfigure[\scriptsize {{\em DBLP\_0.2}}]
	{\includegraphics[scale=0.19,angle=270]{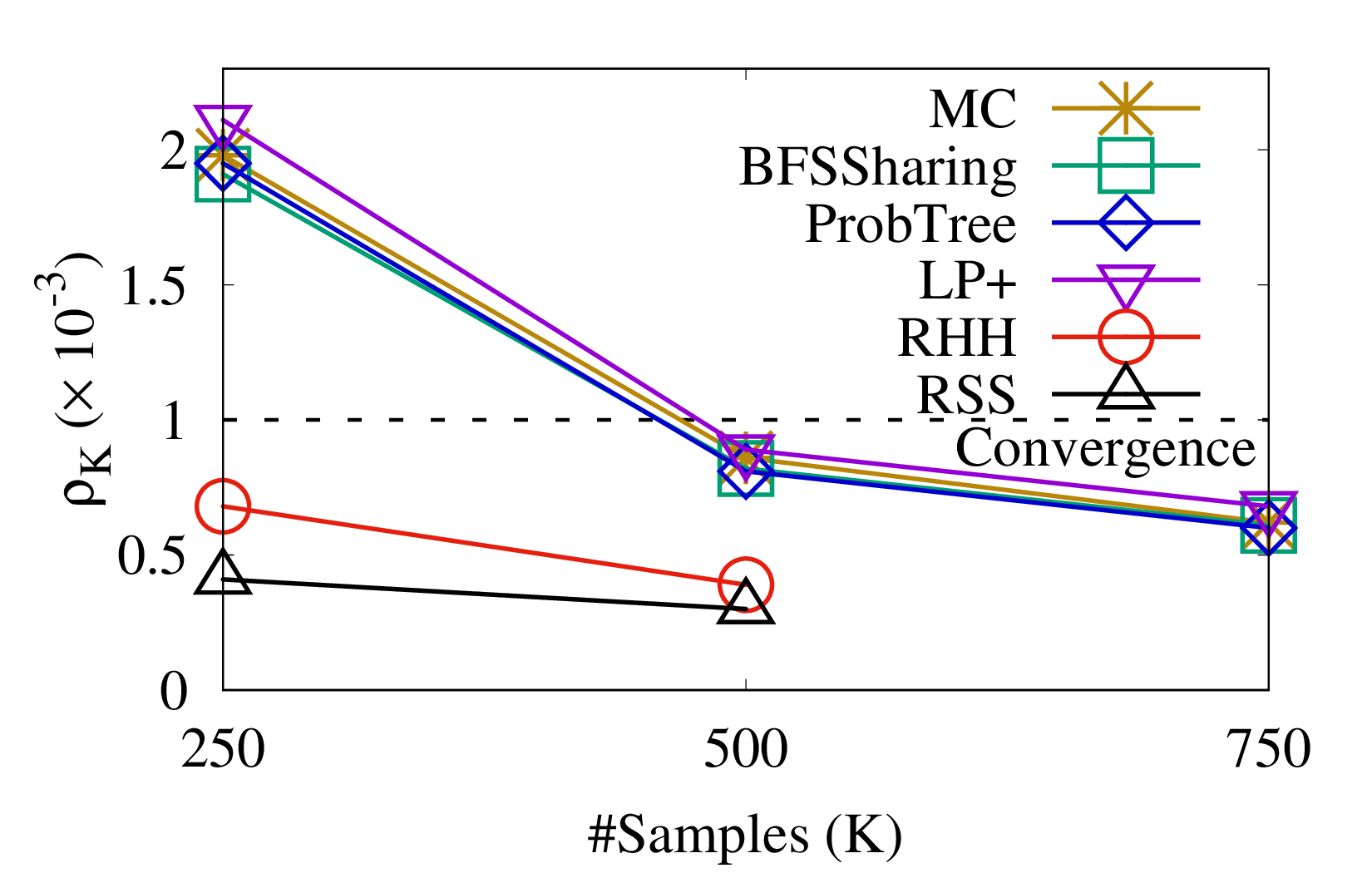}
		\label{fig:var_dblp02}}
	\subfigure[\scriptsize {{\em DBLP\_0.05}}]
	{\includegraphics[scale=0.19,angle=270]{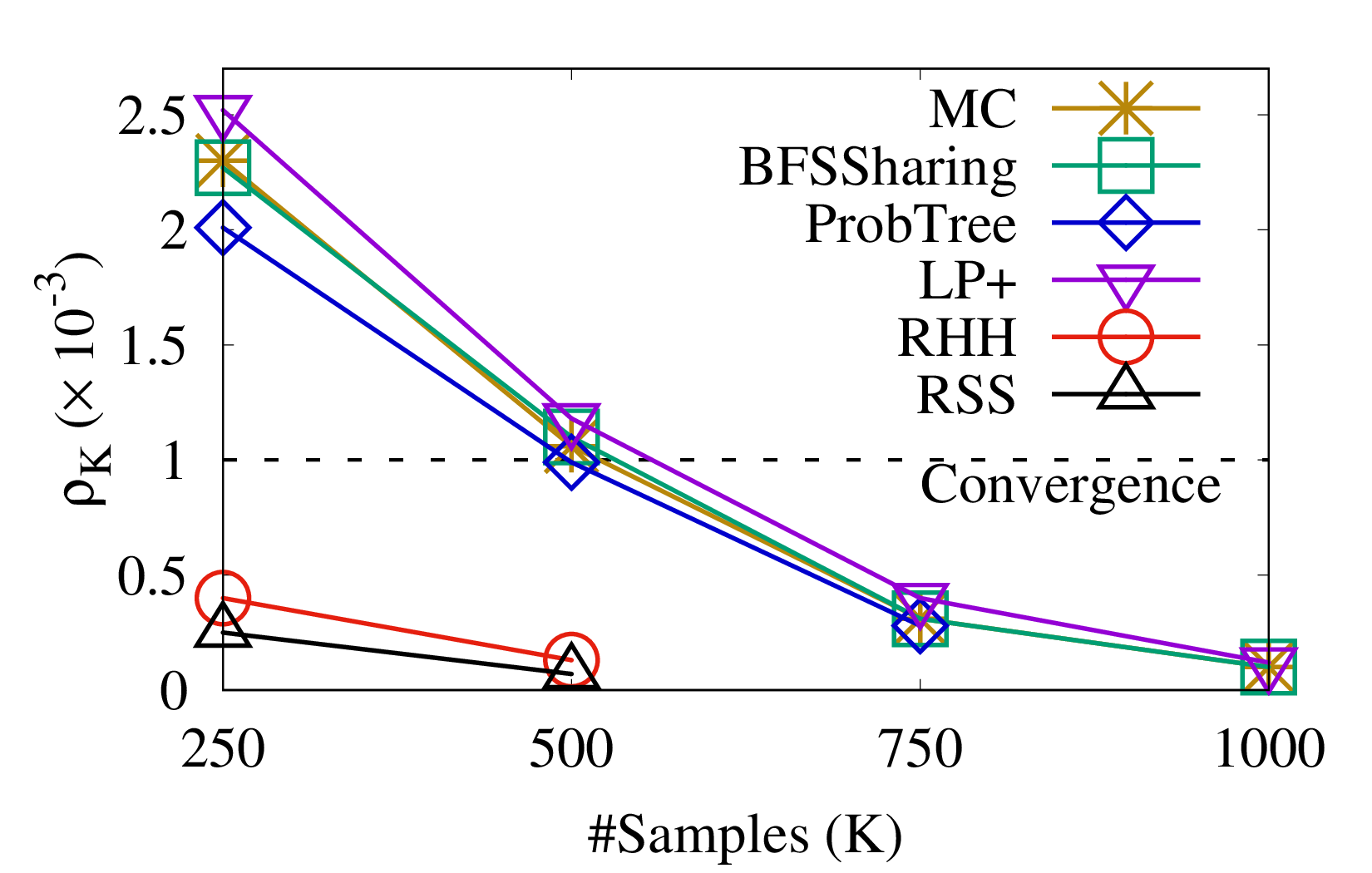}
		\label{fig:var_dblp005}}
	\subfigure[\scriptsize {{\em BioMine}}]
	{\includegraphics[scale=0.19,angle=270]{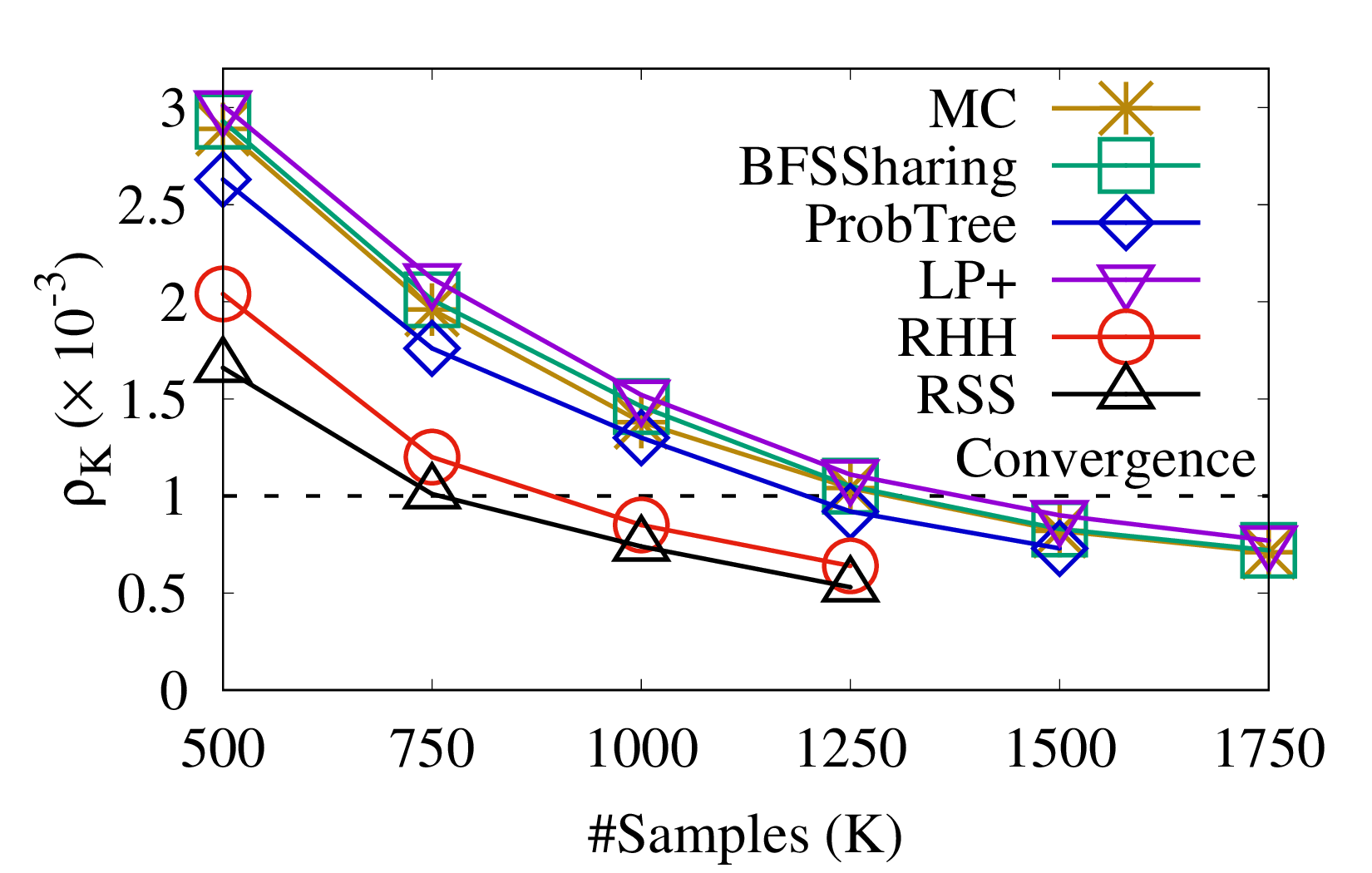}
		\label{fig:var_bio}}
	\vspace{-5mm}
	\caption{\small Comparison of estimator variance and convergence. The symbol $\rho_K$ on Y-axis denotes the ratio $\frac{V_K}{R_K}=\frac{\text{average variance at \#samples=$K$}}{\text{average reliability at \#samples=$K$}}$.}
	\label{fig:convergence}
	\vspace{-6mm}
\end{figure*}

\begin{figure}
	\subfigure
	{\includegraphics[scale=0.15,angle=270]{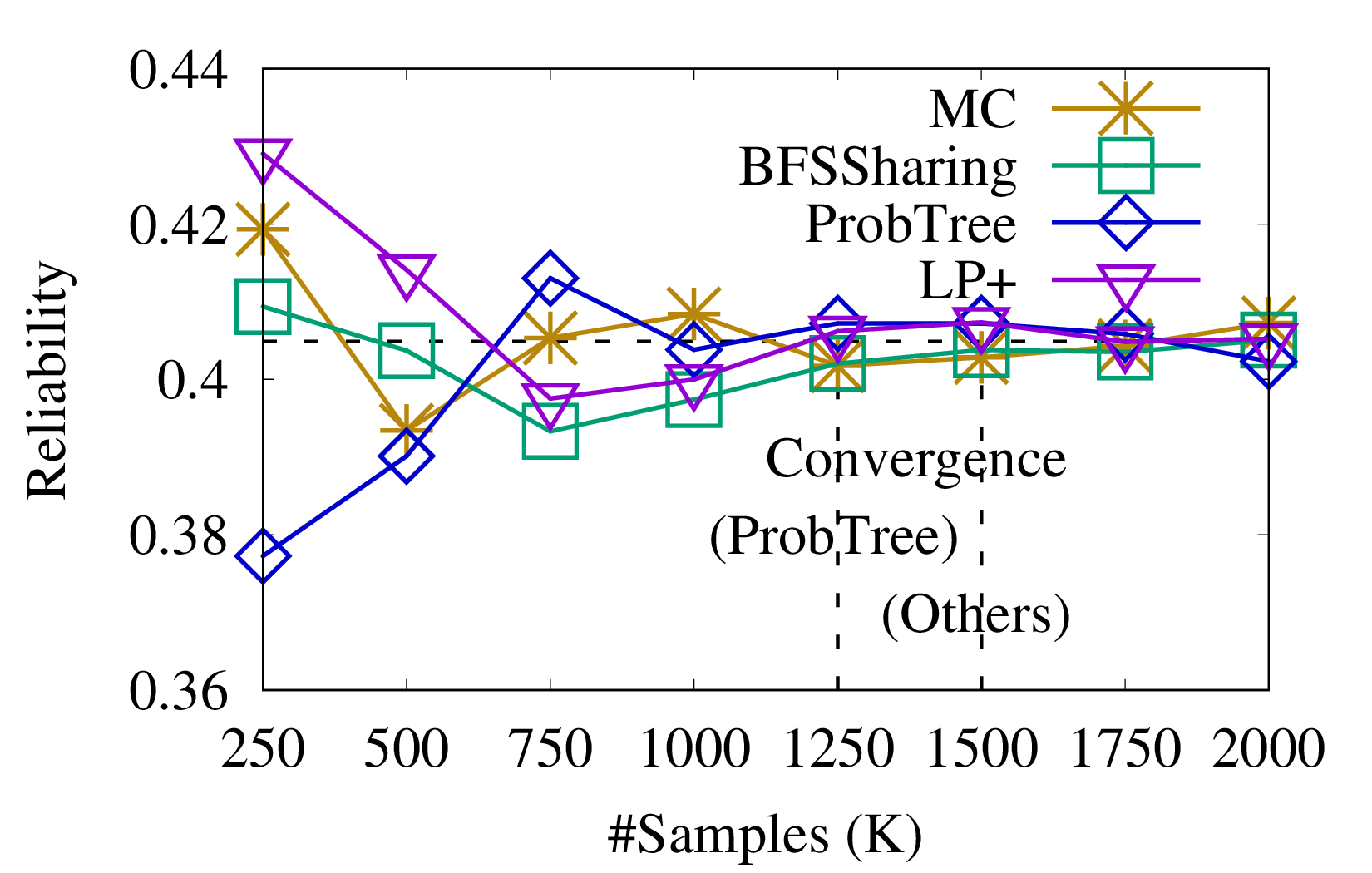}
		\label{fig:er1}}
	\subfigure
	{\includegraphics[scale=0.15,angle=270]{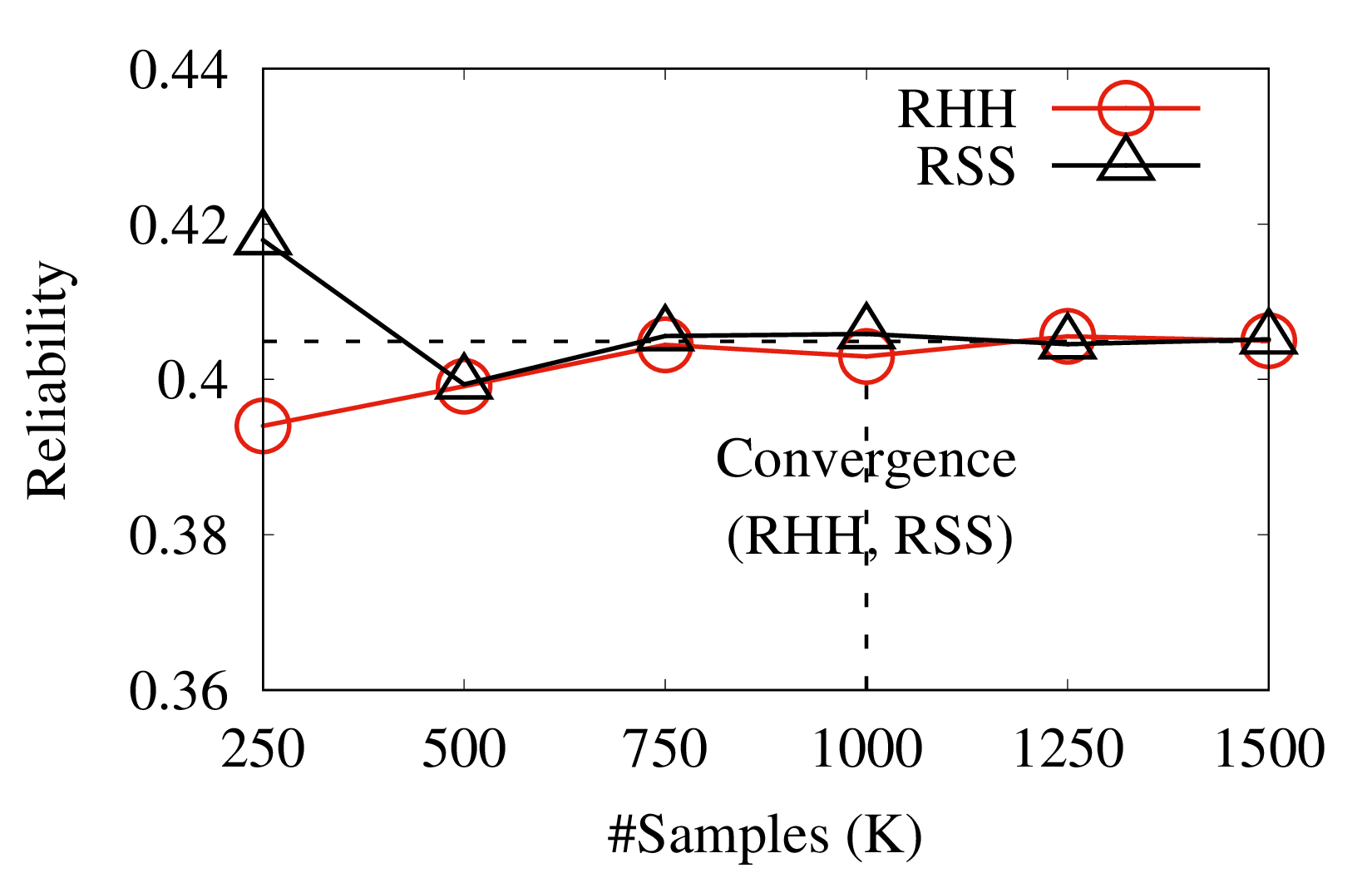}
		\label{fig:er2}}
	\vspace{-6mm}
	\captionof{figure}{\small Comparison of average reliability returned by each estimator in regards to average reliability returned by MC sampling at $K=10\,000$ (shown as horizontal dashed line), {\em BioMine}.}
	\label{fig:er}
	\vspace{-5mm}
\end{figure}

\vspace{-1.5mm}
\subsubsection{Parameters Setting}
\label{sec:para}
\vspace{-1mm}
For each dataset, 100 distinct $s$-$t$ pairs are generated as follows.
For a specific graph, we first select 100 different source nodes, uniformly at random.
For a source node, we next perform BFS up to 2 hops. Among these visited nodes
we select one target node that is 2-hop away from $s$, uniformly at random. These 100
$s$-$t$ pairs are used consistently for all six competing methods over that dataset.
All our results are averaged over these 100 $s$-$t$ pairs.

Notice that we select $s$-$t$ pairs with shortest-path distance = 2 hops, because if they
are closer, their reliability would usually be higher. On the other hand, if some
$s$-$t$ pairs are far apart, their reliability would be naturally small.
Nevertheless, we also demonstrate experimental results by varying the shortest-path distance
between $s$-$t$ pairs in Section~\ref{sec:hop}.

The initial $K$, i.e., \#samples considered in all algorithms is 250.
We then increase $K$ by a step of 250 till convergence is reached.
(we refer to Section \ref{sec:metric} for discussion on convergence).

For recursive estimators \cite{JLDW11,LYMJ14},
a few additional parameters need to be defined. For recursive sampling, we set
the threshold in Algorithm \ref{algo:recursivesampling} to be 5 as per \cite{JLDW11}.
For recursive stratified sampling, we set $r$ = 50 in Algorithm \ref{algo:rss}
as recommended in \cite{LYMJ14}. We find that these parameter values also work well
in our experimental setting. Following \cite{JLDW11,LYMJ14}, we refer to recursive
sampling and recursive stratified sampling as RHH and RSS, respectively.

\vspace{-1.5mm}
\subsubsection{Performance Metrics}
\label{sec:metric}
\vspace{-1mm}
%

\vspace{-1mm}
\spara{Variance:} Unlike MC-based estimators \cite{F86,ZZL15,LFZT17,ManiuCS17}, which report reachability from $s$ to $t$ as 1 or 0 in
each sample; both recursive sampling algorithms \cite{JLDW11,LYMJ14} estimate reliability
in a holistic manner by considering {\em all} $K$ samples. Thus, following \cite{JLDW11,LYMJ14},
we compute variance of an estimator by repeating experiments with the current number of samples ($K$).
Given an estimator and $K$, we repeat querying each $s$-$t$ pair, e.g., $s_i$-$t_i$ for $T$ times;
and we obtain $T$ estimation results for each pair $s_i$-$t_i$:
$R_1(s_i,t_i,K)$, $R_2(s_i,t_i,K)$, $\ldots$, $R_{T}(s_i,t_i,K)$. We calculate the
variance as follows.
\vspace{-2.7mm}
\begin{align}
	V(s_i,t_i,K)=\frac{1}{T-1}\sum_{j=1}^{T}(R_j(s_i,t_i,K)-\overline{R}(s_i,t_i,K))^2
	\label{eq:var}
	\vspace{-5mm}
\end{align}
Here, $\overline{R}(s_i,t_i,K)$ is the average value of these $T$ estimation results,
for a fixed $K$. Following \cite{JLDW11}, we set $T=100$.
We compute this variance for all 100 $s$-$t$ pairs, and the average
variance for this estimator, with given $K$, is:
\vspace{-3mm}
\begin{align}
	V_K=\frac{1}{100}\sum_{i=1}^{100}V(s_i,t_i,K)
	\vspace{-9mm}
	\label{eq:var_r}
\end{align}
Analogously, we define average reliability for an estimator, with given $K$, as:
\vspace{-4mm}
\begin{align}
	R_K=\frac{1}{100}\sum_{i=1}^{100}\overline{R}(s_i,t_i,K)
	\vspace{-9mm}
	\label{eq:var_r}
\end{align}
Based on our experiments, we find that as one increases $K$, the average variance
$V_K$ monotonically decreases. However, we also notice that, with same $K$,
$V_K$ varies for different datasets and estimators. This is primarily because
the average reliability $R_K$ is different for different datasets and estimators,
even for same $K$. Thus, it is {\em difficult} to fix a uniform threshold on the average
variance $V_K$ which could define convergence in all cases.

Instead, we systematically consider the ratio
$\rho_K=\frac{V_K}{R_K}$ to decide convergence of an
estimator over a given dataset. The ratio of variance to mean is also known as the {\bf index of dispersion},
which is a {\em normalized} measure of the dispersion of a dataset. Dispersion is close to zero
if all the data are similar, and increases as the data become more diverse.
Given a dataset and an estimator,
we keep increasing $K$ (in steps of 250), and when $\rho_K=\frac{V_K}{R_K}<0.001$,
we say that the convergence has been reached for that estimator and over that dataset.
For that estimator and dataset, we report average reliability at that specific value of
$K$, since this estimation is robust.

\vspace{-0.3mm}
\spara{Relative error:} By following \cite{ZZL15,ManiuCS17,JLDW11},
we report the relative error (RE) of reliability estimation for an estimator
with respect to MC sampling. This is computed as follows.
\vspace{-2mm}
\begin{align}
	RE_K={\frac{1}{100}}\sum_{i=1}^{100}{\frac{|\overline{R}(s_i,t_i,K)-\overline{R}_{MC}(s_i,t_i,Convergence)|}{\overline{R}_{MC}(s_i,t_i,Convergence)}}
	\vspace{-9mm}
\end{align}
Here, $\overline{R}_{MC}(s_i,t_i,Convergence)$ denotes the reliability of the pair $s_i$-$t_i$, as returned by MC sampling at convergence.
On the other hand, $\overline{R}(s_i,t_i,K)$ denotes the reliability of the same pair, returned by the specific estimator, at \#samples=$K$.

\vspace{-0.7mm}
\spara{Online and offline efficiency and memory usage:}
In regards to online querying, we report (a) total time cost to answer an $s$-$t$ reliability query,
(b) average time cost per sample (i.e., total time cost/\#samples), and (c) memory usage for all algorithms.
In addition, for index-based methods, i.e., BFS sharing \cite{ZZL15} and ProbTree \cite{ManiuCS17},
we report their (i) time costs for creating index, (ii) time costs for
loading index in main memory, and (iii) index sizes.
\vspace{-1mm}
\subsection{Estimator Variance and Convergence}
\label{sec:conv}
\vspace{-1mm}
We report our empirical findings on estimator variance and convergence in Figure \ref{fig:convergence} and Figure \ref{fig:er},
which are summarized below.

\begin{figure*}[t!]
	\vspace{-5mm}
	\centering
	\subfigure[\scriptsize {{\em Relative Error}}]
	{\includegraphics[scale=0.19,angle=270]{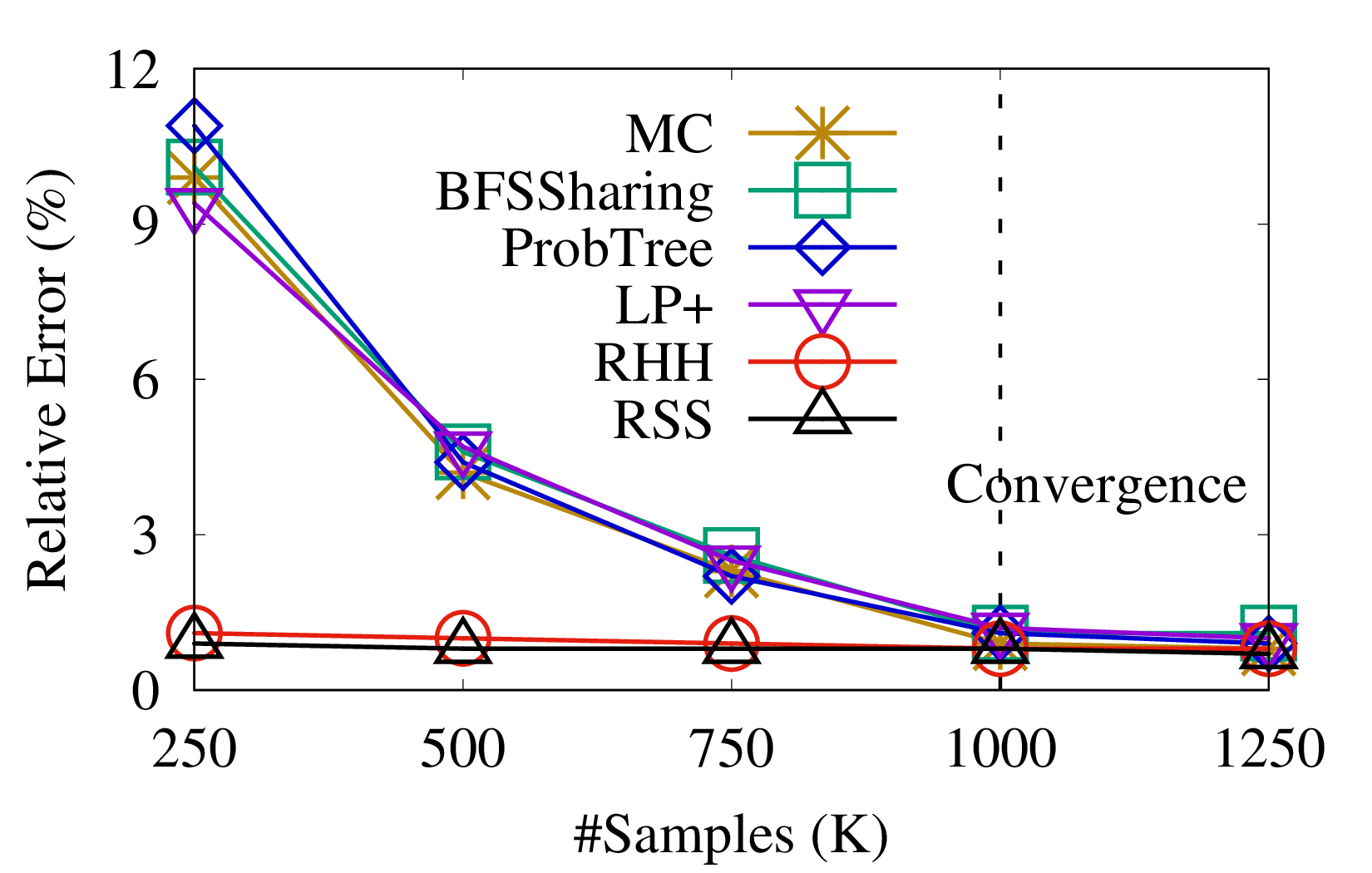}
		\label{fig:last_re}}
	\subfigure[\scriptsize {{\em Running Time}}]
	{\includegraphics[scale=0.19,angle=270]{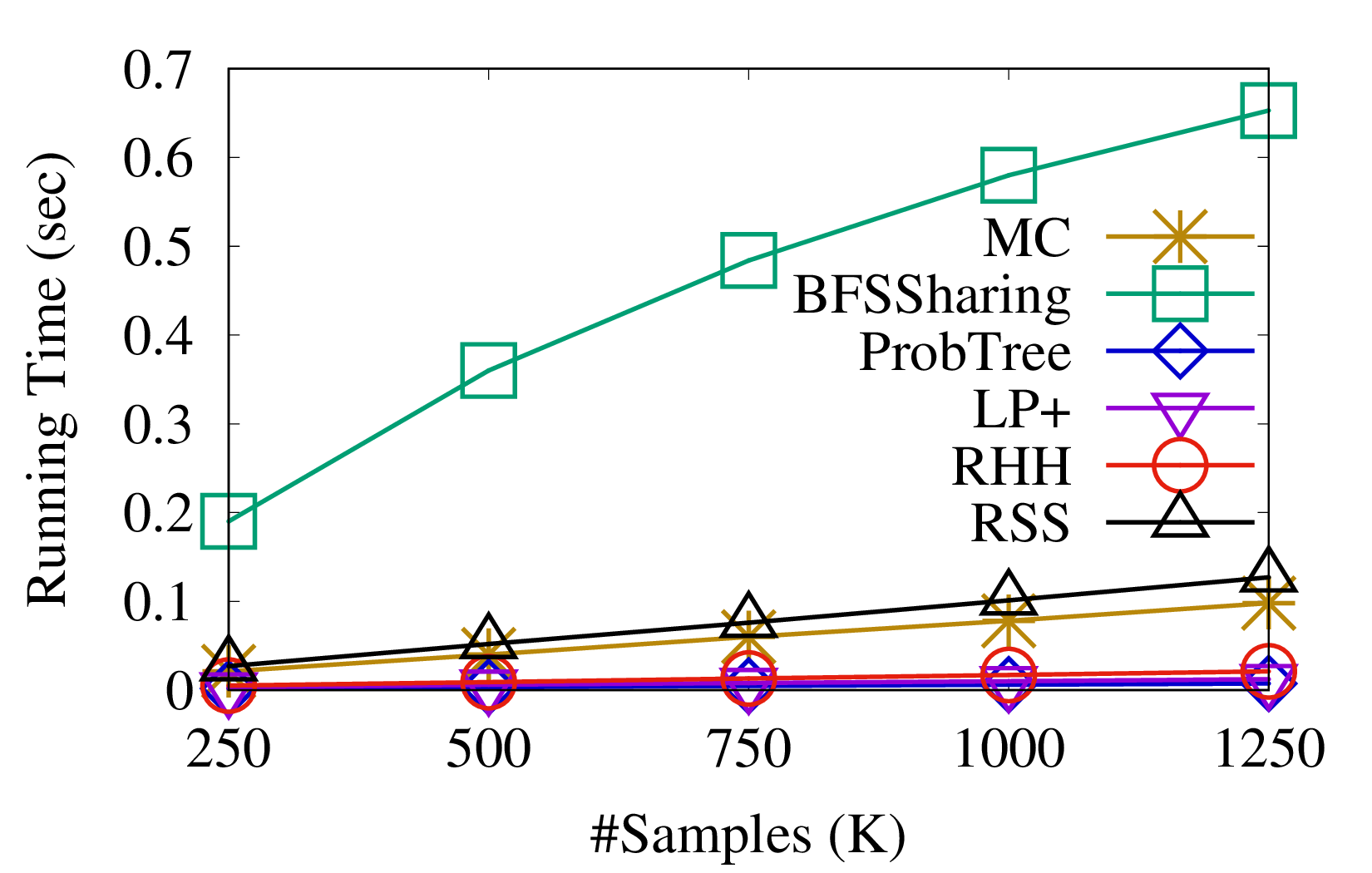}
		\label{fig:last_time}}
	\subfigure[\scriptsize {{\em Memory Usage}}]
	{\includegraphics[scale=0.19,angle=270]{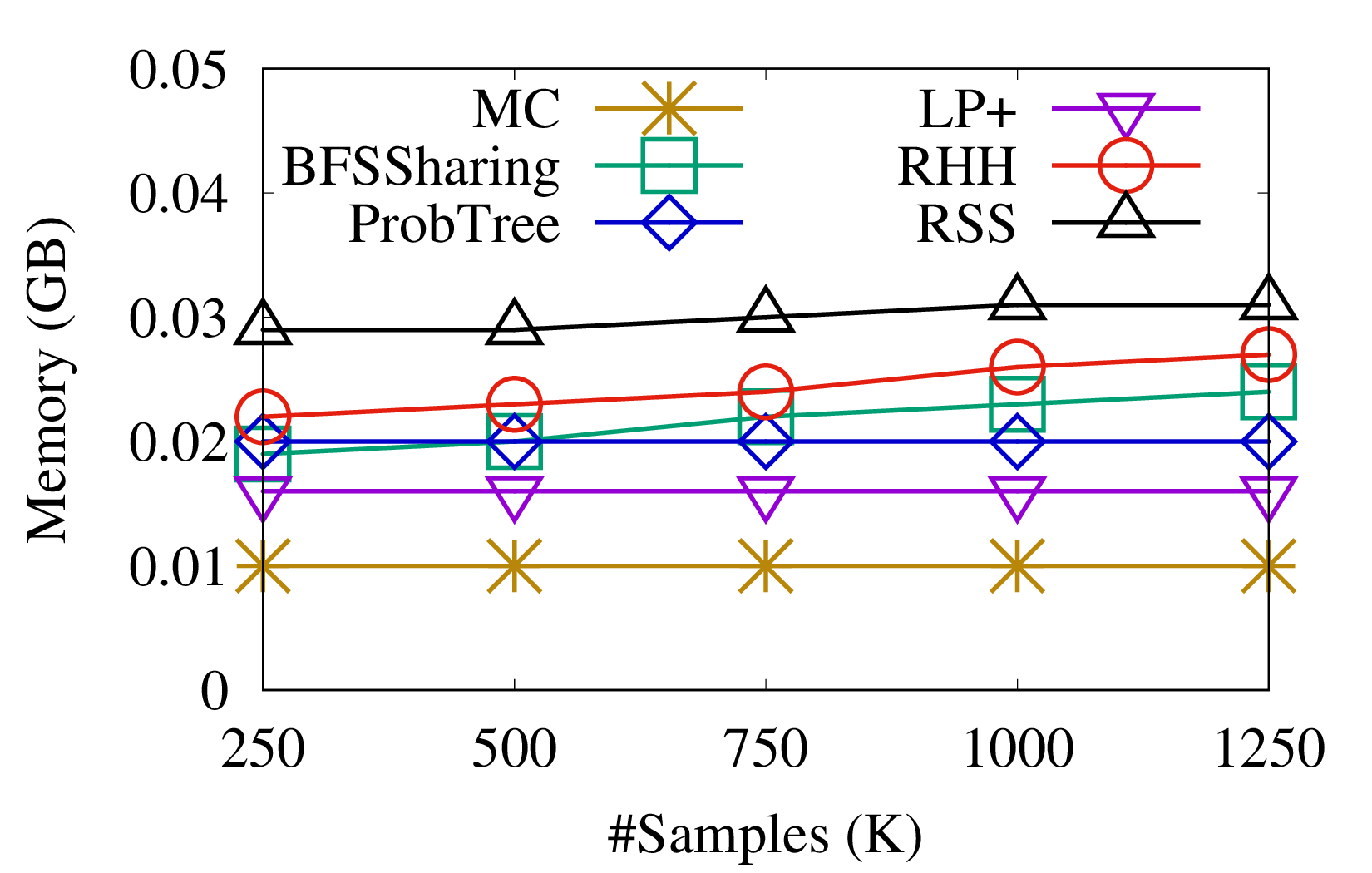}
		\label{fig:last_men}}
	\vspace{-5mm}
	\caption{\small The trade-off between relative error and running time/memory usage, {\em lastFM}.}
	\label{fig:last_RE}
	\vspace{-4mm}
\end{figure*}

\begin{figure*}[t!]
	\centering
	\subfigure[\scriptsize {{\em Relative Error}}]
	{\includegraphics[scale=0.19,angle=270]{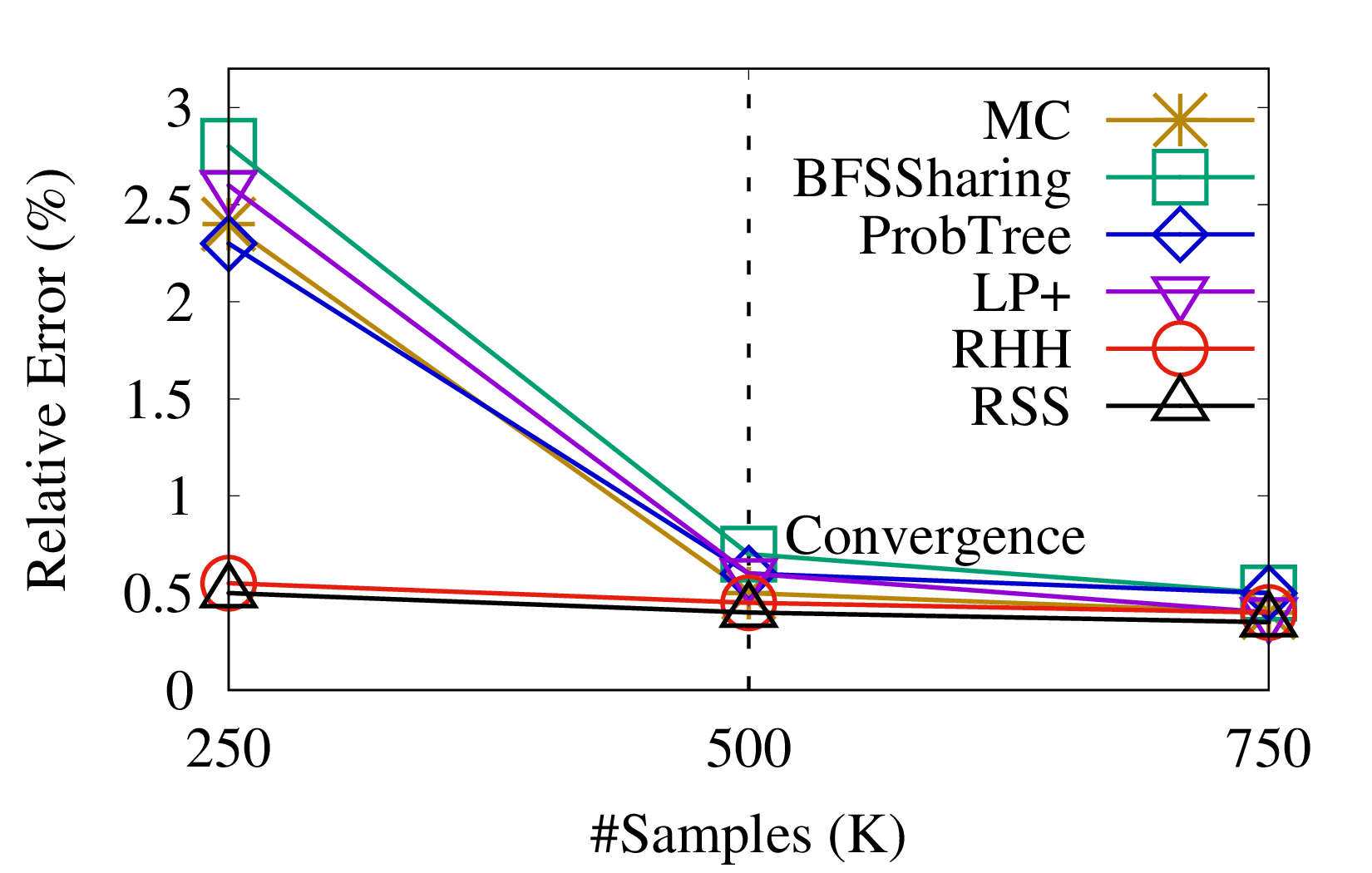}
		\label{fig:as_re}}
	\subfigure[\scriptsize {{\em Running Time}}]
	{\includegraphics[scale=0.19,angle=270]{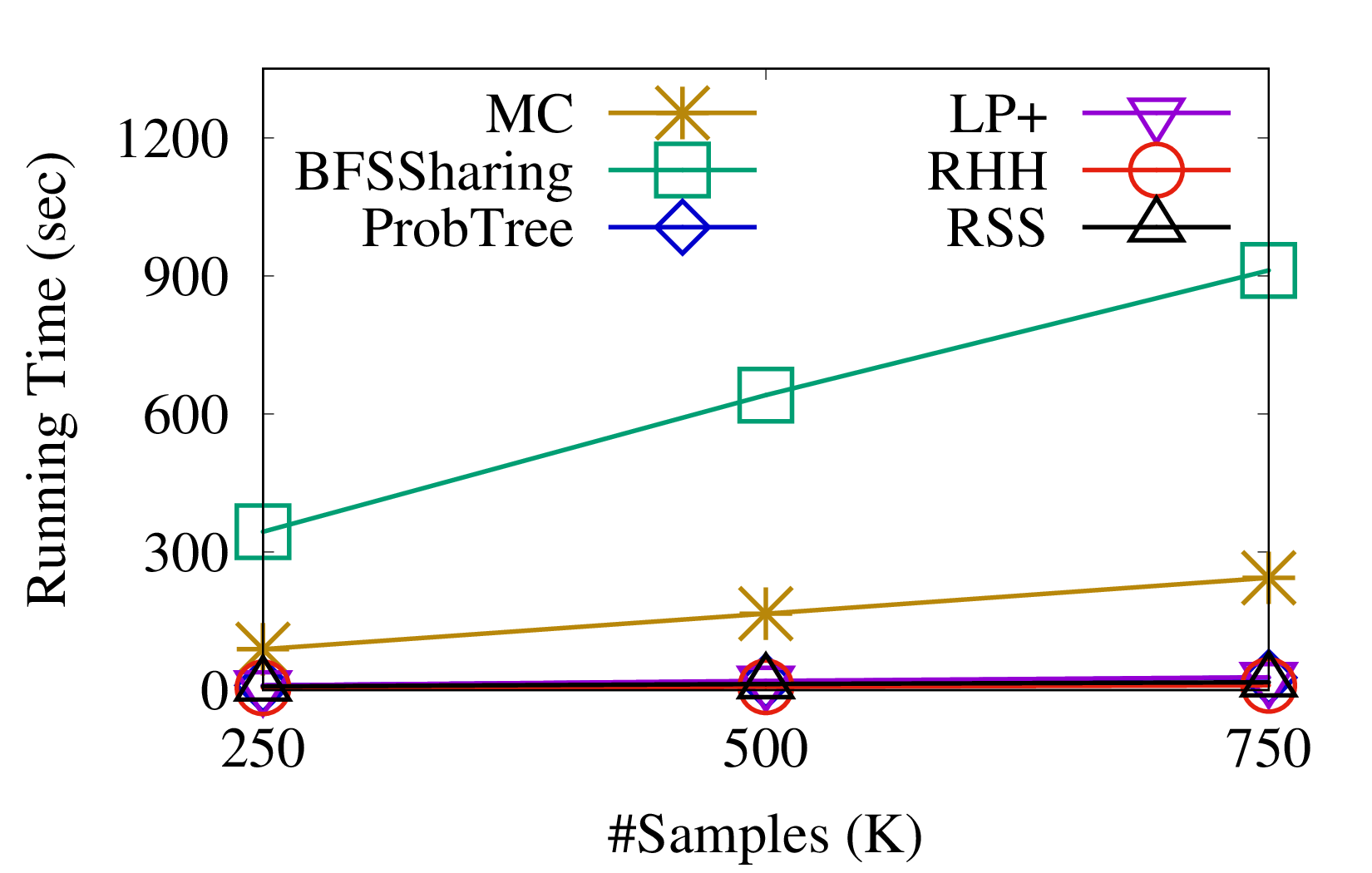}
		\label{fig:as_time}}
	\subfigure[\scriptsize {{\em Memory Usage}}]
	{\includegraphics[scale=0.19,angle=270]{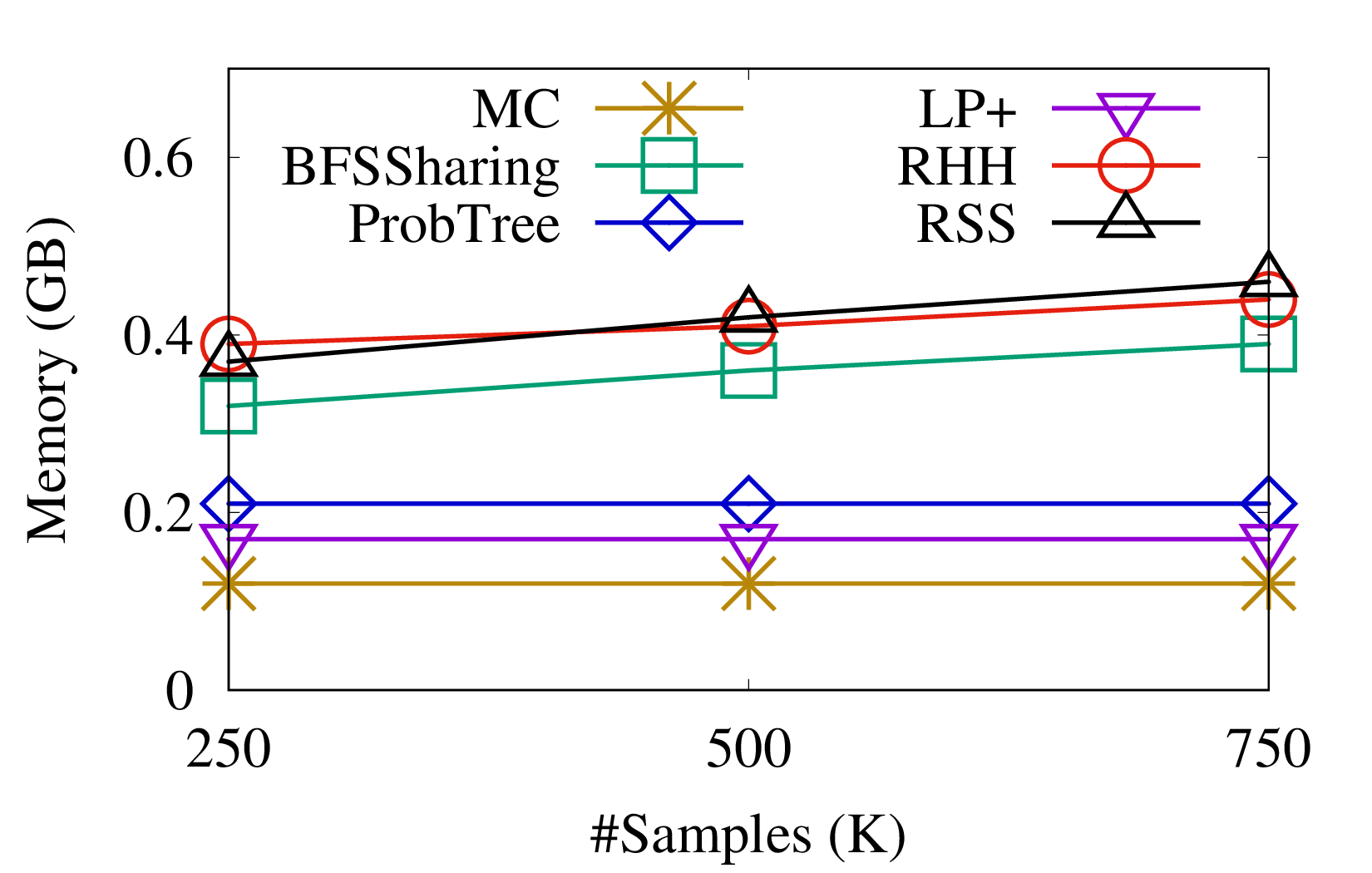}
		\label{fig:as_men}}
	\vspace{-5mm}
	\caption{\small The trade-off between relative error and running time/memory usage, {\em AS\_Topology}.}
	\label{fig:as_RE}
	\vspace{-4mm}
\end{figure*}

\begin{figure*}[t!]
	\centering
	\subfigure[\scriptsize {{\em Relative Error}}]
	{\includegraphics[scale=0.19,angle=270]{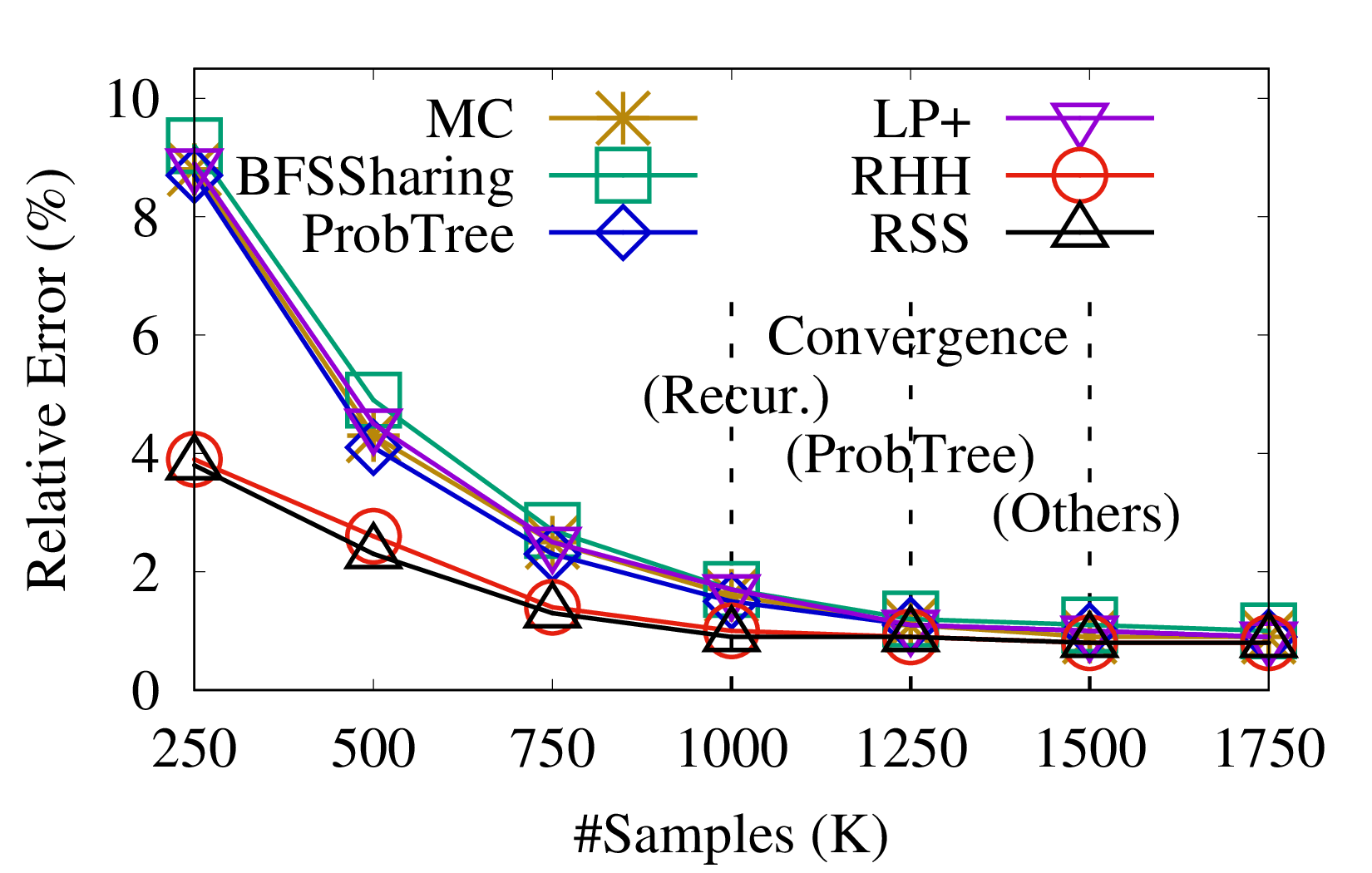}
		\label{fig:bio_re}}
	\subfigure[\scriptsize {{\em Running Time}}]
	{\includegraphics[scale=0.19,angle=270]{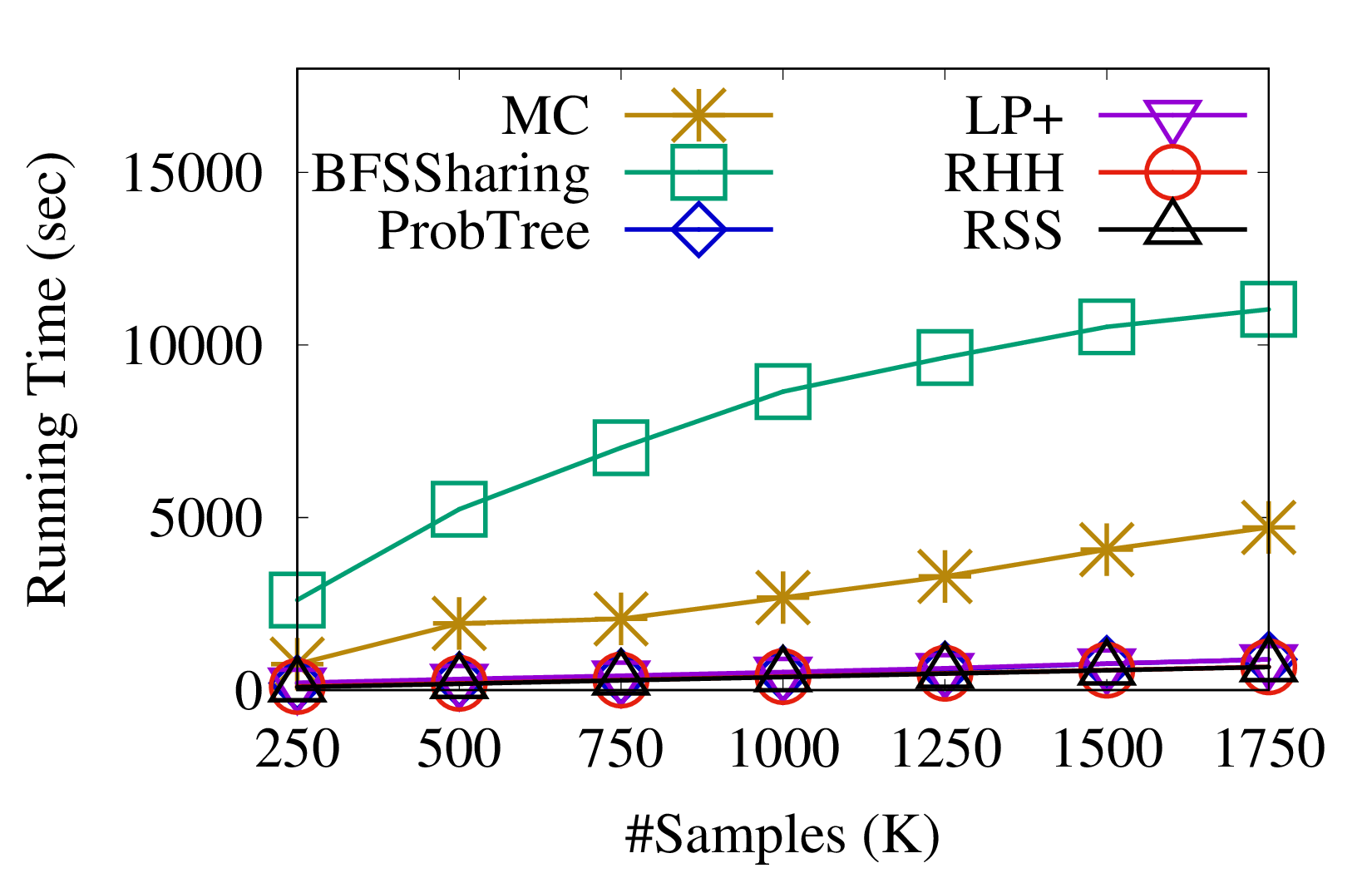}
		\label{fig:bio_time}}
	\subfigure[\scriptsize {{\em Memory Usage}}]
	{\includegraphics[scale=0.19,angle=270]{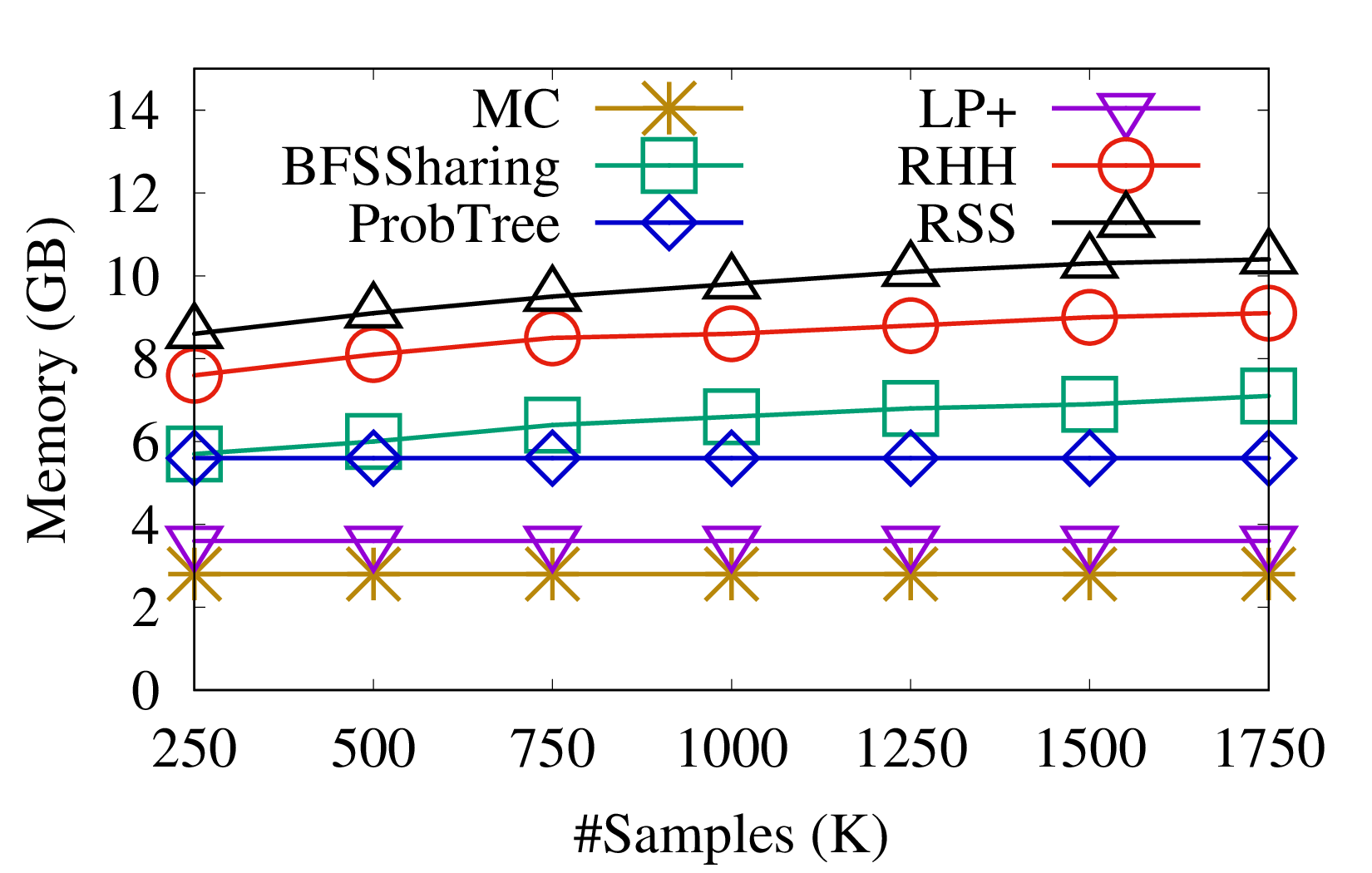}
		\label{fig:bio_men}}
	\vspace{-5mm}
	\caption{\small The trade-off between relative error and running time/memory usage, {\em BioMine}.}
	\label{fig:bio_RE}
	\vspace{-4mm}
\end{figure*}

{\bf (1)} Figure \ref{fig:convergence} depicts that among six competing estimators, four MC-based estimators: Monte Carlo (MC) sampling,
BFS Sharing, ProbTree, and Lazy Propagation (LP+) exhibit nearly similar characteristics in estimator variance.
The other two recursive estimators: Recursive Sampling (RHH) and Recursive Stratified Sampling (RSS) share
similar estimator variance between them.

In particular, BFS Sharing performs the sampling of MC offline by building indexes, and this does not change the estimator variance
compared to MC. Another index-based method, ProbTree decomposes the original graph into ``bags'' and re-organizes them in a tree structure.
When processing an online query, it generates a smaller but equivalent graph from the tree index, by ignoring irrelevant branches. We apply MC
(as the original paper \cite{ManiuCS17} did) to estimate the $s$-$t$ reliability on this smaller but equivalent graph. Hence, the estimator variance
remains same. For LP+, it utilizes the geometric distribution to decide the existence of an edge in samples, which is statistically equivalent \cite{LFZT17}
to MC; therefore results in similar variance. Overall, our empirical results confirm them.

For recursive estimators, \cite{LYMJ14} discussed that RHH is a special case of RSS when $r=1$ ($r$ denotes the number of selected edges in RSS).
In Figure \ref{fig:convergence}, we observe that the curves of RSS and RHH are close, but RSS always has lower variance than RHH.

{\bf (2)} Our experimental results in Figure \ref{fig:convergence} confirm that {\em recursive estimators have lower variance than MC-based estimators}.
By dividing the sample size according to either edge probability (e.g., RHH), or stratum probability (e.g., RSS),
the uncertainty during sampling is reduced, which results in lower variance. It can be noted that {\em some later proposed methods, e.g.,
ProbTree and LP+ do not outperform RHH and RSS in estimator variance}.

{\bf (3)} As discussed in Section \ref{sec:mc_sampling}, larger $K$ ensures more accurate result by MC sampling. In Figure \ref{fig:er}, we compare the
reliability returned by each estimator with respect to that of MC sampling with very large $K$, e.g., $K=10000$ (shown as horizontal dashed line) \cite{WCLYGC15, YMPH16}.
{\em Figure \ref{fig:er} clearly presents that the reliability estimated at variance convergence is very close to that at very large $K$,
which indicates that estimator variance convergence can help find a sample size for high quality estimation}.

{\bf (4)} Past works on uncertain graphs generally employed some standard $K$ (i.e., \#samples) in their experiments, such as
$K$=500 \cite{PBGK10}, 1\,000 \cite{ManiuCS17,JLDW11}, 2\,000 \cite{LYMJ14,ZZL15}, or even 10\,000 \cite{KKT03}, while mentioning
that they observed convergence for those specific $K$ in their experimental setting. For future researchers and
practitioners, we would like to emphasize that {\em there is no single $K$ such that all estimators achieve convergence across all datasets},
which can be confirmed from our empirical findings. The $K$ at convergence for every estimator on each dataset is listed in Tables \ref{tab:AccLast}-\ref{tab:AccBio}
(observed from Figure \ref{fig:convergence}). Interestingly for MC-based methods, $K$ values at convergence are almost different over various datasets.
Recursive estimators reach convergence with smaller $K$, e.g., $K$=250 on three out of five datasets.
However, they require around 750 and 1000 samples on {\em NetHept} and {\em BioMine} datasets, respectively, for convergence.
{\em Recursive methods generally converge with about 500 less samples than MC-based methods on the same dataset.}

{\bf (5)} With a closer look, we find that {\em ProbTree has a slight improvement in estimator variance compared to three other MC-based methods}.
This is because the ProbTree index pre-computes the reliability information contained in a bag's children subtrees, and stores it in the current bag.
By directly applying such pre-computed probabilities, one can reduce the uncertainty of sampling, and lower the estimator variance.
In our experimental results on {\em NetHept} and {\em Biomine}, ProbTree requires around 250 less samples for convergence, when compared with other MC-based estimators.

\begin{table*}[tb!]
	\vspace{-1mm}
	\small
	\parbox{.45\linewidth}{
		\centering
		\caption{\small Comparison of relative error (RE): {\em LastFM} }
		\begin{tabular} {l||l|l|l||l|l||}
			\hline
			\multirow{2}{*}{Estimator}& \multicolumn{3}{c||}{At Convergence} & \multicolumn{2}{c||}{At $K$=1000} \\ \cline{2-6}
			& $K$ & $R_K$ & RE (\%)  & $R_K$ & RE (\%)  \\
			\hline \hline
			MC          & 1000 & 0.1025 & 0.00 &0.1025 & 0.00   \\
			BFS Sharing & 1000 & 0.1030 & {\bf 0.97}  & 0.1030 &{\bf 0.97}  \\
			ProbTree    & 1000 & 0.1007 & 1.77  &0.1007 & 1.77   \\
			Lazy Propagation & 1000 & 0.1052 & 1.84  & 0.1052 & 1.84  \\
			Recursive (RHH) & 250 & 0.1041 & 1.91 &0.1040 & 1.88  \\
			Recur. Stratified (RSS) & 250 & 0.1018 & 1.06  & 0.1020 & 1.07 \\
			\hline \hline
			Pairwise Deviation & \multicolumn{2}{c|}{} & 0.53 & & 0.52 \\
			\hline
		\end{tabular}
		\label{tab:AccLast}
	}
	\hfill
	\parbox{.45\linewidth}{
		\centering
		\caption{\small Comparison of relative error (RE): {\em NetHept} }
		\begin{tabular} {l|l|l||l|l}
			\hline
			\multicolumn{3}{c||}{At Convergence} & \multicolumn{2}{c}{At $K$=1000} \\ \hline
			$K$ & $R_K$ & RE (\%) & $R_K$ & RE (\%) \\
			\hline \hline
			1250 & 0.00190 & 0.00  & 0.00183 & 2.18 \\
			1250 & 0.00194 & 1.83  & 0.00187 & 2.01 \\
			1000 & 0.00187 & {\bf 1.47} & 0.00180 & {\bf 1.47} \\
			1250 & 0.00196 & 1.93  & 0.00180 & 2.41 \\
			750 & 0.00196 & 1.73 & 0.00196 & 1.78  \\
			750 & 0.00192 & {\bf 1.47} &  0.00192 & 1.49 \\
			\hline
			\hline
			\multicolumn{2}{c|}{} & 0.26 & & 0.48\\
			\hline
		\end{tabular}
		\label{tab:AccHepU}
	}
	\vspace{-6mm}
\end{table*}

\begin{table*}[tb!]
	\vspace{-0.5mm}
	\small
	\parbox{.45\linewidth}{
		\centering
		\caption{\small Comparison of relative error (RE): {\em AS\_Topology} }
		\begin{tabular} {l||l|l|l||l|l||}
			\hline
			\multirow{2}{*}{Estimator}& \multicolumn{3}{c||}{At Convergence} & \multicolumn{2}{c||}{At $K$=1000} \\ \cline{2-6}
			& $K$ & $R_K$ & RE (\%)  & $R_K$ & RE (\%)  \\
			\hline \hline
			MC &  500$\ \ $ & 0.5226 & 0.00 & 0.5206 & 0.38\\
			BFS Sharing & 500 & 0.5252 & 0.50 & 0.5248 & 0.42\\
			ProbTree & 500 & 0.5219 & {\bf 0.13} & 0.5217 & 0.17\\
			Lazy Propagation & 500 & 0.5212 & 0.27 & 0.5219 & {\bf 0.13}\\
			Recursive (RHH) & 250 & 0.5241 & 0.29 & 0.5238 & 0.23\\
			Recur. Stratified (RSS) & 250 & 0.5234 & 0.15 & 0.5216 & 0.19\\
			\hline \hline
			Pairwise Deviation & \multicolumn{2}{c|}{} & 0.18 & & 0.13\\
			\hline
		\end{tabular}
		\label{tab:AccAS}
	}
	\hfill
	\parbox{.45\linewidth}{
		\centering
		\caption{\small Comparison of relative error (RE): {\em DBLP\_0.2} }
		\begin{tabular} {l|l|l||l|l}
			\hline
			\multicolumn{3}{c||}{At Convergence} & \multicolumn{2}{c}{At $K$=1000} \\ \hline
			$K$ & $R_K$ & RE (\%) & $R_K$ & RE (\%) \\
			\hline \hline
			500$\ \ $ & 0.6139  & 0.00  & 0.6108 &0.33 \\
			500\quad & 0.6098 &  0.99 & 0.6094 & {\bf 0.93} \\
			500 & 0.6213 & 1.03 & 0.6253 & 1.01 \\
			500 & 0.6164 & 1.08  & 0.6194 & 1.02 \\
			250 & 0.6110 & 1.10  & 0.6104 & 1.20 \\
			250 & 0.6103 &  {\bf 0.97} &  0.6094& 1.01 \\
			\hline
			\hline
			\multicolumn{2}{c|}{} & 0.07 & & 0.11 \\
			\hline
		\end{tabular}
		\label{tab:AccDBLP02}
	}
	\vspace{-6mm}
\end{table*}

\begin{table*}[tb!]
	\vspace{-0.5mm}
	\small
	\parbox{.45\linewidth}{
		\centering
		\caption{\small Comparison of relative error (RE): {\em DBLP\_0.05} }
		\begin{tabular} {l||l|l|l||l|l||}
			\hline
			\multirow{2}{*}{Estimator}& \multicolumn{3}{c||}{At Convergence} & \multicolumn{2}{c||}{At $K$=1000} \\ \cline{2-6}
			& $K$ & $R_K$ & RE (\%)  & $R_K$ & RE (\%)  \\
			\hline \hline
			MC & 750 & 0.2128 & 0.00  & 0.2136 & {\bf 0.17} \\
			BFS Sharing & 750$\ \ $ & 0.2133$\ \ $ & {\bf 1.26}  & 0.2134 & 1.15 \\
			ProbTree & 750 & 0.2156  & 1.37 & 0.2162 & 1.40\\
			Lazy Propagation & 750 & 0.2154 & 1.52 & 0.2153 & 1.49\\
			Recursive (RHH) & 250 & 0.2114 & 1.39 & 0.2108 & 1.32\\
			Recur. Stratified (RSS) & 250 & 0.2124 & 1.36 & 0.2131 & 1.37 \\
			\hline \hline
			Pairwise Deviation & \multicolumn{2}{||c|}{} & 0.11 & & 0.15\\
			\hline
			
		\end{tabular}
		\label{tab:AccDBLP005}
	}
	\hfill
	\parbox{.45\linewidth}{
		\centering
		\caption{\small Comparison of relative error (RE): {\em BioMine} }
		\begin{tabular} {l|l|l||l|l}
			\hline
			\multicolumn{3}{c||}{At Convergence} & \multicolumn{2}{c}{At $K$=1000} \\ \hline
			$K$ & $R_K$ & RE (\%) & $R_K$ & RE (\%) \\
			\hline \hline
			1500 & 0.4019 & 0.00 & 0.4038 & 1.39 \\
			1500 & 0.4040 & 0.85  & 0.4041 &1.62 \\
			1250 & 0.4050 & {\bf 0.79} & 0.4077 &1.43 \\
			1500 & 0.4013 & 1.09 & 0.4068 & 2.47\\
			1000 & 0.4052 & 1.15 & 0.4052 &1.15 \\
			1000 & 0.4047 & 1.08  & 0.4047 & {\bf 1.08}\\
			\hline \hline
			\multicolumn{2}{c|}{} & 0.19 & & 0.65\\
			\hline
		\end{tabular}
		\label{tab:AccBio}
	}
	\vspace{-4mm}
\end{table*}

%

\vspace{-1mm}
\subsection{Trade-off among Relative Error, Running Time, and Memory Usage}
\label{sec:tradeoff}
\vspace{-1mm}

Figures~\ref{fig:last_RE},~\ref{fig:as_RE} and ~\ref{fig:bio_RE} demonstrate the trade-off among estimation error, running time, and memory usage.
The estimation error is provided as the relative error with respect to the reliability returned by MC sampling at variance convergence (as discussed in Section~\ref{sec:metric}).

As shown in each subfigure (a) of Figures~\ref{fig:last_RE},~\ref{fig:as_RE}, and ~\ref{fig:bio_RE}, when reaching the estimator variance convergence, the relative error rates of all six methods are {\bf (1)} very close to each other; {\bf (2)} below 2\%; and {\bf (3)} also converge. The estimator accuracy can benefit little by further increasing the sample size $K$. However, {\em the running time of all estimators grows about linearly with the sample size $K$. The memory usages of estimators are not very sensitive to the sample size.} Memory usage of MC, ProbTree, and LP+ nearly remains the same all the way. With larger number of samples, more indexes are required to be loaded into main memory by BFS Sharing, thus slightly increasing its memory cost. For recursive methods, larger $K$ can allow larger recursion depths, thus more memory is consumed. {\em In summary, the estimator variance convergence can help us to find the sweet point which balances the estimator error and running time/memory consumption.}
More detailed accuracy and efficiency comparison can be found in Sections~\ref{sec:acc} and \ref{sec:eff}.

\vspace{-1mm}
\subsection{Estimator Accuracy}
\label{sec:acc}
We compare relative errors (with respect to MC Sampling at variance convergence, Section \ref{sec:metric}) of all algorithms. Relative errors are reported
(a) at convergence for that estimator, and (b) at $K$=1000. We report relative errors also at $K$=1000, since many prior works \cite{JLDW11,LYMJ14,ManiuCS17,ZZL15}
did the same, irrespective of whether an estimator has converged or not. Our results are given in Tables~\ref{tab:AccLast}-\ref{tab:AccBio}.

{\bf (1)} {\em When the value of $K$ at convergence is larger than 1000 for a method, its relative error at convergence is smaller than that at $K$=1000} (e.g.,
see the results over {\em NetHept} and {\em BioMine}). In contrast, {\em if the value of $K$ at convergence is smaller than 1000 for a method,
the relative error nearly remains the same when increasing $K$ to 1000} (e.g., see the results over {\em lastFM}, {\em AS\_Topology}, {\em DBLP\_0.2}, and {\em DBLP\_0.05} datasets).
Therefore, $K$=1000 is not a fair setting to compare the estimator accuracy across all estimators and datasets. Rather,
$K$ at convergence for that estimator ensures higher accuracy. Similarly, if the estimator has already converged at some $K<$1000, the relative error would not reduce
further, instead only the running time will increase for larger $K$.

{\bf (2)} {\em At convergence, relative errors for all six methods are low ($<$ 2\%) and comparable (no common winner exists),
which indicates that our approach of finding convergence (based on {\bf index of dispersion}) ensures high accuracy}.
The best relative error rate on each dataset is below 1.5\%.

Previous work \cite{JLDW11} compared the relative error at $K$=1000, and concluded that RHH had better accuracy over MC-based
methods. However, this does not hold when we consider $K$ at convergence for respective methods. Empirically we find that $K$=1000
is sufficient for RHH and RSS to achieve convergence, while other methods may still require more samples (e.g., on {\em NetHept} and {\em BioMine})
for their convergence. Only considering $K$=1000 is unfair to them. We notice that at convergence, there does not exist a common winner in regards to relative error
among these six estimators.

We further calculate the pairwise deviation ($D$) of relative errors (RE) across different estimators on each dataset.
\vspace{-2mm}
\begin{align}
	D=\frac{1}{5*6}\sum_{i=1}^{6}\sum_{j=1}^{6}|RE(i)-RE(j)|
	\vspace{-9mm}
\end{align}
Here, $RE(i)$ denotes the relative error of method $i$.
One can observe that if a dataset requires more than 1000 samples for convergence (e.g., {\em NetHept} in Table~\ref{tab:AccHepU} and {\em BioMine}
in Table~\ref{tab:AccBio}), the pairwise deviation of relative errors among estimators over that dataset significantly decreases when increasing $K$ from 1000 to that at convergence
of respective estimators. These results further demonstrate that {\em at convergence, relative errors for all six methods are low and comparable; and measuring relative error
at a specific value of $K$ (e.g., $K$=1000) across all methods can be unfair to certain estimators}.
%
%
%

\begin{table*}[tb!]
	\vspace{-1mm}
	\small
	\parbox{.45\linewidth}{
		\centering
		\caption{\small Comparison of running time: {\em LastFM} }
		\begin{tabular} {l||l|l||l||l||}
			\hline
			\multirow{2}{*}{Estimator}& \multicolumn{2}{c||}{At Convergence} & \multicolumn{1}{c||}{At $K$=1000} & Time Per \\ \cline{2-4}
			& $K$ & Time (sec)  & Time (sec) &  Sample (ms)  \\
			\hline \hline
			MC          & 1000 & 0.078 &0.078 &0.078 \\
			BFS Sharing & 1000 & 0.593 &0.593 &0.593 \\
			ProbTree    & 1000 & 0.006 & {\bf 0.006}& {\bf 0.006}  \\
			Lazy Propagation & 1000 & 0.010 & 0.010 & 0.010  \\
			Recursive (RHH) & 250 & {\bf 0.004} & 0.017 & 0.016  \\
			Recur. Stratified (RSS) & 250 & 0.026 & 0.101 &  0.104 \\
			\hline
		\end{tabular}
		\label{tab:EffLast}
	}
	\hfill
	\parbox{.45\linewidth}{
		\centering
		\centering
		\caption{\small Comparison of running time: {\em NetHept} }
		\begin{tabular} {l|l||l||l}
			\hline
			\multicolumn{2}{c||}{At Convergence} & \multicolumn{1}{c||}{At $K$=1000} & Time Per \\ \cline{1-3}
			$K$ & Time (sec)  & Time (sec) &  Sample (ms)  \\
			\hline \hline
			1250 & 0.027 &	0.022 &	0.021  \\
			1250 & 0.123  & 0.104 & 0.098 \\
			1000 & 0.006   & 0.006 & 0.006 \\
			1250 & 0.010  & 0.008 & 0.008 \\
			750 & {\bf 0.004}   &  {\bf 0.004} & {\bf 0.005} \\
			750 & 0.013 &  0.016 & 0.017 \\
			\hline
		\end{tabular}
		\label{tab:EffHepU}
	}
	\vspace{-6mm}
\end{table*}
\begin{table*}[tb!]
	\vspace{-0.5mm}
	\small
	\parbox{.45\linewidth}{
		\centering
		\caption{\small Comparison of running time: {\em AS\_Topology} }
		\begin{tabular} {l||l|l||l||l||}
			\hline
			\multirow{2}{*}{Estimator}& \multicolumn{2}{c||}{At Convergence} & \multicolumn{1}{c||}{At $K$=1000} & Time Per \\ \cline{2-4}
			& $K$ & Time (sec)  & Time (sec) &  Sample (ms)  \\
			\hline \hline
			MC   & 500$\ \ $ & 166 & 327 & 332  \\
			BFS Sharing &  500 & 641 & 1235 & 1282\\
			ProbTree    &  500 & 19 & {\bf 38} & {\bf 38}\\
			Lazy Propagation & 500 & 20 & 39 & 40\\
			Recursive (RHH) & 250 & {\bf 12} & 45 & 48\\
			Recur. Stratified (RSS) & 250 & 14 & 55 & 56\\
			\hline
		\end{tabular}
		\label{tab:EffAS}
	}
	\hfill
	\parbox{.45\linewidth}{
		\centering
		\centering
		\caption{\small Comparison of running time: {\em DBLP\_0.2} }
		\begin{tabular} {l|l||l||l}
			\hline
			\multicolumn{2}{c||}{At Convergence} & \multicolumn{1}{c||}{At $K$=1000} & Time Per \\ \cline{1-3}
			$K$ & Time (sec)  & Time (sec) &  Sample (ms)  \\
			\hline \hline
			500$\ \ $ & 315   & 622 & 629 \\
			500 & 2220  & 4408 & 4441  \\
			500 & 55  & 107 & 109 \\
			500 & 43   & {\bf 85} & {\bf 86} \\
			250 & {\bf 38}  & 150& 152  \\
			250 & 39  & 151 & 156 \\
			\hline
		\end{tabular}
		\label{tab:EffDBLP02}
	}
	\vspace{-6mm}
\end{table*}
\begin{table*}[tb!]
	\vspace{-0.5mm}
	\small
	\parbox{.45\linewidth}{
		\centering
		\caption{\small Comparison of running time: {\em DBLP\_0.05} }
		\begin{tabular} {l||l|l||l||l||}
			\hline
			\multirow{2}{*}{Estimator}& \multicolumn{2}{c||}{At Convergence} & \multicolumn{1}{c||}{At $K$=1000} & Time Per \\ \cline{2-4}
			& $K$ & Time (sec)  & Time (sec) &  Sample (ms)  \\
			\hline \hline
			MC  & 750$\ \ $ & 670  & 899 & 893\\
			BFS Sharing & 750 & 2503 & 3094 & 3337  \\
			ProbTree    & 750& 105  & 134 & 140  \\
			Lazy Propagation & 750 & 81  & {\bf 106} & {\bf 108}\\
			Recursive (RHH) & 250 & {\bf 38}  & 150& 152  \\
			Recur. Stratified (RSS) & 250 & 45  & 150 & 179  \\
			\hline
		\end{tabular}
		\label{tab:EffDBLP005}
	}
	\hfill
	\parbox{.45\linewidth}{
		\centering
		\centering
		\caption{\small Comparison of running time: {\em BioMine} }
		\begin{tabular} {l|l||l||l}
			\hline
			\multicolumn{2}{c||}{At Convergence} & \multicolumn{1}{c||}{At $K$=1000} & Time Per \\ \cline{1-3}
			$K$ & Time (sec)  & Time (sec) &  Sample (ms)  \\
			\hline \hline
			1500 & 4070  & 2678 & 2660 \\
			1500 & 12723  & 8644 & 8482 \\
			1250 & 600  & 482 & 480 \\
			1500 & 770 & 520 & 513  \\
			1000 & 389   & 389 & 389 \\
			1000 & {\bf 375}  & {\bf 375} & {\bf 375}\\
			\hline
		\end{tabular}
		\label{tab:EffBio}
	}
	\vspace{-4mm}
\end{table*}

\vspace{-1mm}
\subsection{Estimator Efficiency}
\label{sec:eff}
\vspace{-1mm}
We present online running times for all methods both at convergence and at $K$=1000 in Tables \ref{tab:EffLast}-\ref{tab:EffBio}.
The time cost per sample is reported in milliseconds at convergence. We summarize our observations below.

{\bf (1)} {\em At convergence, RSS and RHH are the faster estimators}. RSS is similar or even faster than RHH on large datasets.
However, it is slower than RHH on smaller datasets. {\em ProbTree and LP+ are in the middle range}, and are often comparable to RHH.
{\em MC and BFS Sharing are much slower}. BFS Sharing consumes about $4\times$ running time when compared with MC.

High efficiency of RHH and RSS (one $s$-$t$ query finishes within 400 seconds for all datasets) is due to the following reasons.
(i) When finally conducting sampling in RHH and RSS, the graph has been simplified. This is because certain parts of the graph will no
longer be connected to the source node with the absence of some selected edges. Moreover, these estimators can avoid sampling if
there exists a path connecting $s$ and $t$ with selected edges.
(ii) Due to faster convergence, the sample size $K$ required in {\em RHH} and {\em RSS} is smaller than that for other methods.

ProbTree and LP+ reduces about 80\% of the running time of MC. ProbTree simplifies the graph with pre-computed indexes.
LP+ speeds up the sampling procedure by avoiding unnecessary probing of edges, which is implemented with geometric distribution.
Unlike RSS and RHH, they still require larger number of samples; and hence, their running times are in the middle range.

MC can be terminated early when the target node is visited. But for BFS Sharing, no early termination is possible (see our discussion in Section \ref{sec:bfs_sharing}).
In summary, {\em though BFS Sharing was proposed after MC, it can be $4\times$ slower than MC in regards to $s$-$t$ reliability estimation.
Moreover, ProbTree and LP+ were developed after recursive methods, but both recursive methods are generally more efficient due to their faster
convergence}.

{\bf (2)} When $K$=1000, there is no common winner in running time. RHH, RSS, ProbTree, and LP+ are comparable, and each of them wins once or twice out of all our datasets.
This is because the advantage of requiring less samples (to reach convergence) does not hold here for RHH and RSS. Therefore, {\em comparing these methods at a fixed $K$
(e.g., $K$=1000 as it is done in \cite{ManiuCS17}) is unfair to both RHH and RSS}. We notice that even at a fixed $K$=1000, MC and BFS Sharing are
much slower than other estimators.

{\bf (3)} We observe that {\em except for BFS Sharing, the time cost per sample is about the same at convergence and at $K$=1000, which
indicates that their running time is linear in the sample size $K$}. For BFS Sharing, a small decrease in running time per sample can be viewed
with increasing $K$, which implies that sharing BFS among samples reduces the impact of sample size $K$. However, its online running time is
still not independent of $K$. {\em With larger $K$, the total running time of BFS Sharing still increases (which is evident in Tables~\ref{tab:EffHepU}-\ref{tab:EffBio}).
Therefore, we do not agree with the claim in \cite{ZZL15} that their online running time is independent of $K$}. In fact, as we already
discussed, its theoretical time complexity is still $\bigO(k(n+m))$, which is due to cascading updates (see the paragraph ``Our correction in complexity analysis''
in Section \ref{sec:bfs_sharing}).

\begin{figure}
	\subfigure
	{\includegraphics[scale=0.15,angle=270]{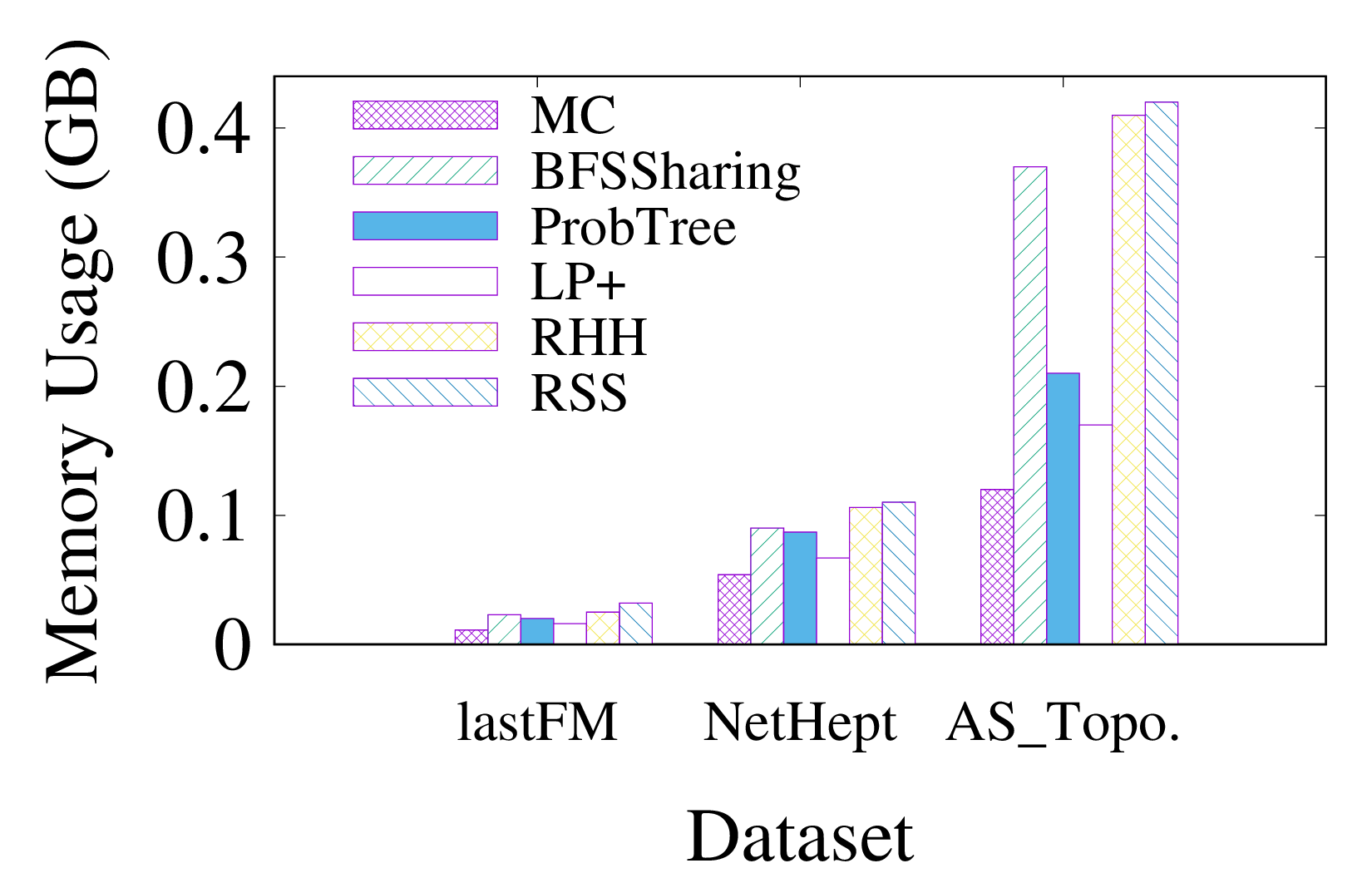}
		\label{fig:mem_small}}
	\subfigure
	{\includegraphics[scale=0.15,angle=270]{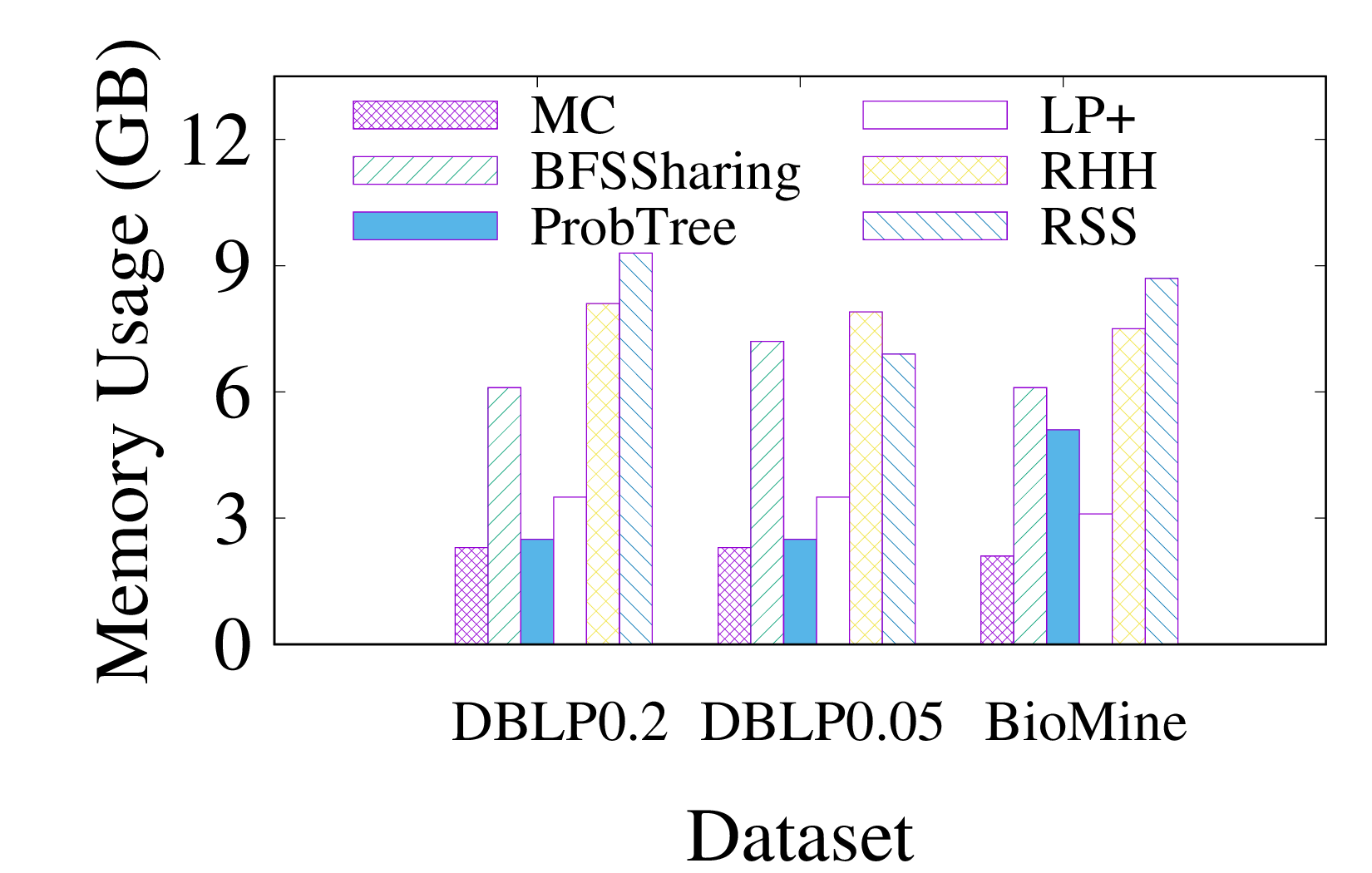}
		\label{fig:mem_big}}
	\vspace{-6mm}
	\captionof{figure}{\small Online memory usage comparison}
	\label{fig:mem}
	\vspace{-5mm}
\end{figure}

\vspace{-1mm}
\subsection{Estimator Memory Usage}
\vspace{-1mm}
Figure \ref{fig:mem} presents the online memory usage for each algorithm. Since the memory usage to reach
convergence is similar to that at $K$=1000, we only report the memory cost at convergence
for every estimator. {\em The general increasing order of memory usage is:
MC $<$ LP+ $<$ ProbTree $<$ BFS Sharing $<$ RHH $\approx $ RSS.}
The memory usage of both RSS and RHH is about $4\times$ to that of MC, and reaches up to 10GB over our larger datasets.
RHH and RSS consume space for recursion stack, selected edge sets and their existence statuses, and for the
simplified graph instance. MC only stores the graph and BFS status variables. Compared to MC, LP+ requires a global
counter for each node and a geometric random instance heap for its neighbors. Both ProbTree and BFS Sharing build indexes,
and due to efficiency reasons, we load their indexes into memory. The index size of BFS Sharing is larger than that of ProbTree, and
BFS Sharing additionally maintains a state vector for each node online. In spite of that, BFS Sharing and ProbTree require less memory
than recursive estimators: RSS and RHH.
\begin{figure*}[t!]
	\vspace{-5mm}
	\centering
	\subfigure[\scriptsize {{\em Index building time}}]
	{\includegraphics[scale=0.205,angle=270]{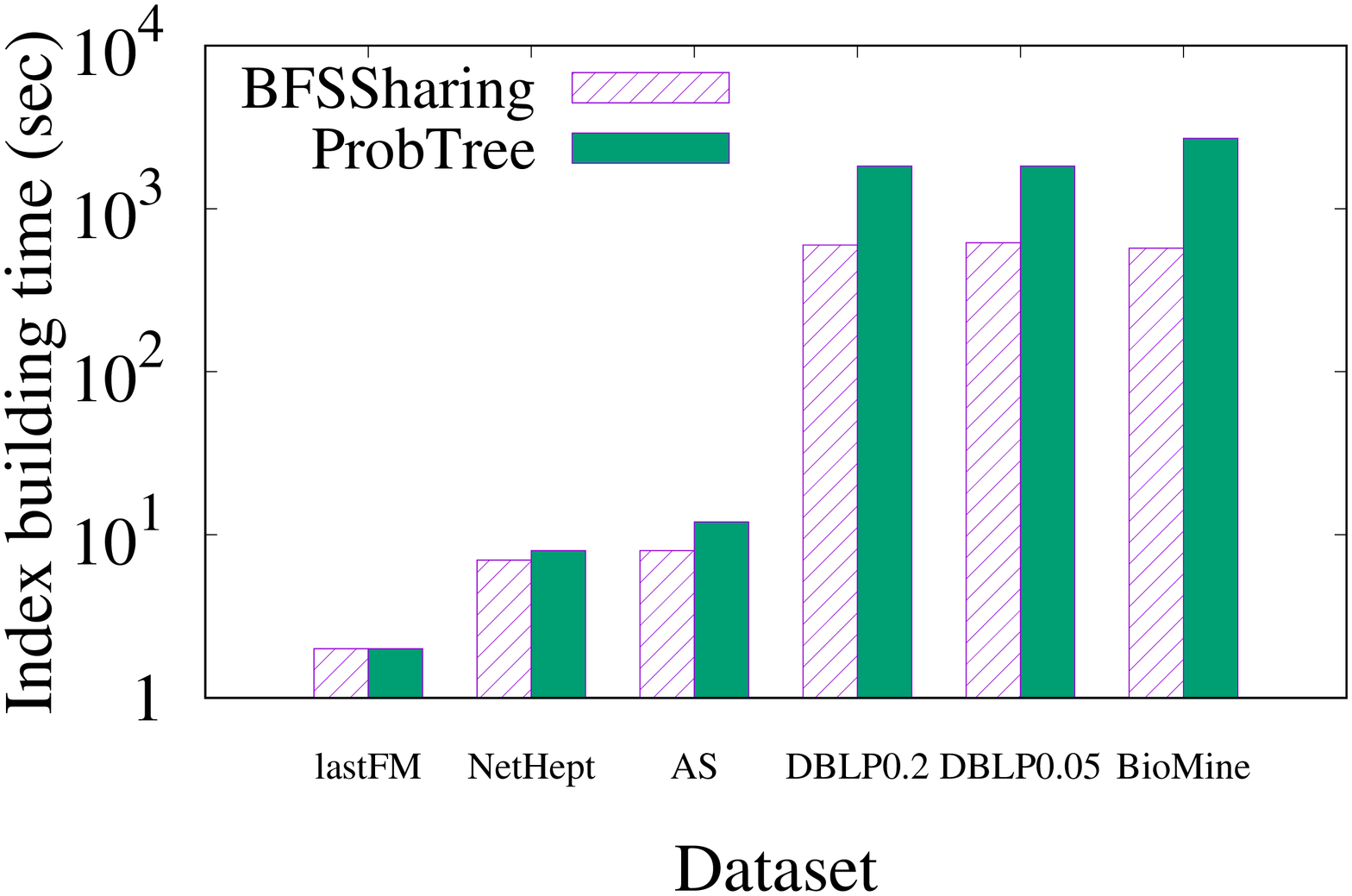}
		\label{fig:build}}
	\subfigure[\scriptsize {{\em Index size}}]
	{\includegraphics[scale=0.205,angle=270]{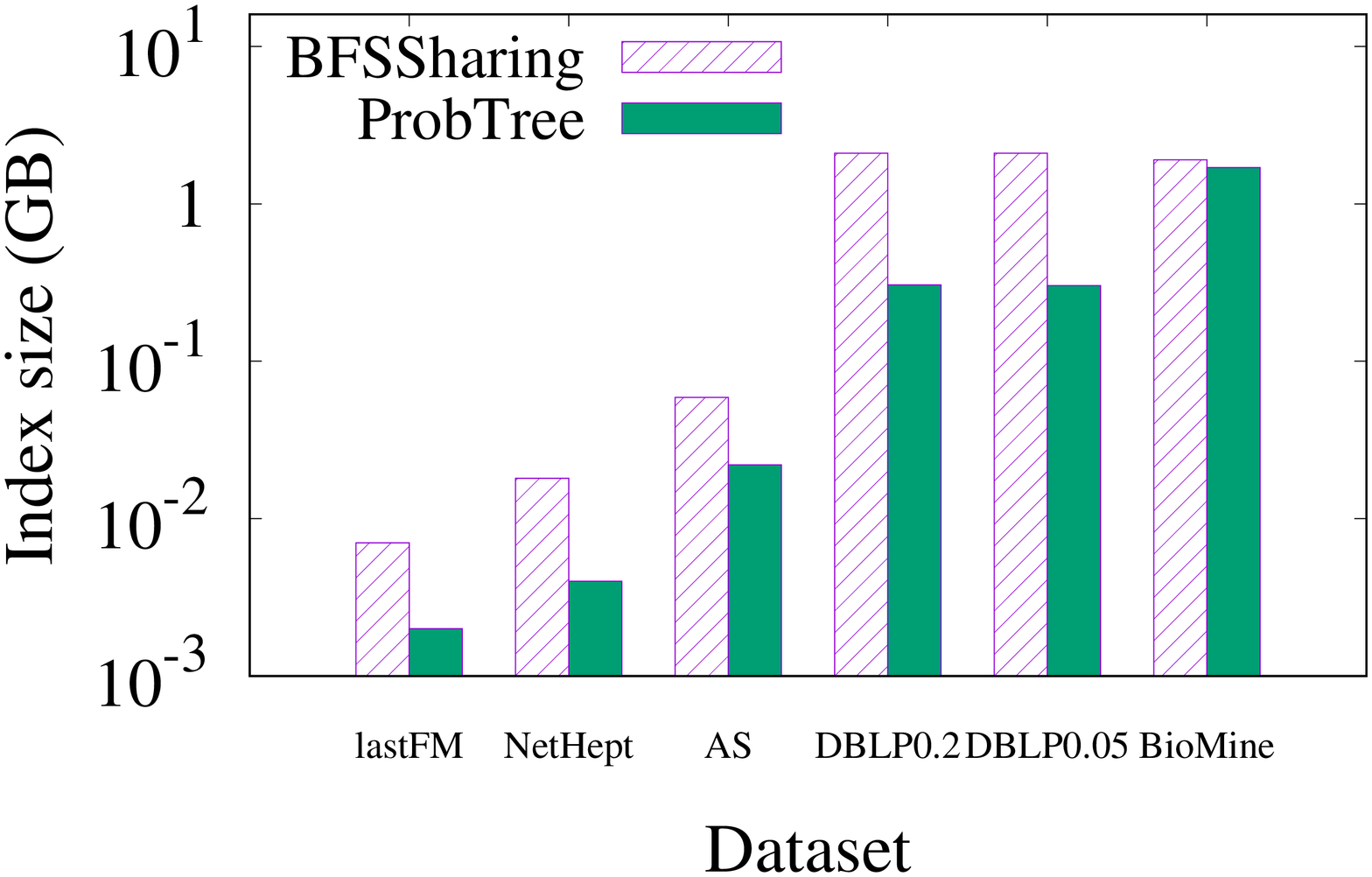}
		\label{fig:size}}
	\subfigure[\scriptsize {{\em Index loading Time}}]
	{\includegraphics[scale=0.205,angle=270]{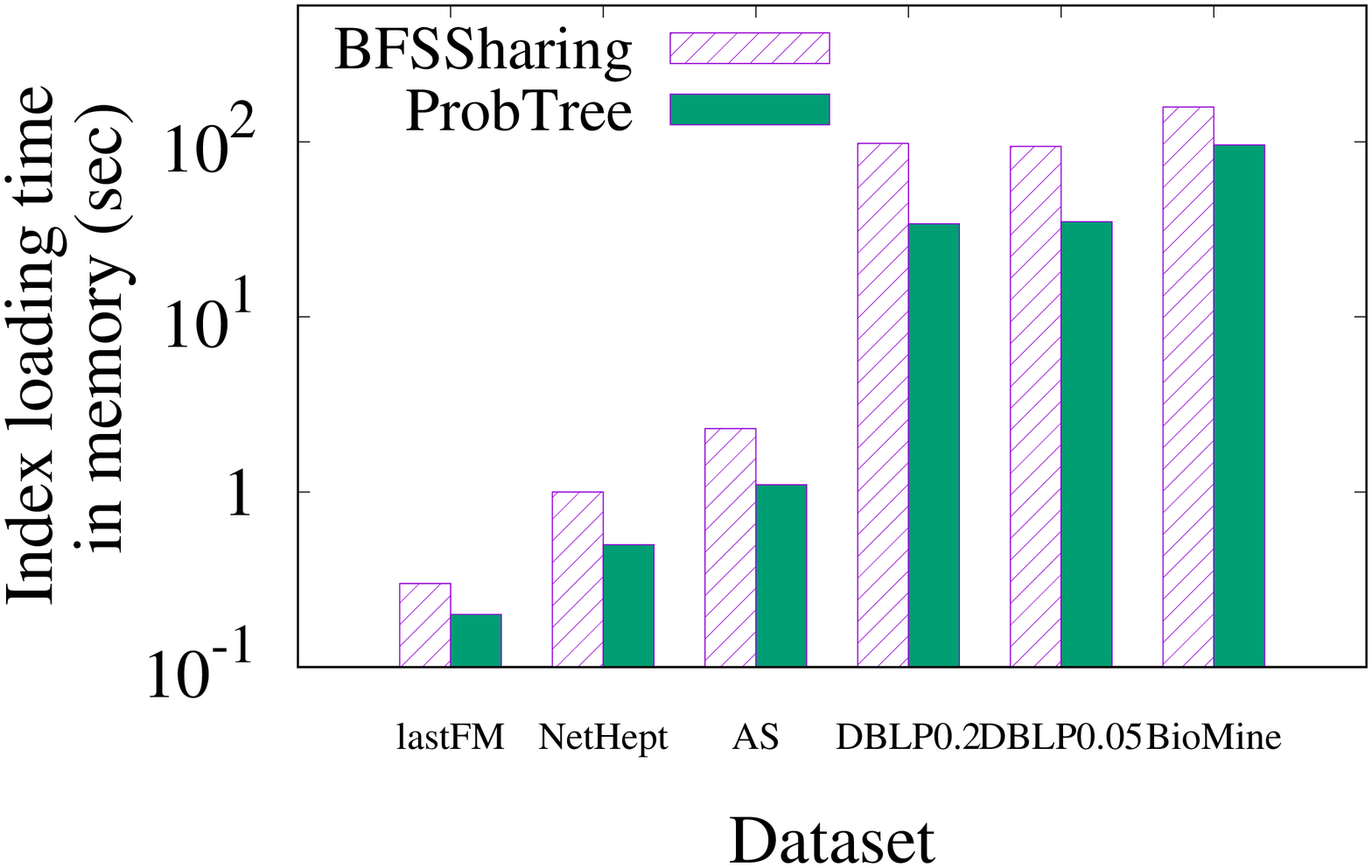}
		\label{fig:load}}
	\vspace{-5mm}
	\caption{\small Offline index cost comparison for {\em ProbTree} and {\em BFS Sharing}}
	\label{fig:add}
	\vspace{-5.5mm}
\end{figure*}

\vspace{-1mm}
\subsection{Indexing Time and Space}
\vspace{-1mm}
Since ProbTree and BFS Sharing rely on graph indexes, we evaluate offline cost for building, storing, and loading of their indexes
(Figure \ref{fig:add})
{\bf (1)} ProbTree index is independent of sample size $K$. It decomposes the graph into ``bags'' and stores them in a tree structure.
Its index size and building time depends on how many bags can be decomposed from the graph, and the depth of index tree.
The maximum index size is 1.8GB over our largest {\em BioMine} dataset, and the corresponding maximum index loading time is 98 seconds,
which is trivial when comparing with the time cost of $s$-$t$ reliability estimation. However, index building requires about one hour
over {\em BioMine}.
{\bf (2)} The index size of BFS Sharing is linear in the sample size $K$.
As $K$ at convergence is not known apriori, a length-$L$ binary vector is attached to each edge in BFS Sharing index to
represent the existence of this edge across $L$ samples. In our experiments, we set $L$=1500 as a safe bound. We find that
{\em the index building time of BFS Sharing is smaller than that of ProbTree}, since former just simply samples each edge $L$ times.
{\em However, its index size can be larger than that of ProbTree, and as a result the index loading time is also higher than that of ProbTree}.
Still index loading time for BFS Sharing is within 200 seconds in most cases. {\em Unlike ProbTree, those edge indexes used by BFS Sharing to answer a
query need to be re-sampled before processing the next query, in order to maintain inter-query independence}. We conduct 1000 successive queries with BFS Sharing and present the additional time cost (per query) for index updating in Table~\ref{tab:up}.

\begin{table}
	\centering
	\caption{\small Additional time cost per query of BFS Sharing index update while answering 1000 successive queries}
	\begin{tabular} {l||c}
		\hline
		Dataset & Time Cost (sec)  \\
		\hline \hline
		{\em lastFM} & 0.02\\
		{\em NetHept} & 0.05\\
		{\em AS\_Topology} &0.11\\
		{\em DBLP\_0.2} & 6.14\\
		{\em DBLP\_0.05} & 5.64\\
		{\em BioMine} &6.98\\
		\hline
	\end{tabular}
	\label{tab:up}
	\vspace{-4mm}
\end{table}

\begin{figure}[tb!]
	\centering
	\subfigure[\scriptsize {\#Samples for convergence}]
	{\includegraphics[scale=0.146,angle=270]{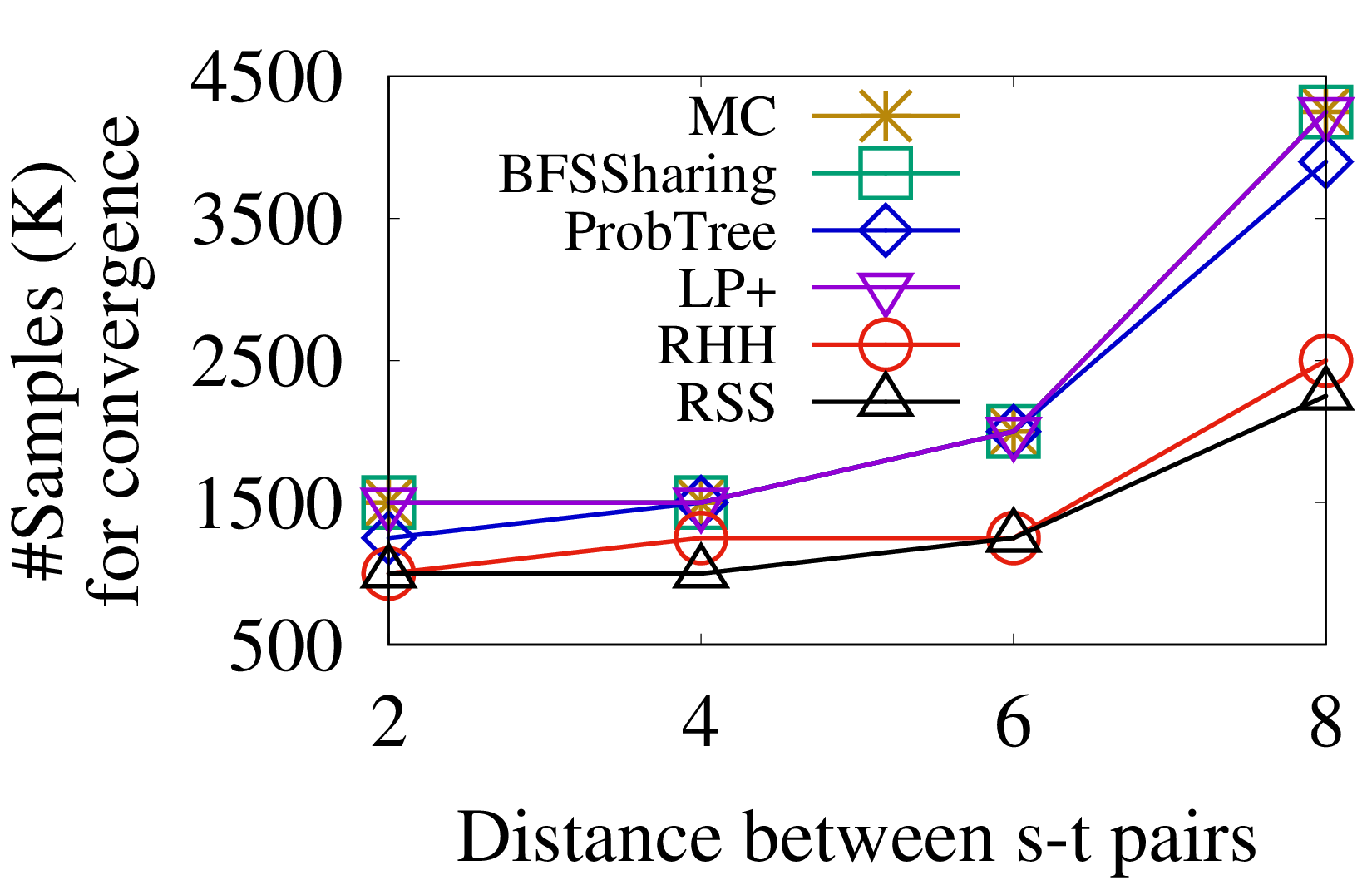}
		\label{fig:hop_k}}
	\subfigure[\scriptsize {Relative error comparison}]
	{\includegraphics[scale=0.146,angle=270]{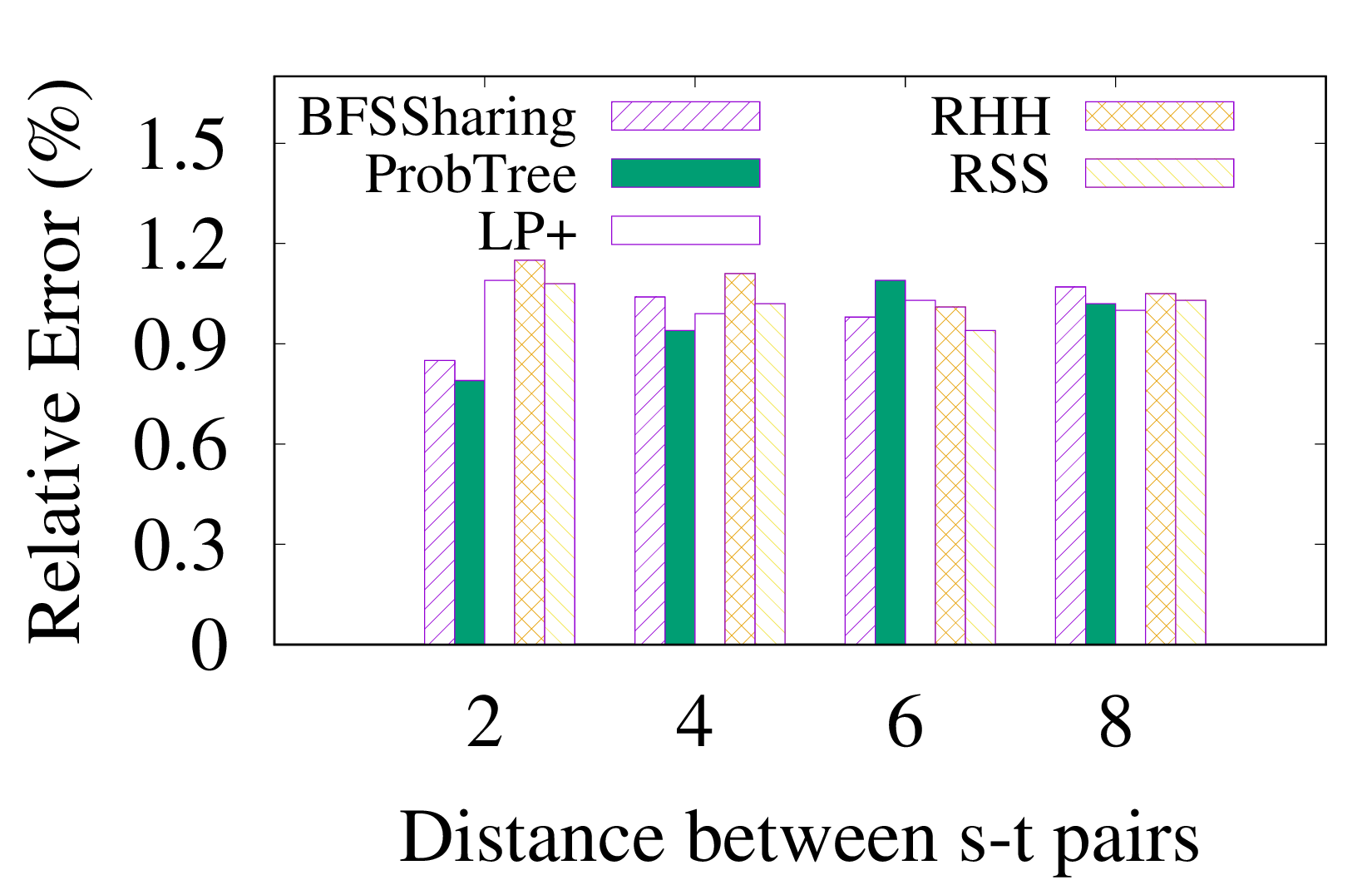}
		\label{fig:hop_error}}
	\vspace{-5mm}
	\caption{\small Sensitivity with varying $s$-$t$ distances}
	\label{fig:hop_sensitivity}
	\vspace{-4mm}
\end{figure}

\vspace{-1mm}
\subsection{Coupling ProbTree with \\ Efficient Estimators}
\label{sec:prob_more}

ProbTree index decomposes the graph and pre-compute the reliability information during index building. When answering online queries, a smaller but equivalent graph is generated from index. The  sampling procedure is conducted on this simplified graph, thus the efficiency is improved.
In previous sections, the ProbTree index is analysed only with MC sampling (as the original paper \cite{ManiuCS17} did).
As presented in Table~\ref{tab:prob}, ProbTree is able to support other estimators and even improve the efficiency by 10-30\%.

\begin{table}
	\centering
	\caption{\small Additional analysis for ProbTree with efficient estimators}
	\begin{tabular} {l||c|c|c}
		\hline
		\multirow{2}{*}{Method}&
		\multicolumn{3}{c}{Running Time at Convergence (sec)} \\ \cline{2-4}
		 & {\em lastFM} & {\em AS\_Topology} & {\em BioMine} \\ \hline \hline
		(1) LP+ & {\bf 0.01} & 20 & 770\\
		(2) ProbTree+LP+ &{\bf 0.01} & {\bf 16} & {\bf 663}\\ \hline
		(1) RHH & 0.004 & 12 & 389\\
		(2) ProbTree+RHH & {\bf 0.003} & {\bf 10} & {\bf 356}\\ \hline
		(1) RSS & 0.026 & 14 & 375\\
		(2) ProbTree+RSS & {\bf 0.012} & {\bf 10} & {\bf 321}\\ \hline \hline
	\end{tabular}
	\label{tab:prob}
	\vspace{-4mm}
\end{table}

\vspace{-1mm}
\subsection{Sensitivity Analysis with $s$-$t$ Pair Distances}
\label{sec:hop}
\begin{figure}[tb!]
	\centering
	\subfigure[\scriptsize {Faster estimators}]
	{\includegraphics[scale=0.146,angle=270]{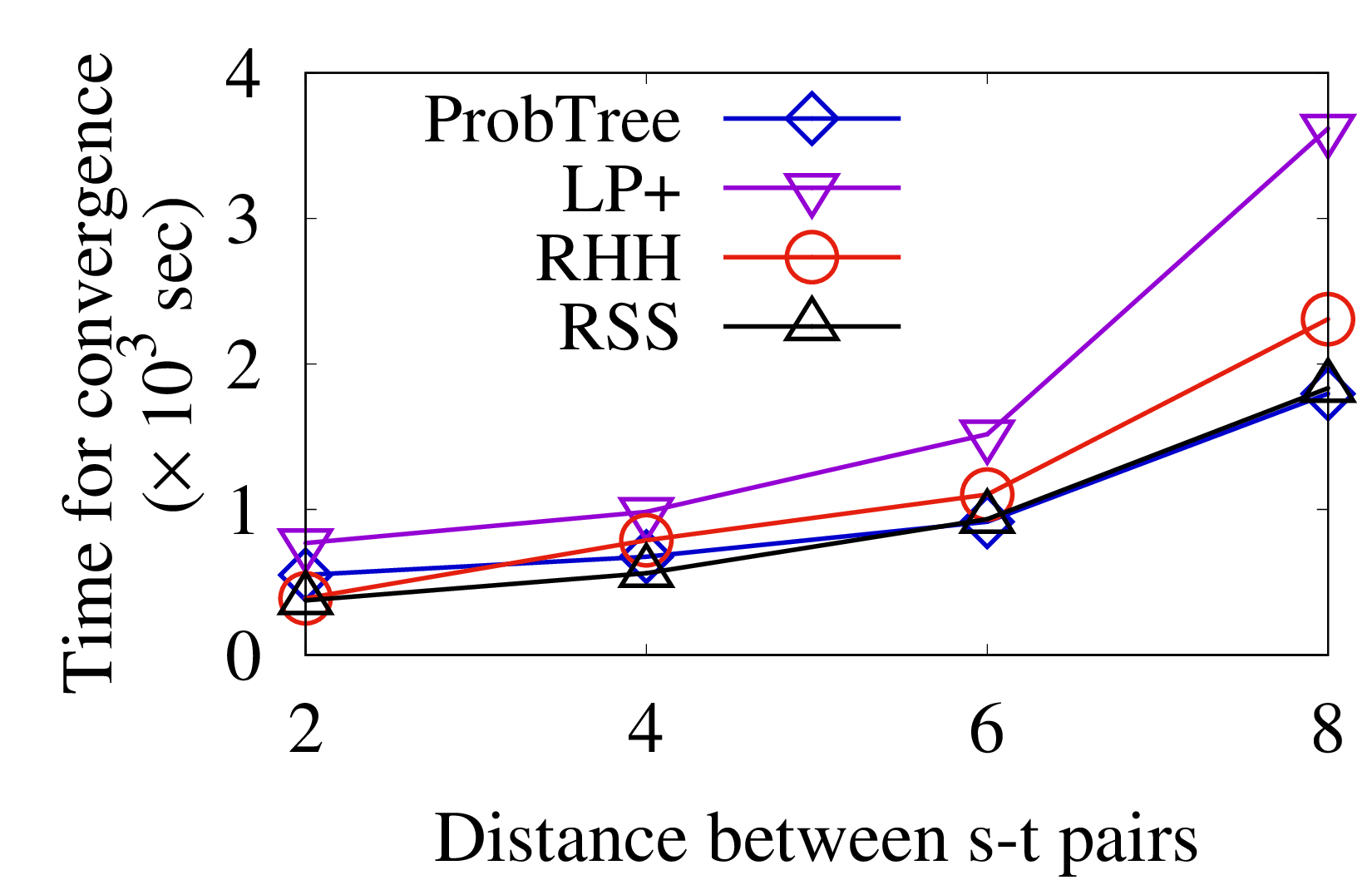}
		\label{fig:hop_time_small}}
	\subfigure[\scriptsize {Slower estimators}]
	{\includegraphics[scale=0.146,angle=270]{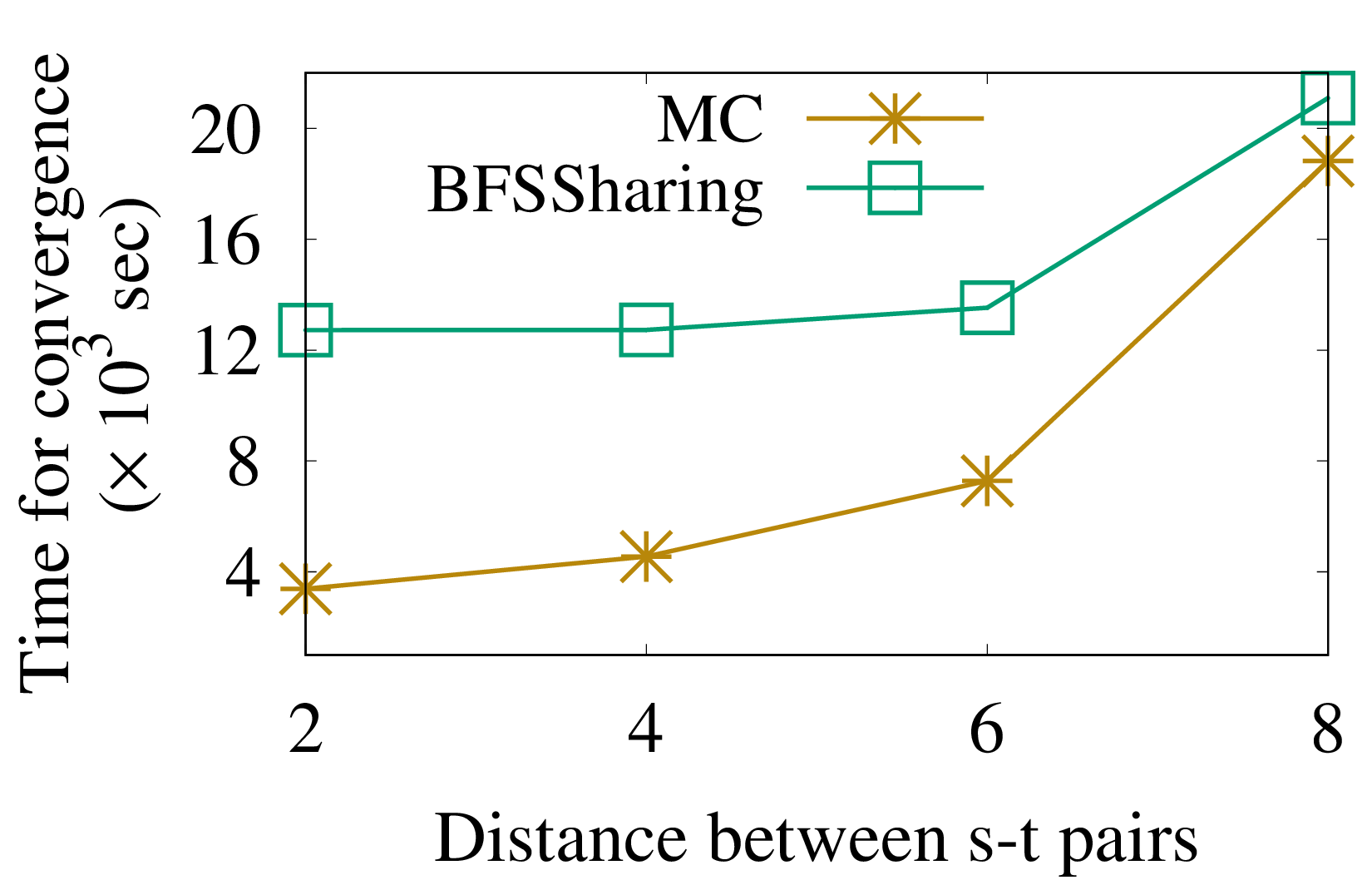}
		\label{fig:hop_time_big}}
	\vspace{-5mm}
	\caption{\small Running time comparison with varying $s$-$t$ distances}
	\label{fig:hop_time}
	\vspace{-5mm}
\end{figure}

In previous experiments, all $s$-$t$ pairs have shortest-path distance $h$ = 2 hops.
In Figures~\ref{fig:hop_sensitivity}, \ref{fig:hop_time} we vary shortest-path distance $h$, and analyze sensitivity
of each metric for every method on the largest {\em BioMine} dataset.
Intuitively, if $s$ and $t$ are close, reliability between them tends to be high. The average reliability via MC at convergence over {\em BioMine}
is 0.4019 for $h$=2, 0.0184 for $h$=4, 0.0041 for $h$=6, and 0.0002 for $h$=8. Such a low reliability when $h>$6 is less important in practical applications.
However, we still study the effect of $h$ up to 8, and summarize our findings.

{\bf (1)} {\em The sample size $K$ necessary for convergence nearly stays the same when $h\leq$6, and sharply increases for $h>$ 6. This holds for all estimators}.
Recall that we adopted the convergence criteria based on index of dispersion, i.e., $\frac{V_K}{R_K}<0.001$ (Section \ref{sec:metric}). This
indicates that when query nodes $s$ and $t$ are not far away ($h\leq$6), the estimator variance decreases in the same order with the decrease in reliability.
However, when further increasing the distance between $s$ and $t$, the reliability decreases much faster.

{\bf (2)} {\em ProbTree and RSS have the best performance in running time with larger $h$}.
With larger $h$, the BFS search depth from $s$ to $t$ increases.
Therefore, the running time of MC, LP+, and RHH increases, even though the sample size
required for convergence remains the same ($h\leq$6).
The running time of BFS Sharing does not change if sample size stays the same,
because it evaluates all nodes' reliability from $s$, regardless of the distance between $s$ and $t$.
ProbTree has a modest increasing rate in running times w.r.t. increase in $h$. Many edges in its pre-computed index are aggregated,
allowing it to build a much simple graph for answering queries. The efficiency of RSS is mostly due to the small sample size required.

{\bf (3)} It can be observed from Figure \ref{fig:hop_error} that {\em the relative error rate is not sensitive to the distance between $s$ and $t$}.

\begin{figure}[tb!]
	\vspace{-1mm}
	\centering
	\subfigure[\scriptsize {Variance}]
	{\includegraphics[scale=0.146,angle=270]{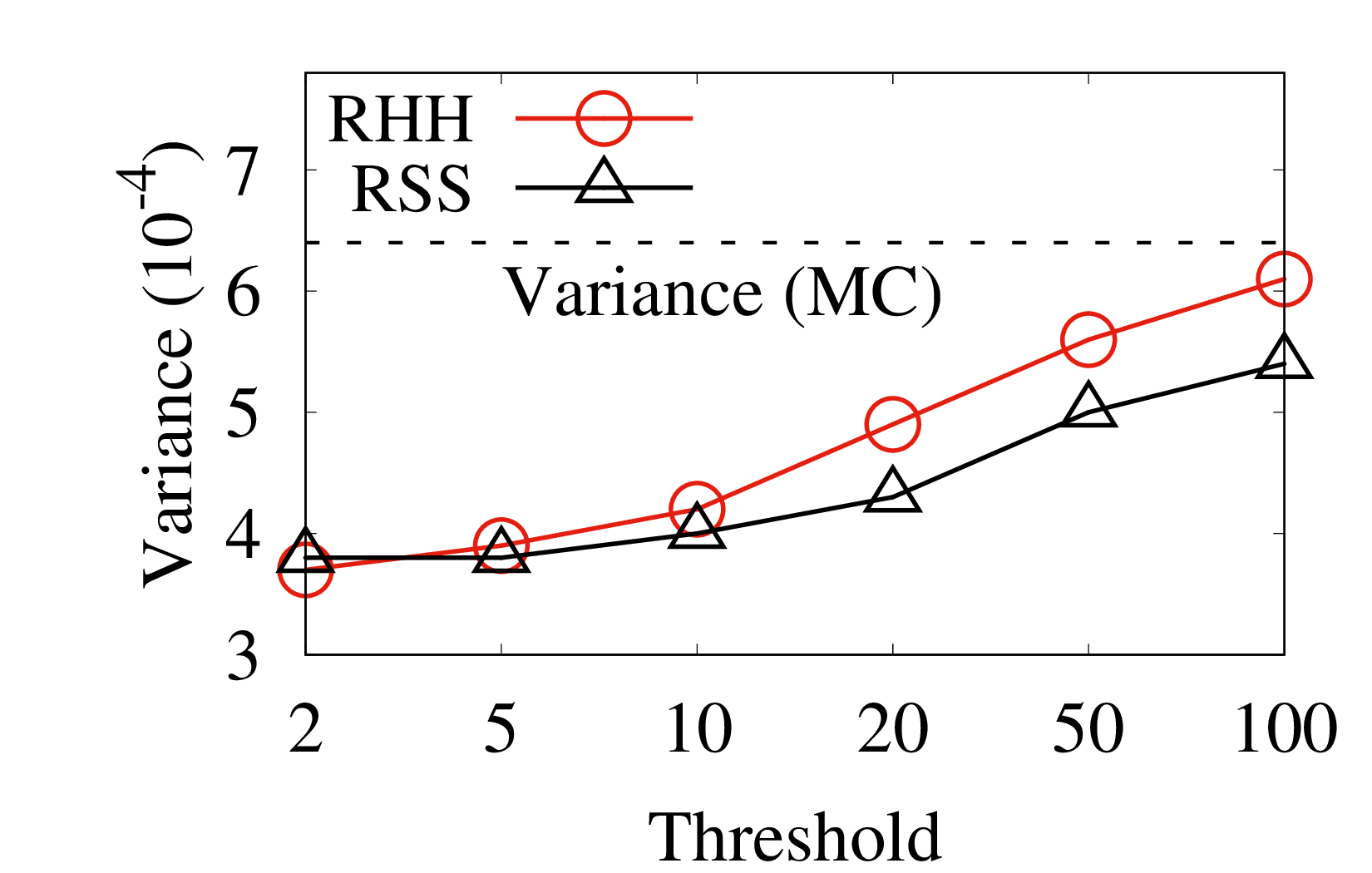}
		\label{fig:var_th}}
	\subfigure[\scriptsize {Running Time}]
	{\includegraphics[scale=0.146,angle=270]{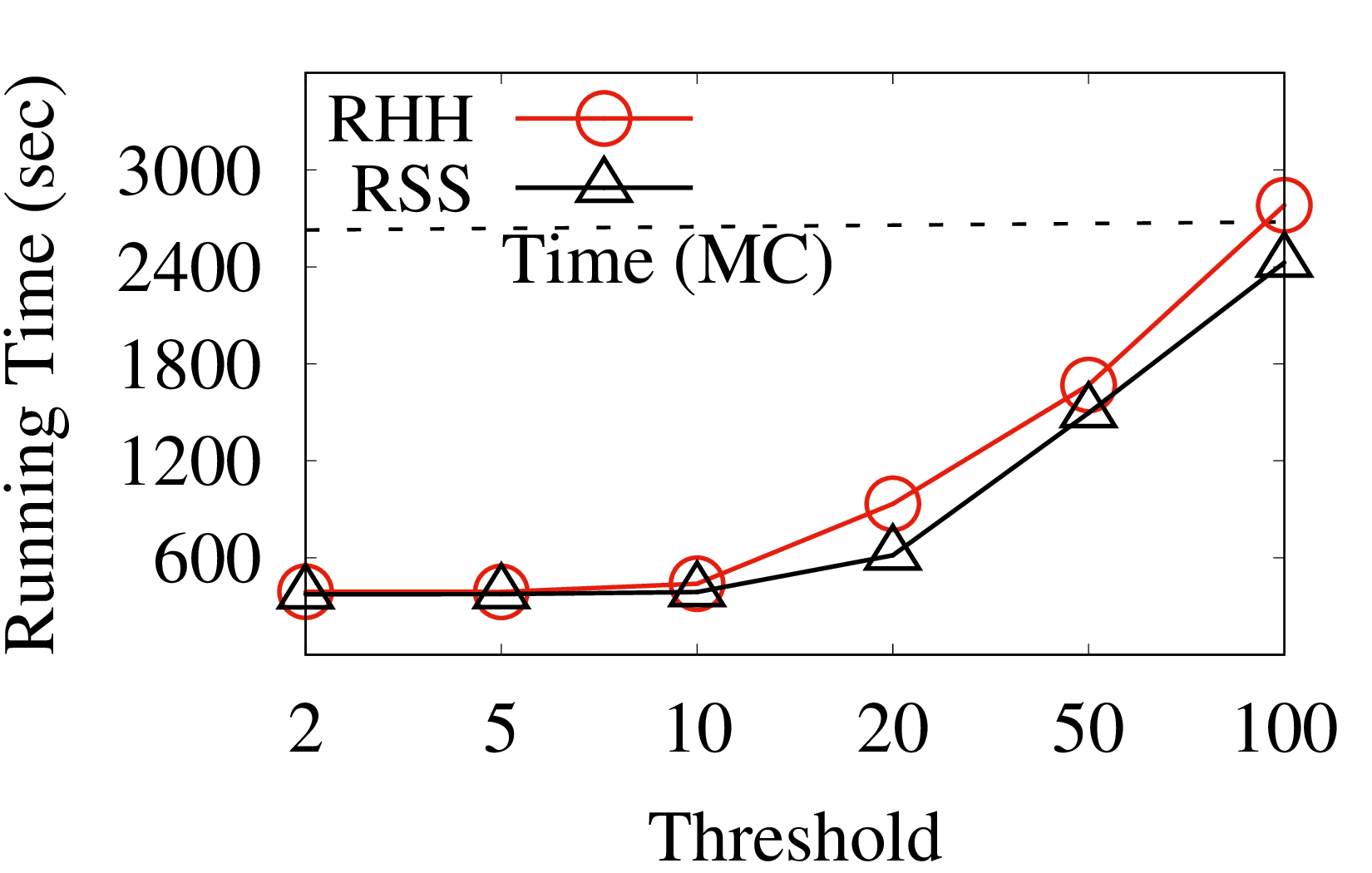}
		\label{fig:time_th}}
	\vspace{-6mm}
	\caption{\small Sensitivity with sample size threshold, $K=1000$, BioMine}
	\label{fig:thr}
	\vspace{-3mm}
\end{figure}

\begin{figure}[tb!]
	\vspace{-1mm}
	\centering
	\subfigure[\scriptsize {Variance}]
	{\includegraphics[scale=0.146,angle=270]{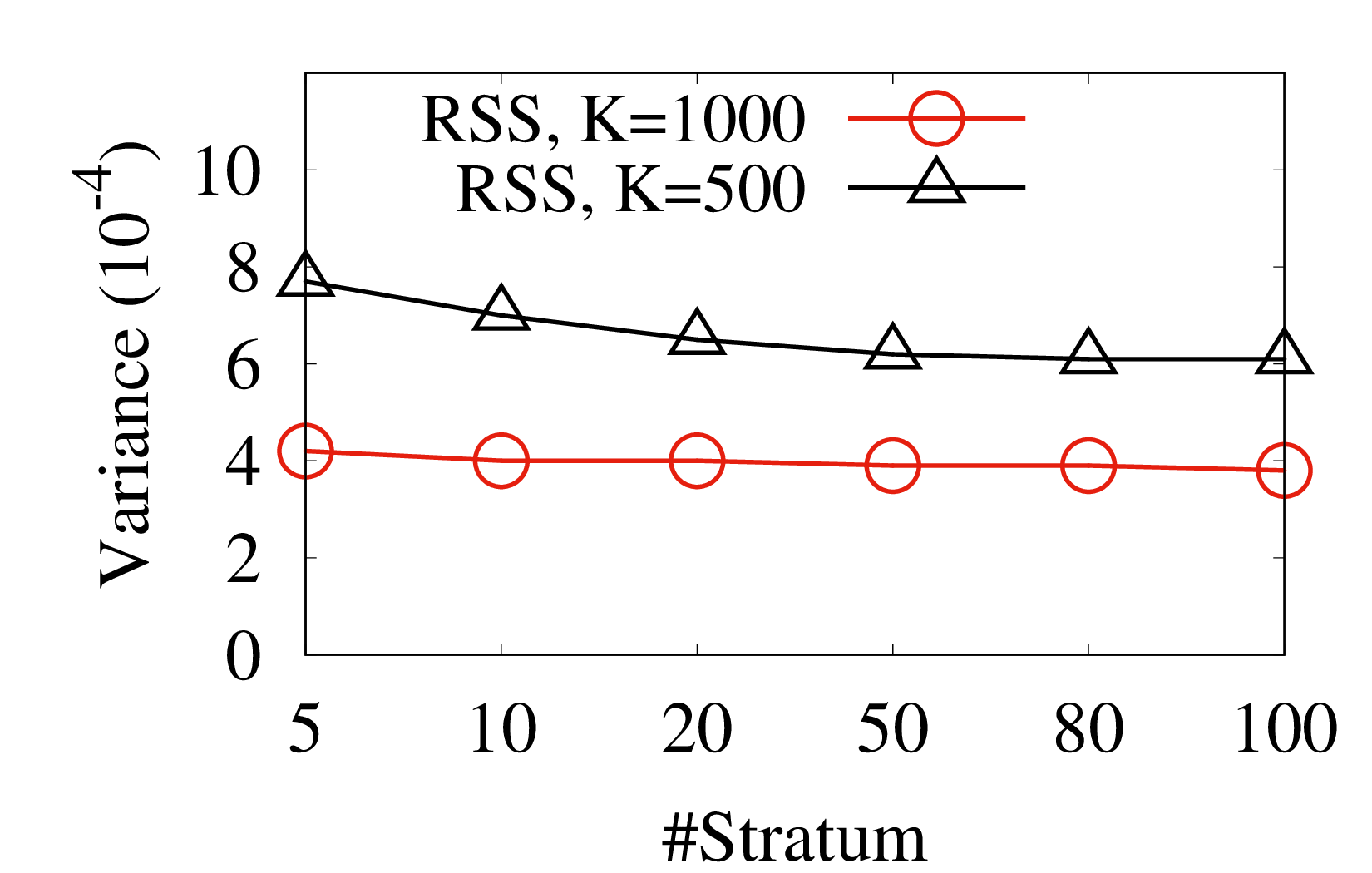}
		\label{fig:var_r_th}}
	\subfigure[\scriptsize {Running Time}]
	{\includegraphics[scale=0.146,angle=270]{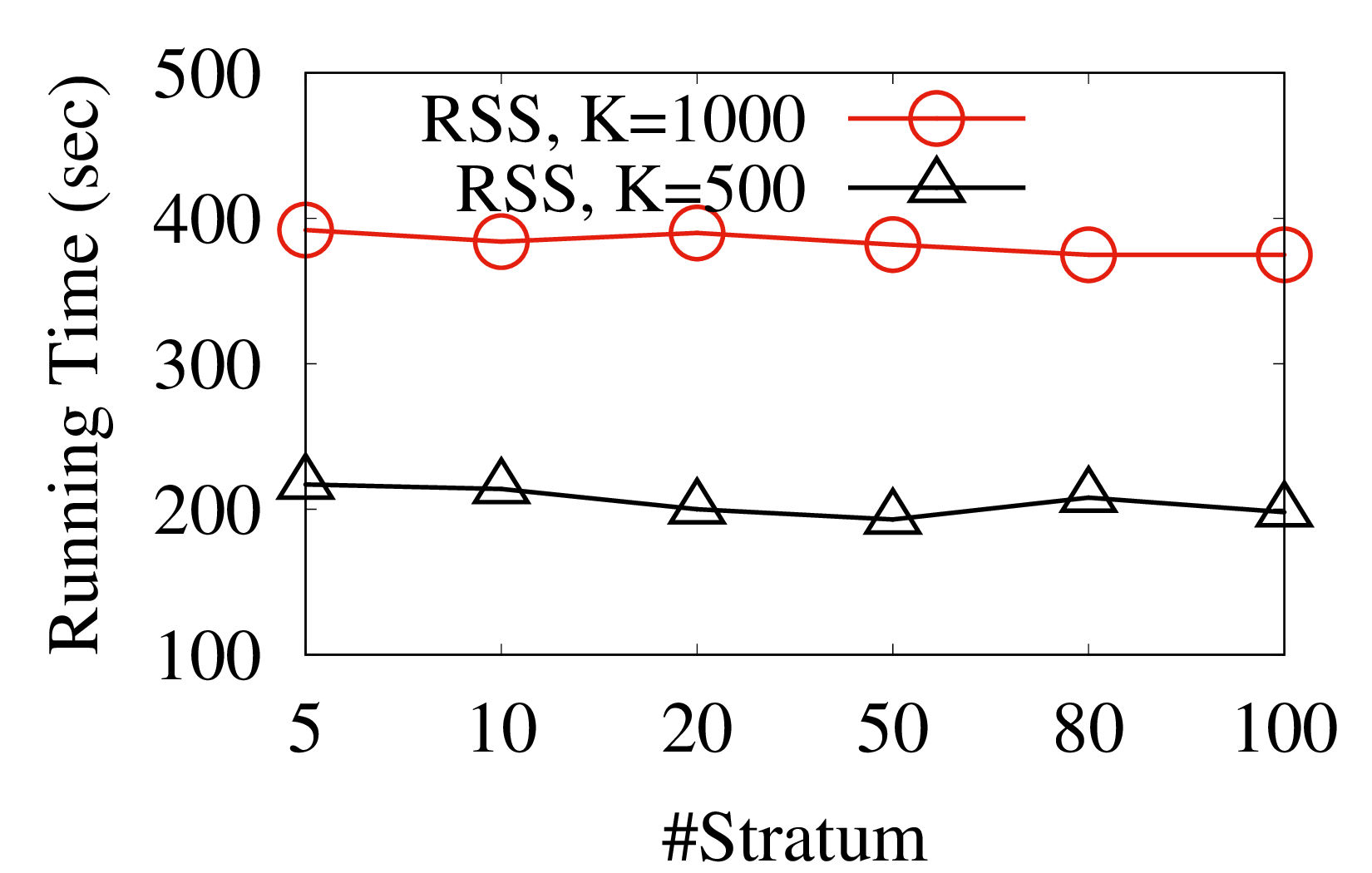}
		\label{fig:time_r_th}}
	\vspace{-6mm}
	\caption{\small Sensitivity with number of stratum $r$, BioMine}
	\label{fig:r}
	\vspace{-5mm}
\end{figure}

\vspace{-1mm}
\subsection{Sensitivity Analysis for Parameters \\ in Recursive Methods}
\vspace{-1mm}
\label{sec:sen_para}

\spara{Sample size threshold for termination.} In both recursive methods: RSS and RHH, the sample size is
divided into multiple parts in each round. If the new sample size is smaller than a given threshold,
the reliability is estimated by a non-recursive method (e.g., MC Sampling) with this sample size.
Several interesting findings from our experiments are listed as follows:

{\bf (1)} As demonstrated in Figure~\ref{fig:thr}, a larger threshold (close to 100) prevents the recursive methods to go deep,
and non-recursive MC sampling is directly applied to a graph comparable to the original one.
This leads to almost the same estimator variance as compared to that of MC sampling.
On the other hand, when the threshold is smaller than 5, the variance of both recursive methods
does not keep decreasing.

{\bf (2)} A smaller threshold also helps reduce the running time of both recursive methods, since the graph can be simplified more before the
non-recursive MC sampling. However, when the threshold is small enough (e.g., smaller than 5), the benefit of decreasing threshold is
not significant. This is because the intermediate graph is often simplified to a single node before reaching this sample size threshold.

{\bf (3)} RSS is more robust with larger threshold than RHH. The suitable thresholds found for two recursive methods are both 5.

\spara{Number of stratum $r$.} The sensitivity of the stratum number $r$ for RSS is shown in Figure~\ref{fig:r}, and is summarized below.

{\bf (1)} {\em In general, the estimator variance of RSS decrease with larger stratum number $r$.} However, this tendency is more obviuos when number of samples $K$ is not sufficient for variance convergence. As discovered in previous experiments, both RSS and RHH reach convergence when $K=1000$. Therefore, at $K=1000$, variance of RSS only has a slight reduction with larger $r$.
We observe that by setting $r=50$, the RSS estimator variance reduces by 25\%, when $K=500$. {\em Moreover, when $r$ is larger than 50, the variance of RSS does not keep decreasing.}

{\bf (2)} The running time of RSS is not sensitive to $r$. Therefore, we adopt $r=50$ as our default setting.

\begin{table}
	\vspace{-1mm}
	\centering
	\caption{\small Summary and recommendation}
	\begin{tabular} {l||c|c|c|c}
		\hline \hline
		\multicolumn{5}{c}{Online Query Processing} \\ \hline \hline
		Method & Variance & Accuracy & Running Time & Memory  \\
		\hline
		MC & $\bigstar$ & $\bigstar \bigstar \bigstar$ & $\bigstar \bigstar$  & $\bigstar \bigstar \bigstar \bigstar$ \\
		BFS Sharing & $\bigstar$& $\bigstar \bigstar \bigstar$ & $\bigstar$ & $\bigstar \bigstar$ \\
		ProbTree &$\bigstar$ & $\bigstar \bigstar \bigstar$ & $ \bigstar \bigstar \bigstar$ & $\bigstar \bigstar \bigstar$ \\
		LP+ &$\bigstar$& $\bigstar \bigstar \bigstar$ & $\bigstar \bigstar \bigstar$ & $\bigstar \bigstar \bigstar \bigstar$ \\
		RHH &$\bigstar \bigstar \bigstar \bigstar$  & $\bigstar \bigstar \bigstar \bigstar$ & $\bigstar \bigstar \bigstar \bigstar$ & $\bigstar$ \\
		RSS &$\bigstar \bigstar \bigstar \bigstar$ & $\bigstar \bigstar \bigstar \bigstar$ & $\bigstar \bigstar \bigstar \bigstar$ & $\bigstar$  \\
		\hline \hline
		\multicolumn{5}{c}{Index-related} \\ \hline \hline
		\multirow{2}{*}{Method} & Time & Time & Time & \multirow{2}{*}{Size}  \\
		& (build) & (load) & (update) & \\
		\hline
		BFS Sharing & $\bigstar \bigstar \bigstar \bigstar$& $\bigstar \bigstar \bigstar$ & $\bigstar$ & $ \bigstar \bigstar \bigstar$ \\
		ProbTree &$\bigstar \bigstar \bigstar $ & $\bigstar \bigstar \bigstar \bigstar$ & $\bigstar \bigstar \bigstar \bigstar$ & $\bigstar \bigstar \bigstar \bigstar$ \\
		\hline \hline
	\end{tabular}
	\label{tab:sum}
	\vspace{-3mm}
\end{table}

\begin{figure}[tb!]
	\vspace{-2mm}
	\centering
	\includegraphics[scale=0.34]{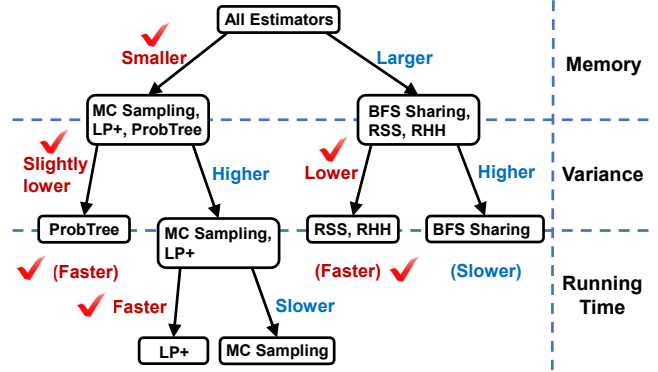}
	\vspace{-7mm}
	\caption{\small The decision tree for selecting a proper reliability estimator under different scenarios}
	\label{fig:rec}
	\vspace{-6mm}
\end{figure}

\vspace{-2mm}
\section{Discussion and Conclusion}
\label{sec:conclusion}
\spara{Summary.} In this work we investigated six state-of-the-art sequential algorithms for $s$-$t$ reliability estimation,
corrected certain issues in these algorithms to further improve their performance, and conducted a thorough experimental evaluation.

For estimator variance, both recursive methods: RHH and RSS exhibit significantly better performance than other four MC-based approaches: MC,
BFSSharing, ProbTree, and LP+. Methods in the same category share very similar variance. In general, RSS is the best regarding variance,
and achieves fastest convergence. To achieve convergence, there is no single sample size $K$ that can be used across various
datasets and estimators. Usually recursive methods require about 500 less samples than MC-based methods on the same dataset.

For accuracy, all methods have similar relative error ($<$1.5\%) at convergence. If $K$ is set as 1000 for all estimators,
some of them might not reach convergence, thus their relative errors can further be reduced by using larger $K$ until convergence.

For efficiency, RHH and RSS are the fastest when running time is measured at convergence. When $K$ is set as 1000, there is no common winner
in terms of running time. Overall, the running times of RHH, RSS, ProbTree, and LP+ are comparable. BFSSharing is
$4\times$ slower than MC, since it estimates all nodes' reliability from the source node.

The memory usage ranking (in increasing order of memory) is: MC $<$ LP+ $<$ ProbTree $<$ BFSSharing $<$ RHH $\approx $ RSS.

\vspace{-1mm}
\spara{Recommendation.} Table~\ref{tab:sum} summarizes the recommendation level of each method according to different performance metrics. The scale is from 1 to 4 stars, and larger star number stands for higher ranking. Clearly, there is no single winner. Considering various trade-offs, in conclusion we recommend ProbTree for $s$-$t$ reliability estimation.
It provides good performance in accuracy, online running time, and memory cost. Its index can slightly reduce the variance,
compared to other MC-based estimators.
Notably, we adopted MC as ProbTree's reliability estimating component (as the original paper \cite{ManiuCS17} did).
However, one can replace this with any other estimator (e.g., recursive estimators: RHH and RSS)
to further improve ProbTree's efficiency and to reduce its variance (as demonstrated in Section~\ref{sec:prob_more}). 

The decision tree shown in Figure~\ref{fig:rec} demonstrates our recommended strategy for estimator selection
under different constraints. Following the branch with red tick, the better estimator(s) under the current condition 
can be determined. Notice that the path from the root to the leaf of ProbTree consists of all red ticks,
indicating its applicability and trade-off capacity considering various factors.

\vspace{-2mm}
{
\bibliographystyle{abbrv}
\bibliography{ref,ref_rel,ref_leroy}
}


\end{document}